\documentclass[twoside,12pt,a4paper]{article}
\usepackage[top=25mm,bottom=25mm,left=25mm,right=25mm]{geometry}
\usepackage{amsmath,amssymb,color,graphics,graphicx,bm}
\usepackage{epstopdf}% EPS to PDF
\usepackage{graphicx,array}
\usepackage{dcolumn}% Align table columns on decimal point
\usepackage{setspace}
\usepackage[utf8]{inputenc}
\usepackage{authblk}

\usepackage{scalerel}
\usepackage{hyperref}
\usepackage{relsize}
\usepackage{multicol}
\usepackage{blindtext, subfig}
\usepackage{float} % For option 'H' in \begin{figure}[H]
\usepackage{natbib} % to use \cite, \citep, \citet and \citealt
\usepackage[toc]{appendix}
\usepackage{enumerate}
\numberwithin{equation}{section}

\pdfoutput=1
\title{Contact mechanics of adhesive beams. \\ Part II: Low to high indentation.}
\date{}
\author{V. S. Punati\thanks{Electronic address: punati@iitk.ac.in; punati.iitk@gmail.com} , I. Sharma and P. Wahi}
\affil{Mechanics \& Applied Mathematics Group, Department of Mechanical Engineering, \\ Indian Institute of Technology Kanpur,  Kanpur, Uttar Pradesh, India - 208016.}

\begin{document}
\maketitle
%% Abstract
%%\doublespacing

\begin{abstract}
In this article, we extended our approach, that is mentioned in part 1, to model the indentation of an adhesive beam by a rigid cylindrical punch. We considered clamped and simply supported beams for this study. We first modeled these beams as infinite length elastic layers which obey the kinetic and kinematic constraints imposed by the end supports. The adhesion effects are considered via the Dugdale-Barenblatt model based adhesive zone model. Solving the governing equations of this infinite layer along with its boundary conditions, we obtain a set of coupled Fredholm integral equations of first kind. These integral equations are then solved employing the collocation technique. The results obtained are then compared with finite element (FE) simulations and the previously published results for the non-adhesive case. We also obtained the results for the well-known Johnson-Kendall-Roberts (JKR) approximation of the contact. Finally we investigated the effect of various adhesive strengths on the contact parameters and showed the transition of the results from `Hertzian' to `JKR' approximation. 
\end{abstract}


\section{Introduction}
This article forms the second part of a three part study on the indentation of adhesive beams. In this second part we greatly expand the theoretical framework of part I \citep{Punati2017}, hereafter Paper I. 

Hertz, in 1882, proposed a theory of contact of non-adhesive spheres which addressed the indentation of three dimensional half-spaces by rigid axi-symmetric punhces. Later, \cite{johnson1971surface} and, \cite{derjaguin1934untersuchungen} and \cite{derjaguin1975effect} proposed theories for adhesive axi-symmetric indentation. Finally, \cite{Maugis1992adhesion} demonstrated that JKR \citep{johnson1971surface} and DMT \citep{derjaguin1975effect} theories are two limits of one general theory. Some of the above theories and the corresponding two-dimensional versions are discussed in \cite{alexandrov2001three}, \cite{galin2008contact}, \cite{Gladwell1980contact}, \cite{Johnson1985contact}, \cite{Hills1993mechanics}, and \cite{goryacheva1998contact}. 

\begin{figure}[h!]
\centering
\includegraphics[scale=1]{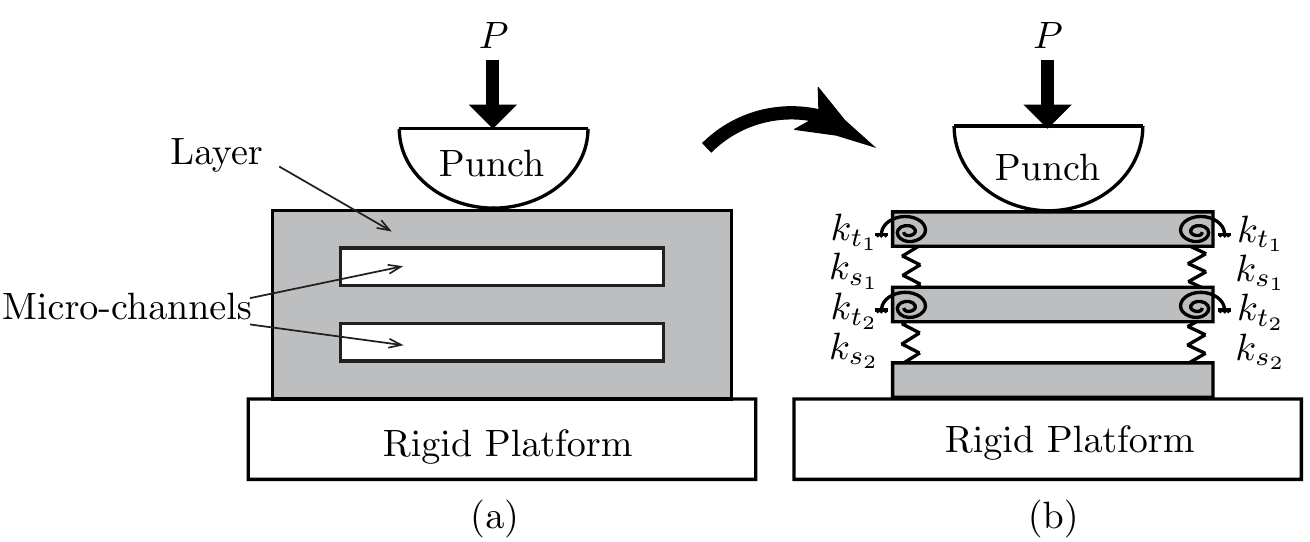}
\caption{(a) Structural adhesive designed by \cite{Arul2008bioinspired}. (b) Mechanical model of the structural adhesive in (a) employing an interconnected stack of adhesive beams. The rigidity of the vertical walls is modeled through torsional (stiffness $k_t$) and vertical translational (stiffness $k_s$) springs, as shown. The system is indented by a rigid punch, pressed down by the force $P$.}
\label{fig:struct_adhesive_model}
\end{figure}

In recent years, indenaion of thin adhesive structures have attracted the attention of researchers because of their applications in electronics and computer industry, see e.g. \cite{barthel2007adhesive} and \cite{Dalmeya2012contact}. Some of the designs for these structural adhesives are inspired from biology, such as the one proposed by \cite{Arul2008bioinspired}; see Fig.~\ref{fig:struct_adhesive_model}. The theoretical modeling and characterisation of such structural adhesives is of great interest. Paper I presented a step towards the modeling of structural adhesives of the type shown in Fig.~\ref{fig:struct_adhesive_model}, by investigating the indentation of adhesive beams resting on flexible supports.

Indentation studies on non-adhesive beams were pursued in the past by \cite{Keer1983smooth}, and \cite{Sankar1983}. They employed integral transforms and Fourier series, respectively. Adhesion was not considered. Recently \cite{Kim2014}, revisited the indention of non-adhesive beams  through approximate techniques. However, extending the methods of these papers to adhesive beams pose difficulties, as they involve several iterated integral transforms and/or asymptotic matching. Paper I presented a formulation which could address non-adhesive and adhesive contact of beams within the same framework.

In Paper I we investigated indentation by a rigid cylindrical punch of non-adhesive and adhesive beams on flexible end supports. However, the mathematical model assumed that, during indentation, the displacement of the beam's bottom surface could be approximated by the deflection of the corresponding Euler-Bernoulli beam under the action of a point load. This approximation limited the application of Paper I's approach to indentations where the extent $a$ of the contact region was less than or equal to the thickness $h$ of the beam.  Here we release the assumption of Paper I in order to extend our framework to indentation with large contact areas. This leads naturally to a set of dual integral equations for the unknown contact pressure and the displacement of the beam's bottom surface. 

In this article, we restrict ourselves to clamped and simply supported beams which, as shown in Paper I, bound the range of behaviors displayed by beams on flexible supports.
\begin{figure}[h!]
\centering
\subfloat[][]{\includegraphics[width=0.25\linewidth]{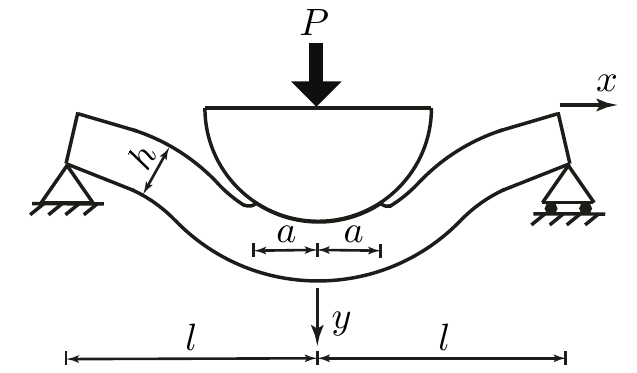}} 
\subfloat[][]{\includegraphics[width=0.5\linewidth]{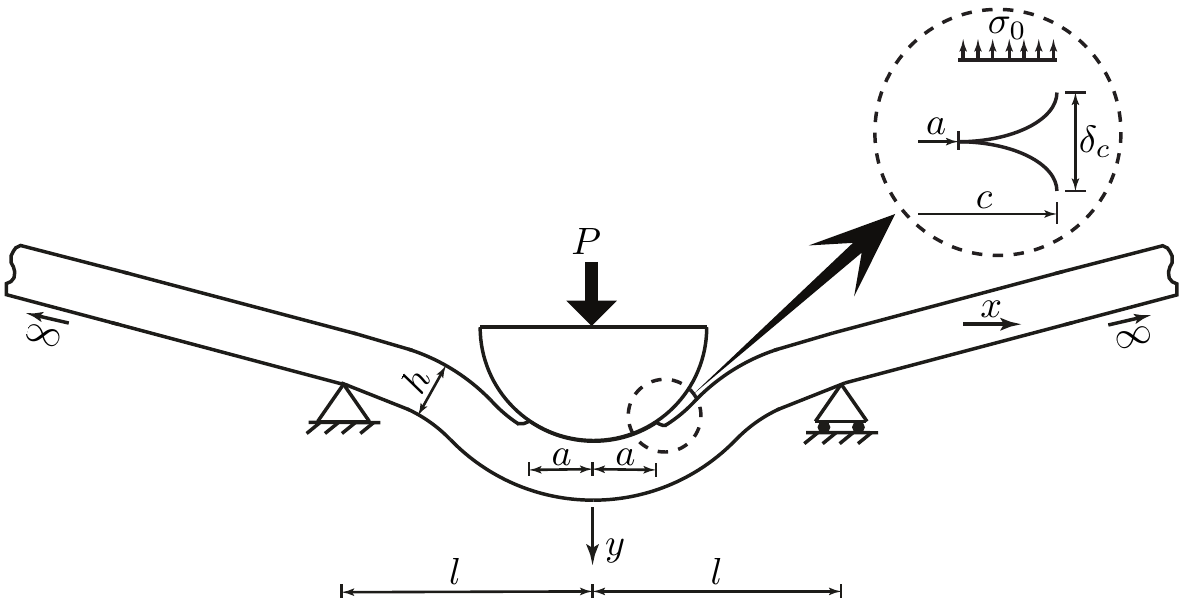}}
\caption{ (a) Indentation by a rigid circular punch with an adhesive simply supported beam. (b) Mathematical model for the simply supported beam indentation by extending the beam to infinity along its slope near the ends. The inset shows the adhesive zone model employed in our mathematical formulation.}
\label{Thinbeams:Indentation} 
\end{figure}
As in Paper 1, adhesion is modeled through the adhesive-zone model, which allows us to study the JKR \citep{johnson1971surface} and DMT \citep{derjaguin1975effect} approximations of adhesive contact by varying a parameter that regulates adhesive strength. Additionally, by setting adhesive strength to zero we obtain results for non-adhesive contact. 

This paper is organised as follows: We first present the mathematical model, which leads to a set of dual integral equations in terms of the contact pressure and the vertical displacement of the beam's lower surface. This is followed by non-dimensionalization and the formulation of the corresponding numerical algorithm to solve the integral equations. We then briefly discuss the FE model employed to study non-adhesive beam indentation. Next, we present and discuss results for different types of adhesive and non-adhesive contacts. Finally, we compare our predictions with preliminary experiments.

\section{Mathematical model}
\label{sec:Thinbeam_Mathematical_model}
We begin, as in Paper I, by extending the beams of Fig.~\ref{Thinbeams:Indentation}(a) beyond the supports to infinity; see Fig.~\ref{Thinbeams:Indentation}(b). This extension is done in keeping with the kinematic and kinetic constraints imposed by the supports. Thus, the beam is extended linearly along the slope at the supports. The beams may now be represented as a linear elastic layer of infinite length with thickness $h$, and with Young's modulus $E$  and Poisson's ratio $\nu$. The top and bottom surfaces of the beam are frictionless. The corresponding elasticity problem is now solved, the details of which are in Sec.~2 of Paper I.  This yields the vertical displacement of the beam's top surface $(y=0)$ and the normal traction acting on the beam's bottom surface $(y=h)$ as, respectively,
\begin{flalign}
&& v \left( x, 0 \right) &=  - \frac{2}{\pi E^*} \int\limits_{0}^{\infty} \bar{P}_c \left( \xi \right)\,  K_1 \left( \xi,  x \right) \, d \xi  + \frac{1}{\pi} \int\limits_{0}^{\infty} \bar{v}_b \left( \xi \right) \,  K_2 \left( \xi,  x \right) \, d \xi \label{thinbeam:toplayer_v_x} &&\\
\text{and}
&&\tau_{yy} \left( x, h \right) &= \frac{1}{\pi} \int\limits_{0}^{\infty} \bar{P}_c \left( \xi \right) \,  K_2 \left( \xi,  x \right) \, d \xi  + \frac{{E^*}}{\pi} \int\limits_{0}^{\infty} \bar{v}_b \left( \xi \right) \,  K_3 \left( \xi,  x \right) \, \cos{\xi x} \, d \xi, \label{thinbeam:bottomlayer_tyy_x} && 
\end{flalign}
where
\begin{equation}
\bar{P}_c \left( \xi \right) = \int\limits_{-\infty}^{\infty} - P_c \left( t \right) \cos{\xi t } \, dt 
\quad \text{and} \quad 
\bar{v}_b \left( \xi \right)  = \int\limits_{-\infty}^{\infty} v_b \left( t \right) \cos{\xi t } \, dt 
\end{equation}
are the Fourier transforms of the normal force $P_c \left( t \right)$ acting on the beam's top surface and the vertical displacement $v_b \left( t \right)$ of the beam's bottom surface, respectively, while
\begin{flalign}
&& 
 K_1 \left( \xi,  x \right) &= \frac{ \sinh^2{\xi h}}{ \xi \left( \xi h + \sinh{\xi h } \cosh{\xi h} \right) } \, \cos{\xi x}, \quad
 K_2 \left( \xi, x \right) = \frac{\sinh{ \xi h} + \xi h \cosh{ \xi h } }{ \xi h + \sinh{ \xi h} \cosh{ \xi h }} \, \cos{\xi x} \nonumber && \\
\text{and} \quad
&&  K_3 \left(  \xi, x \right) &= \frac{\xi}{2} \cdot \frac{\sinh^2{\xi h} - \xi^2 h^2 }{ \xi h + \sinh{ \xi h} \cosh{ \xi h }} \, \cos{\xi x}. \nonumber &&
\end{flalign}
For more details see Appendix \ref{sec:Fourier_tranforms}.

To ease non-dimensionalization we expand \eqref{thinbeam:toplayer_v_x} and \eqref{thinbeam:bottomlayer_tyy_x} by employing definitions of $\bar{P}_c $ and $\bar{v}_b$ to obtain: 
\begin{flalign}
 && v \left( x, 0 \right) &= \frac{2}{\pi E^*} \int\limits_{0}^{\infty} \int\limits_{-\infty}^{\infty} P_c \left( t \right) \cos{\xi t } \, dt \, K_1 \left( \xi,  x \right)  \, d \xi + \frac{1}{\pi} \int\limits_{0}^{\infty} \int\limits_{-\infty}^{\infty} v_b \left( t \right) \cos{\xi t } \, dt \,  K_2 \left( \xi,  x \right) \, d \xi 
\label{thinbeam:toplayer_v_x_expanded}  \\
\text{and} &&
\tau_{yy}  \left( x, h \right) &= \frac{1}{\pi} \int\limits_{0}^{\infty} \int\limits_{-\infty}^{\infty} - P_c \left( t \right) \cos{\xi t } \, dt \, K_2 \left( \xi,  x \right)  \, d \xi  + \frac{{E^*}}{\pi} \int\limits_{0}^{\infty} \int\limits_{-\infty}^{\infty} v_b \left( t \right) \cos{\xi t } \, dt \, K_3 \left( \xi,  x \right) \, d \xi. 
\label{thinbeam:bottomlayer_tyy_x_expanded_temp} 
\end{flalign}
In \eqref{thinbeam:bottomlayer_tyy_x_expanded_temp}, the second integral is singular at $\xi \rightarrow \infty$. This singularity may be eliminated by integrating twice by parts, to find
\begin{align}
\tau_{yy}  \left( x, h \right) &= \frac{1}{\pi} \int\limits_{0}^{\infty} \int\limits_{-\infty}^{\infty} - P_c \left( t \right) \cos{\xi t } \, dt \,  K_2 \left( \xi,  x \right) \, d \xi  - \frac{{E^*}}{\pi} \int\limits_{0}^{\infty} \int\limits_{-\infty}^{\infty} f \left( t \right) \cos{\xi t } \, dt \, \,  \, \frac{1}{\xi^2 } \,   K_3 \left( \xi,  x \right) \, d \xi,
\label{thinbeam:bottomlayer_tyy_x_expanded}
\end{align} 
where $\kappa \left( t \right) = d^2v_b /dt^2$. \\

The contact region's vertical displacement is fixed by the displacement of the punch $\delta$ and the profile $g(x)$ of the punch. So within the contact zone, i.e. $-a \le x \le a$ -- where $a$ locates the contact edge - we set $v(x,0) = \delta - g(x) $. When $a$ and $\delta$ are small compared to the radius $R$ of the punch, we approximate the cylindrical profile of the punch by a parabola. Furthermore, there is no normal traction at the bottom surface of the beam, except at the supports; see Fig.~\ref{Thinbeams:Indentation}. Thus, at the top and bottom surfaces of the layer we have, respectively, 
\begin{flalign}
&& v(x,0) &= \delta-\frac{x^2}{2R} \quad \text{for} \quad  -a \leq x \leq a, \label{thinbeam:contactzone_disp_x} &&\\
\text{and } &&
\tau_{yy} \left( x, h \right) &= 0 \quad \text{for} \quad -l < x <  l. && \label{thinbeam:bottom_traction}
\end{flalign} 

We model adhesive interaction between the punch and the beam through an adhesive-zone \citep{Maugis1992adhesion}. The normal tractions within the adhesive zone follow the Dugdale-Barenblatt model, in which a constant attractive force $\sigma_0$ acts per unit length within the adhesive zone of length $d=c-a$, where $c$ demarcates the adhesive zone's outer edge, see inset in Fig.~\ref{Thinbeams:Indentation}(b). With this we write the normal traction on the top surface as
\begin{equation}
\label{thinbeam:Pfn}
P_c \left( x \right) = \left\{
\begin{array}{ll}
p \left( x \right) &\quad \text{for } \,   |x| \leq a, \\
-\sigma_0 & \quad \text{for } \,  a \leq |x| \leq c, \\
0 & \quad \text{for } \,   |x| > c.
\end{array}
\right.
\end{equation}
An adhesive-zone model resolves stress singularities at the contact edges ($x=\pm a$) inherent in the JKR approximation by requiring the normal traction be continuous there, i.e.,
\begin{equation}
\label{thinbeam:contact_end_pressure}
\lim_{ x  \rightarrow \pm a^{\mp}}  p \left( x \right) = -\sigma_0.
\end{equation}
An adhesive zone introduces the extra variable $c$ into the contact problem. The required additional equation is obtained by equating the energy release rate $G$ -- computed employing the $J-$Integral \citep{rice1968path} -- to the work of adhesion $w$, to obtain the energy balance
\begin{equation}
\label{thinbeam:Griffith_eqn}
\sigma_0 \delta_c = w,
\end{equation}
where 
\begin{equation}
\delta_c = \left( c^2/2R \right) - \delta + v_c
\end{equation}
is the air gap at the end of the adhesive zone (see inset in Fig.~\ref{Thinbeams:Indentation}(b)) and $v_c = v\left( c,0 \right)$ is the vertical displacement of the top surface at $x=c$.

In non-adhesive indentation, $\sigma_0 = 0 =w$, and \eqref{thinbeam:Griffith_eqn} is automatically satisfied. When the JKR approximation is invoked, $\sigma_0 \rightarrow \infty$ and $c \rightarrow a$, so that employing the Griffith's criterion \eqref{thinbeam:Griffith_eqn} becomes
\begin{equation}
\label{thinbeam:fracture_Griffith}
\frac{K_1^2}{2E^*} = w,
\end{equation}
where
\begin{equation}
\label{thinbeam:stress_intensity_factor}
K_1=-\lim_{x\rightarrow a^{-}} \sqrt{2 \pi \left( a-x \right)}  p \left( x \right)
\end{equation}
is the \emph{stress intensity factor}; see e.g. \citet[p.~168]{kanninen1985advanced}. Note that the continuity condition \eqref{thinbeam:contact_end_pressure} is redundant for the JKR approximation.

Finally, the total load acting on the punch is
\begin{equation}
\label{thinbeam:total_load}
P = \int\limits_{-\infty}^{\infty} P_c \left( x \right) \, dx = \int\limits_{-a}^{a} p \left( x \right) dx - 2 \, \sigma_0  \left( c - a \right).
\end{equation}

\section{Non-dimensionalization}
We generally follow the non-dimensionalization of Paper I: 
\begin{gather*}
A = \frac{a}{l}; \quad 
\varphi \left( \bar{\tau} \right) = \frac{a R l}{K h^3} \, p\left( a \bar{\tau} \right);  \quad
\bar{P} = \frac{P R l}{K h^3}; \\
\Delta = \frac{\delta R}{l^2}; \quad 
L = \frac{l}{R}; \quad
\lambda = 2 \sigma_0 \left( \frac{R}{\pi w K^2} \right)^{1/3};  \quad  
m = \left( \frac{\pi w}{RK} \right)^{1/3},
\end{gather*}
where $K=4 \, E^*/3$. The variables are scaled as 
\begin{gather*}
\left\{ \bar{x} , \bar{\tau}, \bar{c}, \bar{l}, \bar{\gamma}  \right\} = \frac{1}{a} \left\{ x, t, c,  l, h \right\}; \quad  
\left\{ \hat{\tau}, \hat{\gamma}  \right\} =  \frac{1}{l} \left\{  t,  h \right\};  \quad
\left\{ \omega, \bar{\omega}, \hat{\omega}  \right\} = \left\{ \xi h, \frac{\omega}{\bar{\gamma}},  \frac{\omega}{\hat{\gamma}} \right\}; \\
\Phi \left( \tau \right) = \frac{a R l}{K h^3} \, P \left( a \bar{\tau} \right);  \quad
\vartheta_b \left(  \hat{\tau} \right) = v_b \left( \hat{\tau} \right) \, \left( \frac{R}{l^2} \right); \quad
 \kappa_b \left( \hat{\tau} \right) =\frac{d^2 \vartheta_b \left( \hat{\tau} \right)}{d \hat{\tau}^2};  \quad 
% \text{and } \quad
 \mathcal{T} = \frac{\tau_{yy} \left(\hat{x},h \right) }{K}  \frac{8 \hat{\gamma}}{3L}. \nonumber
\end{gather*}

Employing the above, the non-dimensional vertical displacement of the top surface \eqref{thinbeam:toplayer_v_x_expanded} and the  normal traction at the bottom surface \eqref{thinbeam:bottomlayer_tyy_x_expanded} become, respectively,
\begin{flalign}
\vartheta \left( \bar{x}, 0 \right) &=  \frac{8 \hat{\gamma}^3}{3 \pi} \int\limits_{0}^{\infty} \int\limits_{-\infty}^{\infty} \Phi \left(  \bar{\tau} \right) \cos \left( \bar{\omega} \tau \right) \, d  \bar{\tau} \,  K^{t}_1 \left( \bar{\omega}, \bar{x} \right) d \omega  + \frac{1}{\pi \hat{\gamma}} \int\limits_{0}^{\infty} \int\limits_{-\infty}^{\infty} \vartheta_b \left( \hat{\tau} \right) \cos \left( \hat{\omega}  \hat{\tau} \right) \, d \hat{\tau} \, K^{t}_2 \left( \bar{\omega}, \bar{x} \right) d \omega 
\label{thinbeam:toplayer_final_2} \\
\text{and} \quad
\mathcal{T} &=  \frac{8 \hat{\gamma}^3}{3 \pi} \int\limits_{0}^{\infty} \int\limits_{-\infty}^{\infty} - \Phi \left( \bar{\tau} \right) \cos \left( \bar{\omega} \bar{\tau} \right)  \, d \bar{\tau}  \,  K^{b}_{1} \left(  \hat{\omega}, \hat{x} \right) d \omega  - \frac{\hat{\gamma}}{\pi}  \int\limits_{0}^{\infty} \int\limits_{-\infty}^{\infty} \kappa_b \left( \hat{\tau} \right) \cos \left(  \hat{\omega} \hat{\tau} \right)  \, d \hat{\tau} \,  K^{b}_{2} \left(  \hat{\omega}, \hat{x} \right) d \omega, \label{thinbeam:bottomlayer_final_2} 
\end{flalign}
where the kernels
\begin{align}
 K^{t}_1 \left( \bar{\omega}, \bar{x} \right) &= \frac{ \sinh^2{\omega}}{ \omega \left( \omega + \sinh{\omega } \cosh{\omega} \right) } \, \cos \left( \bar{\omega} \bar{x}  \right), &\quad
 K^{t}_2 \left( \bar{\omega}, \bar{x} \right) &= \frac{\sinh{ \omega} + \omega \cosh{ \omega } }{ \omega + \sinh{ \omega} \cosh{ \omega }} \, \cos \left( \bar{\omega} \bar{x}  \right), \nonumber \\
 K^{b}_{1} \left(  \hat{\omega}, \hat{x} \right) &= \frac{ \sinh{ \omega} + \omega \cosh{ \omega }}{  \omega + \sinh{\omega } \cosh{\omega} } \, \cos \left(   \hat{\omega}  \hat{x}  \right) \quad \text{and} &\quad
 K^{b}_{2} \left(  \hat{\omega}, \bar{x} \right) &= \frac{1}{\omega} \cdot \frac{\sinh^2{\omega} - \omega^2 }{ \omega + \sinh{ \omega} \cosh{ \omega }} \, \cos \left(   \hat{\omega} \hat{x}  \right). \label{Part2_Kernels}
 %\nonumber
\end{align}
Non-dimensionalizing \eqref{thinbeam:contactzone_disp_x} -- \eqref{thinbeam:Griffith_eqn} yields
\begin{flalign}
&& \vartheta (\bar{x},0) &= \Delta  - \frac{1}{2} \, \bar{x}^2 A^2 \quad \text{for} \quad -1 \leq \bar{x} \leq 1, \label{thinbeam:contactzone_disp_nondim} &&\\
&& \mathcal{T} \left( \hat{x},h \right) & = 0 \quad \text{for} \quad -1 < \hat{x} < 1,  \label{thinbeam:bottom_traction_nondim} && \\
&& \Phi \left( \bar{\tau} \right) &= \left\{
\begin{array}{ll}
\varphi \left( \bar{\tau} \right), & \quad -1 \leq \bar{\tau} \leq 1 \\
-\lambda A m / 2 \hat{\gamma}^3 L, & \quad 1 \leq |\bar{\tau}| \leq \bar{c} \\
0, & \quad  |\bar{\tau}| > \bar{c}
\end{array} 
\right. \label{thinbeam:Pfn_nondim} && \\
&& \varphi \left( \pm 1 \right) &= - \frac{\lambda A m}{2 \hat{\gamma}^3 L} \label{thinbeam:contact_end_pressure_nondim} && \\
 \text{and} &&
1 &= \frac{\pi \lambda L^2}{2 m^2} \,  \left( \frac{1}{2} \, \bar{c}^2 \, A^2  - \Delta  +  \vartheta_c \right), \label{thinbeam:Griffith_Eqn_nondim} &&
\end{flalign}
where $\vartheta \left( \bar{c} \right) = \vartheta \left( \bar{c},0 \right)$, and $\Delta  =  \vartheta \left( 0,0 \right)$. 
In the JKR approximation, we replace \eqref{thinbeam:contact_end_pressure_nondim} and \eqref{thinbeam:Griffith_Eqn_nondim} by the non-dimensional Griffith's criterion, obtained from \eqref{thinbeam:fracture_Griffith} and \eqref{thinbeam:stress_intensity_factor}:
\begin{equation}
\label{fracture_Griffith_nondim}
\lim_{\bar{x} \rightarrow 1^{-}} \sqrt{\left( 1 - \bar{x} \right)}  \varphi \left(  \bar{x} \right) = -\frac{m}{2 \pi L} \left( \frac{l}{h} \right)^3 \sqrt{\frac{3 A m}{L}}.
\end{equation}
Substituting \eqref{thinbeam:Pfn_nondim} in  \eqref{thinbeam:toplayer_final_2} and \eqref{thinbeam:bottomlayer_final_2} yields
\begin{align}
\vartheta \left( \bar{x}, 0 \right) =& - \frac{8 \hat{\gamma}^3}{3 \pi} \int\limits_{0}^{\infty} \bar{\varphi} \left( \bar{\omega} \right) K^{t}_1 \left( \bar{\omega},\bar{x}  \right)
 d \omega  - \frac{8 \lambda A m}{3 \pi L}   \int\limits_{0}^{\infty}  \bar{\varphi}_0 \left( \bar{\omega} \right) K^{t}_1 \left(\bar{\omega}, \bar{x}  \right) d \omega  +  \frac{1}{\pi \hat{\gamma}} \int\limits_{0}^{\infty} \hat{\vartheta}_b \left( \hat{\omega} \right) K^{t}_2 \left( \bar{\omega},\bar{x}  \right) d \omega,
\label{thinbeam:Int_eqn_toplayer}
\end{align} 
and  
\begin{align}
\mathcal{T}  =& \frac{8 \hat{\gamma}^3}{3 \pi} \int\limits_{0}^{\infty} \bar{\varphi} \left( \bar{\omega} \right) K^{b}_{1} \left( \hat{\omega},\hat{x}  \right) d \omega  + \frac{8 \lambda A m}{3 \pi L}   \int\limits_{0}^{\infty}  \bar{\varphi}_0 \left( \bar{\omega} \right) K^{b}_{1} \left(\hat{\omega}, \hat{x}  \right) d \omega   - \frac{\hat{\gamma}}{\pi} \int\limits_{0}^{\infty}  \hat{\kappa}_b \left( \hat{\omega} \right) K^{b}_{2} \left( \hat{\omega}, \hat{x}  \right) d \omega,
 \label{thinbeam:Int_eqn_bottomlayer}
\end{align}
with
%\begin{subequations}
\begin{gather}
\label{integrals_in_omega}
\begin{alignat}{3}
\bar{\varphi} \left( \bar{\omega}  \right) & = - \int\limits_{-1}^{1} \varphi \left( \bar{\tau} \right) \cos{\bar{\omega}  \, \bar{\tau}} \, d\bar{\tau},  &\quad
\bar{\varphi}_0 \left( \bar{\omega}  \right) &= \int\limits_{1}^{\bar{c}}\cos{ \bar{\omega}  \, \bar{\tau} } \, d\bar{\tau}, \nonumber \\
\hat{\vartheta}_b \left( \hat{\omega} \right) &= \int\limits_{-\infty}^{\infty} \vartheta_b \left( \hat{\tau} \right) \cos{ \hat{\omega}   \hat{\tau} } \, d \hat{\tau} \quad \text{and } & \quad 
\hat{\kappa}_b \left( \hat{\omega} \right) &= \int\limits_{-\infty}^{\infty} \kappa_b \left( \hat{\tau} \right) \cos{ \hat{\omega}   \hat{\tau} } \, d \hat{\tau}. 
\end{alignat}
\end{gather}
%\end{subequations}

Then, combining \eqref{thinbeam:Int_eqn_toplayer} with \eqref{thinbeam:contactzone_disp_nondim}, and \eqref{thinbeam:Int_eqn_bottomlayer} with \eqref{thinbeam:bottom_traction_nondim}, we obtain
\begin{flalign}
&& \label{thinbeam:toplayer_Int_eqn_final}
\Delta  - \frac{1}{2} \, \bar{x}^2 A^2  = & - \frac{8 \hat{\gamma}^3}{3 \pi} \int\limits_{0}^{\infty} \bar{\varphi} \left( \bar{\omega}  \right) \, K^{t}_1 \left( \bar{\omega} ,\bar{x}  \right)
 \,  \, d \omega  - \frac{8 \lambda A m}{3 \pi L}   \int\limits_{0}^{\infty}  \bar{\varphi}_0 \left( \bar{\omega}  \right) \,  K^{t}_1 \left(\bar{\omega} ,\bar{x}  \right) \, \, d \omega  \nonumber && \\
&& &  + \frac{1}{\pi\hat{\gamma}} \int\limits_{0}^{\infty} \hat{\vartheta}_b \left( \hat{\omega} \right) \,  K^{t}_2 \left( \bar{\omega} ,\bar{x}  \right) \, \, d \omega \quad \text{for } -1 \le \bar{x} \le 1  &&\\
 \text{and} &&
\label{thinbeam:bottomlayer_Int_eqn_final}
0 =& - \frac{8 \hat{\gamma}^3}{3 \pi} \int\limits_{0}^{\infty} \bar{\varphi} \left( \bar{\omega} \right) \, K^{b}_{1} \left( \hat{\omega} ,\hat{x}  \right)
 \,  \, d \omega  - \frac{8 \lambda A m}{3 \pi L}   \int\limits_{0}^{\infty}  \bar{\varphi}_0 \left( \bar{\omega} \right) \,  K^{b}_{1} \left(\hat{\omega}, \hat{x}  \right) \, \, d \omega \nonumber &&\\ 
 && & + \frac{\hat{\gamma}}{\pi}  \int\limits_{0}^{\infty} \hat{\kappa}_b \left( \hat{\omega} \right) \,  K^{b}_{2} \left( \hat{\omega}, \hat{x}  \right) \, \, d \omega \quad \text{for } -1 < \hat{x} < 1. &&
\end{flalign}
Finally, the total non-dimensional load acting on the punch is given by
\begin{equation}
\bar{P} = \int\limits_{-1}^{1} \varphi \left( \bar{\tau} \right) \, d \bar{\tau} - \frac{\lambda A m}{\hat{\gamma}^3 L} \left( \bar{c} - 1 \right).
\end{equation}

Equations  \eqref{thinbeam:toplayer_Int_eqn_final} and \eqref{thinbeam:bottomlayer_Int_eqn_final} are coupled Fredholm integral equations of the first kind; see \citet[p. 573]{polyanin2008handbook}. These, along with \eqref{thinbeam:contact_end_pressure_nondim} and \eqref{thinbeam:Griffith_Eqn_nondim}, are to be solved for $\bar{\varphi}$, $\bar{\vartheta}$ and $\Delta$ for a given contact area $A$. The numerical algorithm employed for this is discussed next.

\section{Numerical solution}
\label{sec:Thinbeam_Numerical_solution}
The dual integral equations \eqref{thinbeam:toplayer_Int_eqn_final} and \eqref{thinbeam:bottomlayer_Int_eqn_final} cannot be solved in closed form due to the presence of complex kernels; cf. \ref{Part2_Kernels}. We, therefore employ a numerical solution.

We begin by approximating the contact pressure $\varphi$ as
\begin{equation}
\label{thinbeam:phi_cheby}
\varphi \left( \bar{\tau} \right) = \frac{-\lambda A m}{2 \hat{\gamma}^3 L}+\frac{1}{\sqrt{\left( 1- \bar{\tau}^2 \right)}} \sum_{n=0}^{N} b_{2n} T_{2n} \left( \bar{\tau} \right) \quad \text{for } -1 \le \bar{\tau} \le 1,
\end{equation}
where $T_{2n} \left( \bar{\tau} \right)$ are Chebyshev polynomials of the first kind and $b_{2n} $ are constants that are to be determined. Only even Chebyshev polynomials are considered as the indentation is symmetric about $\bar{\tau}=0$. The constant term is chosen  to account for the contact pressure at the contact edge in the adhesive-zone model explicitly.  Evaluating the integrals $\bar{\varphi} \left( \bar{\omega}  \right)$ and $\hat{\varphi} \left( \bar{\omega} \right)$ in \eqref{integrals_in_omega} after employing  \eqref{thinbeam:phi_cheby} yields
\begin{equation}
\bar{\varphi} \left( \bar{\omega} \right) = \frac{\lambda A m}{\hat{\gamma}^3 L} \, \frac{\sin{ \bar{\omega} }}{\bar{\omega}} - \sum_{n=0}^{N} b_{2n} \alpha_{2n} \left( \bar{\omega} \right)
\quad \text{and} \quad
\bar{\varphi}_0 \left( \bar{\omega} \right) = \frac{1}{\bar{\omega}} \, \left( -\sin{ \bar{\omega} } +\sin{ \bar{\omega}  \,\bar{c} }  \right),
 \label{thinbeam:phi_calcs}
\end{equation}
where
\begin{equation}
\label{thinbeam:alpha_cheby_2}
\alpha_{2n} \left( \bar{\omega} \right) = \int\limits_{-1}^{1} \frac{1}{\sqrt{ \left( 1-\bar{\tau}^2 \right)}} T_{2n} \left( \bar{\tau} \right) \cos{ \bar{\omega} \bar{\tau} }\: \: d \bar{\tau}.
\end{equation}
The evaluation of the integrals $\alpha_{2n} \left( \bar{\omega} \right)$ at different $n$ are available in the appendix of Paper I.

Next, we approximate the displacement of the bottom surface $\vartheta_b \left( \hat{\tau} \right)$ in a series of the natural mode shapes $S_n \left( \hat{\tau} \right)$ of the beam:
\begin{equation}
\vartheta_b \left( \hat{\tau} \right) = d_0 +  \sum_{n=1}^{N} d_n \, S_n \left( \hat{\tau} \right). \label{part2_displacement_approx}
\end{equation}
We note that  $S_n \left( \hat{\tau} \right) = \cos \left( n \pi \hat{\tau} \right)$ and $S_n \left( \hat{\tau} \right) = \sin \left\{ \left( 2n-1 \right) \pi \left( \hat{\tau}+1 \right) /2 \right\}$ for clamped and simply supported beams, respectively. After these approximations are made to satisfy the beam's end conditions, the curvature $\kappa_b$ of the beam is calculated from $\vartheta_b$. The details of these calculations are available in Appendices \ref{sec:Appendix_thinbeam_displacements} and \ref{sec:Appendix_thinbeam_vomega_calc}. From \eqref{thinbeams:v_omega_derivation} we find that the Fourier transforms $\hat{\vartheta}_b \left( \hat{\omega} \right)$ and $\hat{\kappa}_b  \left( \hat{\omega} \right)$, for both clamped and simply supported beams, may be written as
\begin{gather}
\hat{\vartheta}_b  \left( \hat{\omega} \right) = \sum_{n=1}^{M}  d_n \hat{\beta}_{n} \left( \hat{\omega} \right)  \quad \text{and} \quad
\hat{\kappa}_b  \left( \hat{\omega} \right) =  \sum_{n=1}^{M} d_n \hat{\kappa}_{n} \left( \hat{\omega} \right), \label{thinbeams:v_omega} 
\end{gather}
where the expressions for $\hat{\beta}_{n} $ and $\hat{\kappa}_{n} $ are provided in Appendix \ref{sec:Appendix_thinbeam_vomega_calc}.

Substituting \eqref{thinbeam:phi_calcs} and \eqref{thinbeams:v_omega} in the integral equations  \eqref{thinbeam:toplayer_Int_eqn_final} and \eqref{thinbeam:bottomlayer_Int_eqn_final} yields, respectively,
\begin{flalign}
&& \Delta  - \frac{1}{2} \, \bar{x}^2 A^2 &= \frac{8 \hat{\gamma}^3}{3 \pi} \sum_{n=0}^{N} b_{2n} \mathcal{J}^{t}_{2n} \left( \bar{x}  \right) - \frac{8 \lambda A m}{3 \pi L}   \mathcal{J}^{t} \left( \bar{x}  \right) + \frac{1}{\pi\hat{\gamma}}  \sum_{n=1}^{M} d_{n} \mathcal{Q}^{t}_{n} \left( \bar{x}  \right) \label{thinbeam:toplayer_Int_eqn_numerical} &&\\
\text{and }
&& 0 &= \frac{8 \hat{\gamma}^3}{3 \pi} \sum_{n=0}^{N} b_{2n} \mathcal{J}^{b}_{2n} \left( \hat{x}  \right) - \frac{8 \lambda A m}{3 \pi L}   \mathcal{J}^{b} \left( \hat{x}  \right) + \frac{\hat{\gamma}}{\pi} \sum_{n=1}^{M} d_{n} \mathcal{Q}^{b}_{n} \left( \hat{x}  \right), 
\label{thinbeam:bottomlayer_Int_eqn_numerical} &&
\end{flalign}
where
\begin{align}
\mathcal{J}^{t}_{2n} \left( \bar{x}  \right) &= \int\limits_{0}^{\infty} \alpha_{2n} \left( \bar{\omega} \right) \, K^{t}_{1} \left( \bar{\omega},\bar{x}  \right) \,  \, d \omega,  & \quad 
\mathcal{J}^{t} \left( \bar{x}  \right) &= \int\limits_{0}^{\infty} \frac{1}{\bar{\omega}} \, \sin \left( \bar{\omega} \,\bar{c} \right)  \,  K^{t}_{1} \left( \bar{\omega}, \bar{x}  \right) \, \, d \omega, \nonumber \\
\mathcal{Q}^{t}_{n} \left( \bar{x}  \right) &= \int\limits_{0}^{\infty} \hat{\beta}_{n} \left( \hat{\omega} \right) \,  K^{t}_{2} \left( \bar{\omega}, \bar{x}  \right) \, \, d \omega,  & \quad 
\mathcal{Q}^{b}_{n} \left( \hat{x}  \right) &= \int\limits_{0}^{\infty} \hat{\kappa}_{n} \left( \hat{\omega} \right) \,  K^{b}_{2} \left( \hat{\omega}, \hat{x}  \right) \, \, d \omega, \nonumber \\
\mathcal{J}^{b}_{2n} \left( \hat{x}  \right) &= \int\limits_{0}^{\infty} \alpha_{2n} \left( \bar{\omega} \right) \, K^{b}_{1} \left( \hat{\omega},\hat{x}  \right) \,  \, d \omega  &  \text{and}  \quad \quad
\mathcal{J}^{b} \left( \bar{x}  \right) &= \int\limits_{0}^{\infty} \frac{1}{\bar{\omega}} \, \sin \left( \bar{\omega} \,\bar{c} \right)  \,  K^{b}_{1 } \left( \hat{\omega}, \hat{x}  \right) \, \, d \omega. \nonumber
\end{align}
The above integrals may be evaluated at any $\bar{x}$ or $\hat{x}$ through the Clenshaw-Curtis quadrature; see \citet[p. 196]{press1992numerical}.

Employing \eqref{thinbeam:phi_cheby}, the constraint \eqref{thinbeam:contact_end_pressure_nondim} on the contact pressure at the ends of the contact region provides
\begin{equation}
\label{thinbeam:end_pressure}
b_{0} + b_{2} + \cdots + b_{2N} = 0.
\end{equation}
Utilizing the approximations \eqref{thinbeam:phi_calcs} and \eqref{thinbeams:v_omega}, the energy balance equation \eqref{thinbeam:Griffith_Eqn_nondim} yields
\begin{equation}
\label{thinbeam:Griffith_Eqn_numerical}
\frac{\pi \lambda L^2}{2 m^2} \,  \left( \frac{1}{2} \, \bar{c}^2 \, A^2  - \Delta  +  \vartheta_c\right) = 1,
\end{equation}
with
\begin{align}
\vartheta \left( \bar{c} \right) = &
\frac{8 \hat{\gamma}^3 }{3 \pi} \sum_{n=0}^{N} b_{2n} \mathcal{J}^{t}_{2n} \left( \bar{c}  \right) - \frac{8 \lambda A m}{3 \pi L}   \mathcal{J}^{t} \left( \bar{c}  \right) + \frac{1}{\pi\hat{\gamma}}  \sum_{n=1}^{M} d_{n} \mathcal{Q}^t_n \left( \bar{c} \right),
\label{thinfixedbeam:V_c_eqn_numerical} 
\end{align}
where $\Delta$ is the displacement of the punch. We recall that the energy balance \eqref{thinbeam:Griffith_Eqn_numerical} is redundant for the case of adhesionless contact or when the JKR approximation is invoked.

Finally, the total load acting on the punch becomes after employing \eqref{thinbeam:bottom_traction} and \eqref{thinbeam:phi_cheby}:
\begin{equation}
\label{thinbeam:total_load_cheby}
\bar{P} = \pi b_0 - \frac{\lambda A m}{\hat{\gamma}^3 L} \bar{c}.
\end{equation}

\section{Algorithm}
\label{sec:Thinbeam_solution_procedure}
We need to solve \eqref{thinbeam:toplayer_Int_eqn_numerical} -- \eqref{thinbeam:Griffith_Eqn_numerical} for the $N+M+3$ unknowns $b_{2n}$, $d_n$, $\Delta$ and $\bar{c}$ at any given contact area $A$. These are solved  through the collocation technique \citep[p. 135]{Atkinson1997Inteqns}, which provides the necessary $N+M+3$ algebraic equations.

In the collocation method, \eqref{thinbeam:toplayer_Int_eqn_numerical} and \eqref{thinbeam:bottomlayer_Int_eqn_numerical} are required to hold exactly at, respectively, $N+1$ and $M$ \textit{collocation points}. The collocation points for \eqref{thinbeam:toplayer_Int_eqn_numerical} and \eqref{thinbeam:bottomlayer_Int_eqn_numerical} are selected to be
\begin{flalign}
&& \bar{x}_i &= \cos \left\{ \frac{ \left( 2i-1\right) \pi}{2 \left( N+1 \right) }\right\} \text{ for } i=1, \cdots, N+1, \nonumber && \\
\text{and}
&& \hat{x}_k &= \frac{k-1}{M} \text{ for } k=1, \cdots, M, \nonumber &&
\end{flalign}
respectively. Here, $\bar{x}_i$ are the $N+1$ zeros of the  (Chebyshev) polynomials $T_{2N+2}(\bar{x}_i)$ \citep[p. 19]{mason2003book}, while $\hat{x_k}$ are simply equally spaced points lying between $0$ and $1$. At these collocation points \eqref{thinbeam:toplayer_Int_eqn_numerical} and \eqref{thinbeam:bottomlayer_Int_eqn_numerical} become, respectively, 
\begin{flalign}
&& \Delta  - \frac{1}{2} \, \bar{x}^2_i A^2 =& \frac{8 \hat{\gamma}^3}{3 \pi} \sum_{n=0}^{N} b_{2n} \mathcal{J}^{t}_{2n} \left( \bar{x}_i  \right) - \frac{8 \lambda A m}{3 \pi L}   \mathcal{J}^{t} \left( \bar{x}_i  \right) + \frac{1}{\pi\hat{\gamma}}  \sum_{n=1}^{M} d_{n} \mathcal{Q}^{t}_{n} \left( \bar{x}_i  \right)
\label{thinbeam:toplayer_Int_eqn_coll} &&\\
\text{and} &&
0 =&\frac{8 \hat{\gamma}^3}{3 \pi} \sum_{n=0}^{N} b_{2n} \mathcal{J}^{b}_{2n} \left( \hat{x}_k \right) - \frac{8 \lambda A m}{3 \pi L}   \mathcal{J}^{b} \left( \hat{x}_k  \right) + \frac{\hat{\gamma}}{\pi}  \sum_{n=1}^{M} d_{n} \mathcal{Q}^{b}_{n} \left( \hat{x}_k  \right),
\label{thinbeam:bottomlayer_Int_eqn_coll} &&
\end{flalign} 
with $i= 1, \cdots, N+1$ and $k=1, \cdots, M$. Thus, we obtain $N+1$ equations from \eqref{thinbeam:toplayer_Int_eqn_coll} and $M$ equations from \eqref{thinbeam:bottomlayer_Int_eqn_coll} for a total of $N+M+1$ equations. Along with \eqref{thinbeam:end_pressure} and \eqref{thinbeam:Griffith_Eqn_numerical}, we finally obtain the required $N+M+3$ equations to solve for the $N+1$ unknowns $b_{2n}$, M unknowns $d_n$, $\Delta$ and $\bar{c}$. This system of non-linear algebraic equations is solved for any given contact area $A$ through the following algorithm:

\begin{enumerate}[Step 1:]
\item  For the given contact area $A$, we make an initial guess for $\bar{c}$.

\item We then write \eqref{thinbeam:toplayer_Int_eqn_coll} and \eqref{thinbeam:bottomlayer_Int_eqn_coll} in matrix notation as
\begin{equation}
\label{thinbeam:Inteqns_using_collocation}
\Delta \, \underline{e} - \underline{f} - \underline{\lambda}  = \underline{\underline{R}} \, \underline{a},
\end{equation}
where
%\begin{gather}
%\underline{e} = \left[ \underline{e}^t, \underline{e}^b \right]^{\text{T}}; \,
%\underline{f} = \left[ \underline{f}^t, \underline{f}^b \right]^{\text{T}}; \,
%\underline{\lambda} = \left[ \underline{\lambda}^t, \underline{\lambda}^b \right]^{\text{T}}; \,
%\underline{a} = \left[ \underline{a}^t, \underline{a}^b \right]^{\text{T}}; 
%\underline{\underline{R}} = \left[ \begin{array}{cc}
%\underline{\underline{\mathcal{J}}}^{t} & \underline{\underline{\mathcal{Q}}}^{t} \\
%\underline{\underline{\mathcal{J}}}^{b} & \underline{\underline{\mathcal{Q}}}^{b} 
%\end{array} \right],
%\end{gather}
\begin{gather}
\underline{e} = \left[ \underline{e}^t, \underline{e}^b \right]^{\text{T}}; \quad
\underline{f} = \left[ \underline{f}^t, \underline{f}^b \right]^{\text{T}}; \quad
\underline{\lambda} = \left[ \underline{\lambda}^t, \underline{\lambda}^b \right]^{\text{T}}; \quad
\underline{a} = \left[ \underline{a}^t, \underline{a}^b \right]^{\text{T}}; \quad
\underline{\underline{R}} = \left[ \begin{array}{cc}
\underline{\underline{\mathcal{J}}}^{t} & \underline{\underline{\mathcal{Q}}}^{t} \\
\underline{\underline{\mathcal{J}}}^{b} & \underline{\underline{\mathcal{Q}}}^{b} 
\end{array} \right],
\end{gather}
with
%\begin{gather}
%a^t_i = b_{2i-2}; \, a^b_k =d_{k} ; \,
%e^t_i = 1; \, e^b_k =0; \,
%f^t_i = \frac{x_i^2  A^2}{2}; \, f^b_k =0; \,
%\lambda^t_i=- \frac{8 \lambda A m}{3 \pi L}  \mathcal{J}^{t} \left( \bar{x}_i \right); \,
%\lambda^b_k= - \frac{8 \lambda A m }{3 \pi L}  \mathcal{J}^{b} \left( \hat{x}_k \right); \nonumber \\
%\mathcal{J}^{t}_{i \, j} = \frac{8 \hat{\gamma}^3}{3 \pi} \mathcal{J}^{t}_{2j-2} \left( \bar{x}_i \right); \,
%\mathcal{Q}^{t}_{i \, r} = \frac{1}{\pi \hat{\gamma}}  \mathcal{Q}^{t}_{r} \left( \bar{x}_i \right); \, 
% \mathcal{J}^{b}_{k \, j} = \frac{8 \hat{\gamma}^3}{3 \pi} \mathcal{J}^{b}_{2j-2} \left( \hat{x}_k \right); \,
% \mathcal{Q}^{b}_{i \, r} = \frac{\hat{\gamma}}{ \pi} \mathcal{Q}^{b}_{r} \left( \hat{x}_k \right),
%\end{gather}
\begin{gather}
a^t_i = b_{2i-2}; \, a^b_k =d_{k} ; \quad
e^t_i = 1; \, e^b_k =0; \quad
f^t_i = \frac{x_i^2  A^2}{2}; \quad 
f^b_k =0; \nonumber \\
\lambda^t_i=- \frac{8 \lambda A m}{3 \pi L}  \mathcal{J}^{t} \left( \bar{x}_i \right); \quad
\lambda^b_k= - \frac{8 \lambda A m }{3 \pi L}  \mathcal{J}^{b} \left( \hat{x}_k \right); \quad
\mathcal{J}^{t}_{i \, j} = \frac{8 \hat{\gamma}^3}{3 \pi} \mathcal{J}^{t}_{2j-2} \left( \bar{x}_i \right); \nonumber \\
\mathcal{Q}^{t}_{i \, r} = \frac{1}{\pi \hat{\gamma}}  \mathcal{Q}^{t}_{r} \left( \bar{x}_i \right); \quad
 \mathcal{J}^{b}_{k \, j} = \frac{8 \hat{\gamma}^3}{3 \pi} \mathcal{J}^{b}_{2j-2} \left( \hat{x}_k \right); \quad
 \mathcal{Q}^{b}_{i \, r} = \frac{\hat{\gamma}}{ \pi} \mathcal{Q}^{b}_{r} \left( \hat{x}_k \right),
\end{gather}
for $i,j=1,2, \cdots, N+1$ and $k,r=1,2, \cdots, M$. Thus, $\underline{e}$, $\underline{f}$, $\underline{\lambda}$, $\underline{a}$ are column vectors of size $N+M+1$, and $\underline{\underline{R}}$ is a matrix of size $(N+M+1) \times (N+M+1)$.

\item Note that the column vector $\underline{a}$ consists of the coefficients occurring in expressions \eqref{thinbeam:phi_cheby} for the contact pressure and \eqref{part2_displacement_approx} for the displacement. We now invert \eqref{thinbeam:Inteqns_using_collocation} to find $\underline{a}$ in terms of $\Delta$:
\begin{equation}
\label{thinbeam:unknown_in_delta}
\underline{a} = \Delta  \underline{E}  - \underline{F}  - \underline{\Lambda}, 
\end{equation}
where
\begin{gather*}
\underline{E}  =\underline{\underline{R}}^{-1} \cdot \underline{e}, \quad
\underline{F}  = \underline{\underline{R}}^{-1} \cdot \underline{f} \quad \text{and} \quad
\underline{\Lambda} = \underline{\underline{R}}^{-1}  \cdot \underline{\lambda} \, . 
\end{gather*}

\item Employing the end condition \eqref{thinbeam:end_pressure} for the contact pressure, we obtain the punch's displacement
\begin{equation}
\label{thinbeam:delta_calc}
\Delta = \frac{p_0 + \sum_{i=1}^{N+1}  F_i + \sum_{i=1}^{N+1} \Lambda_i}{\sum_{i=1}^{N+1} E _i}, 
\end{equation}
where
\begin{equation}
p_0 = \left\{ \begin{array}{l}
0 \quad \text{ if there is no adhesion or an adhesive zone is present}, \\
- \left( {m \sqrt{6A m}}/{2 \pi L^2} \right) \left( {l}/{h} \right)^3 \quad \text{ if the JKR approximation is invoked}.
\end{array} \right.
\end{equation}

\item Once $\Delta$ is known, we evaluate $\underline{a}$ from \eqref{thinbeam:unknown_in_delta} through
\begin{equation}
\label{thinbeam:unknowns}
\underline{a} =  \left(  \frac{ p_0 + \sum_{i=1}^{N+1}  F_i + \sum_{i=1}^{N+1} \Lambda_i}{\sum_{i=1}^{N+1} E_i} \right) \underline{E}  - \underline{F} - \underline{\Lambda}.
\end{equation}

\item  Employing $\Delta$ and $\underline{a}$, we calculate the displacement of the beam's top surface at $\bar{c}$ from \eqref{thinfixedbeam:V_c_eqn_numerical}  and check whether \eqref{thinbeam:Griffith_Eqn_numerical} holds.  If not, then we update $\bar{c}$ employing the Newton-Raphson method \citep{Anindya}. Steps 1-6 are repeated until \eqref{thinbeam:Griffith_Eqn_numerical} is satisfied. Steps 1 and 6 are required only when we employ an adhesive zone. When the Hertzian or JKR approximations are invoked we may conveniently skip this Step 6.

\item We finally proceed to find  the contact pressure distribution $\varphi \left( \bar{\tau} \right)$ and the total load $\bar{P}$ from \eqref{thinbeam:phi_cheby} and \eqref{thinbeam:total_load_cheby}, respectively.
\end{enumerate}

\section{Finite element (FE)  computations}
\label{sec:Thinbeam_FE_model}
Finite element (FE)  computations are carried out for clamped and simply supported beams for adhesionless contact. These are employed to validate our semi-analytical results.

The FE model is prepared in ABAQUS as described in Paper I: the beam is modelled as a linear elastic layer of Young's modulus $E=2000$ MPa, Poisson's ratio $\nu=0.3$, thickness $h=4$ mm, and half-span $l=40$ mm. The rigid punch is modeled as a much stiffer elastic material with Young's modulus $E_p=2 \times 10^{6}$ MPa and radius $R=225$ mm. Plane-strain elements are considered both for the beam and the punch. A concentrated load is applied on the punch. Computations provide the contact pressure $\varphi$, contact area $A$, punch's displacement $\Delta$, and the displacement $\vartheta_b$ of the beam's bottom surface.

\section{Results: Non-adhesive (`Hertzian') contact}
\label{sec:Results_thinbeam_Hertz}
We now report results for the non-adhesive interaction of clamped and simply supported beams with a rigid cylindrical punch.

For non-adhesive interaction, we set $\lambda=0$ in \eqref{thinbeam:toplayer_Int_eqn_numerical} and \eqref{thinbeam:bottomlayer_Int_eqn_numerical} to obtain
\begin{flalign}
&& \Delta  - \frac{1}{2} \, \bar{x}^2 A^2 &= \frac{8 \hat{\gamma}^3}{3 \pi} \sum_{n=0}^{N} b_{2n} \mathcal{J}^{t}_{2n} \left( \bar{x}  \right) + \frac{1}{\pi\hat{\gamma}}  \sum_{n=1}^{M} d_{n} \mathcal{Q}^{t}_{n} \left( \bar{x}  \right) \label{thinbeam:toplayer_Int_eqn_Hertz} && \\
\text{and}
&& 0 &=\frac{8 \hat{\gamma}^3}{3 \pi} \sum_{n=0}^{N} b_{2n} \mathcal{J}^{b}_{2n} \left( \hat{x}  \right) + \frac{\hat{\gamma}}{\pi} \sum_{n=1}^{M} d_{n} \mathcal{Q}^{b}_{n} \left( \hat{x}  \right), 
\label{thinbeam:bottomlayer_Int_eqn_Hertz} &&
\end{flalign}
respectively. In non-adhesive contact, the interacting surfaces detach smoothly at the contact edges, so that the pressure vanishes at the contact edge, and \eqref{thinbeam:end_pressure} holds, i.e.
\begin{eqnarray}
b_{0} + b_{2} + \cdots + b_{2N} &=& 0.
\label{thinbeam:end_pressure_Hertz}
\end{eqnarray}
We proceed to solve \eqref{thinbeam:toplayer_Int_eqn_Hertz} -- \eqref{thinbeam:end_pressure_Hertz} through the procedure of Sec.~\ref{sec:Thinbeam_solution_procedure}. We set $N=5$ and $M=50$ in our computations. Initially to compare our results with FE computations we employ the parameters of Sec.~\ref{sec:Thinbeam_FE_model}. 

\begin{figure}[h!]
\centering
\begin{multicols}{2}
\includegraphics[width=\linewidth]{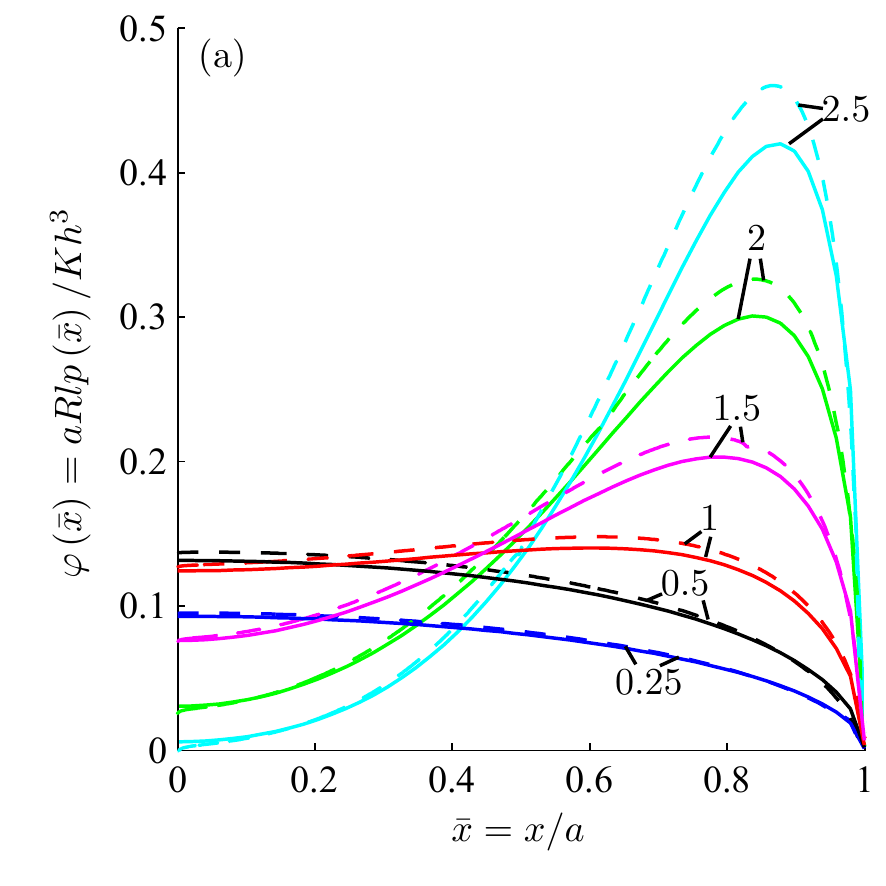} 
\includegraphics[width=\linewidth]{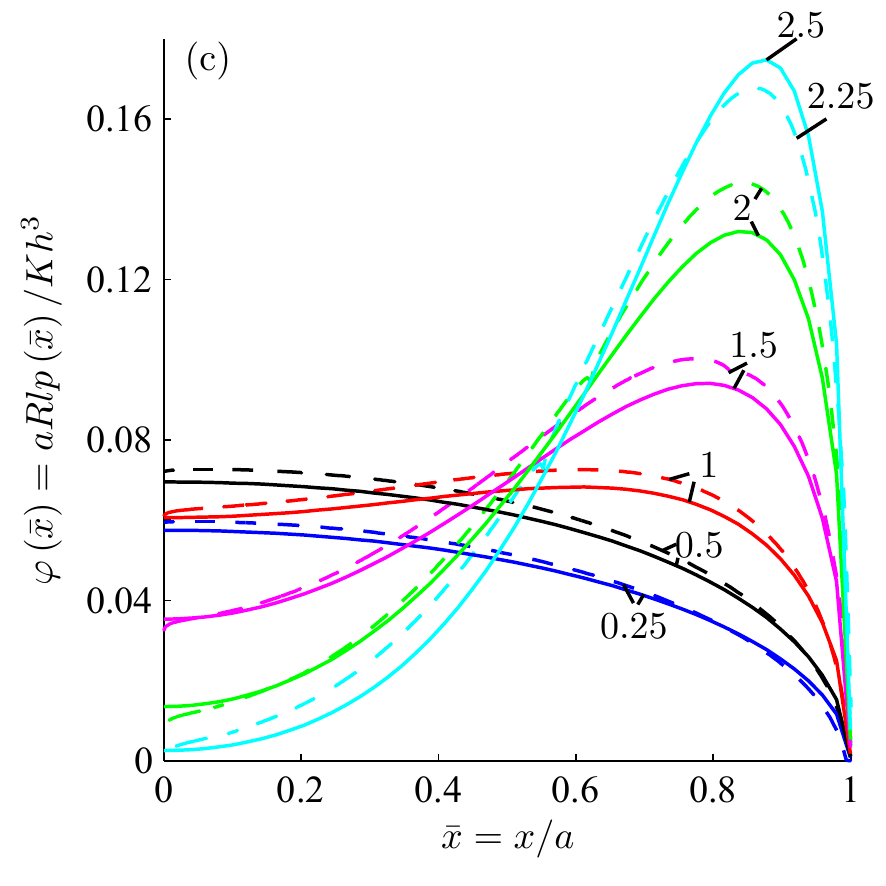} 
\includegraphics[width=\linewidth]{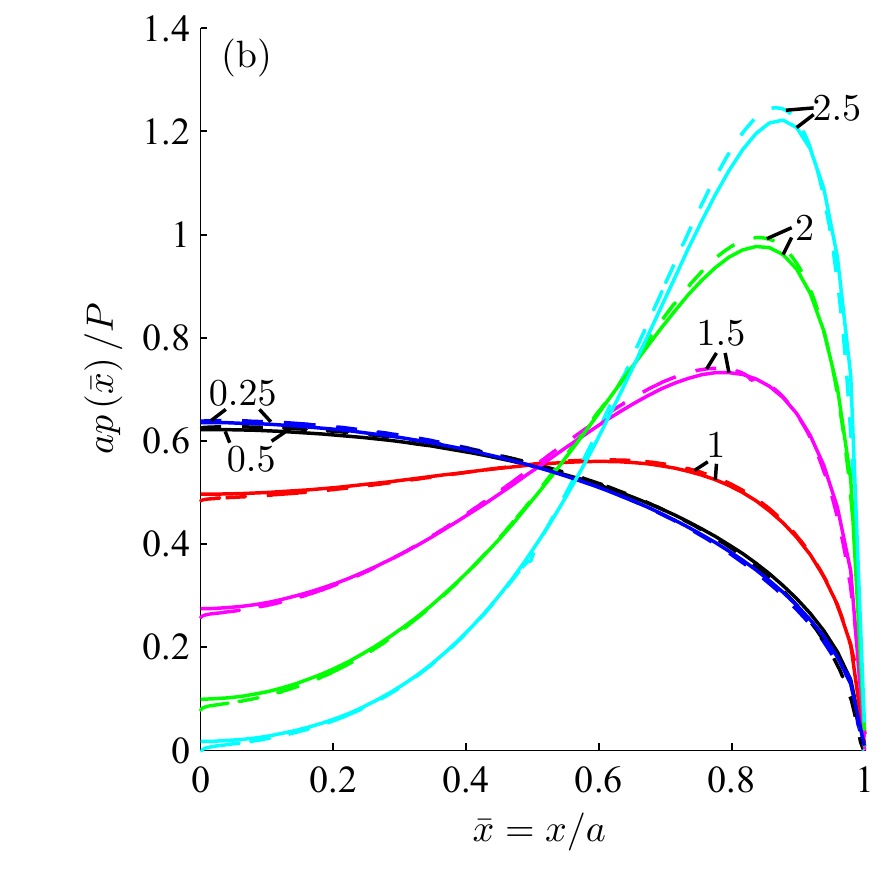} 
\includegraphics[width=\linewidth]{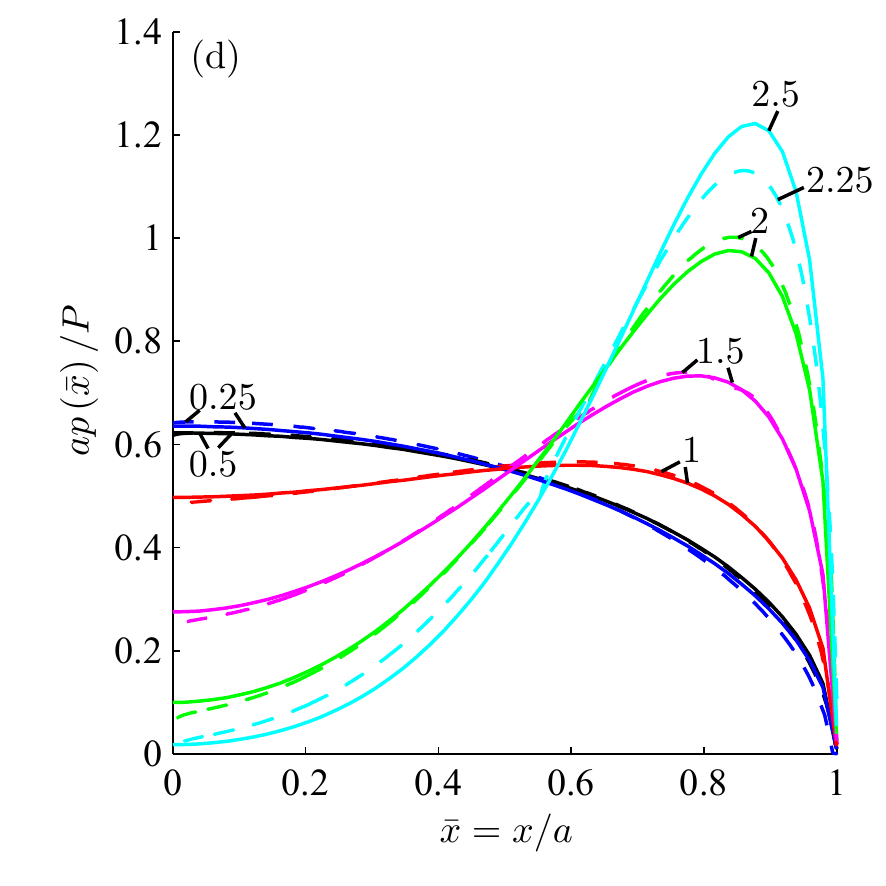} 
\end{multicols}
\caption{The non-dimensional contact pressures $\varphi \left( \bar{x} \right)$ and $a p \left( \bar{x} \right) / P$ during the non-adhesive indentation of a clamped (a and b) and simply supported (c and d) beams. We set $h=4$ mm and $l=40$ mm. Several contact areas $a$ are investigated by varying $a/h$ ratio, which are noted next to then corresponding curves. The solid lines are results obtained from the semi-analytical procedure of Sec.~\ref{sec:Thinbeam_Numerical_solution}, while dashed line represent FE computations.} 
\label{Thinbeams:Hertz_pressures_SA_abaqus_compare}
\end{figure}

\begin{figure}[h!]
\centering
\subfloat[][]{\includegraphics[width=0.5\linewidth]{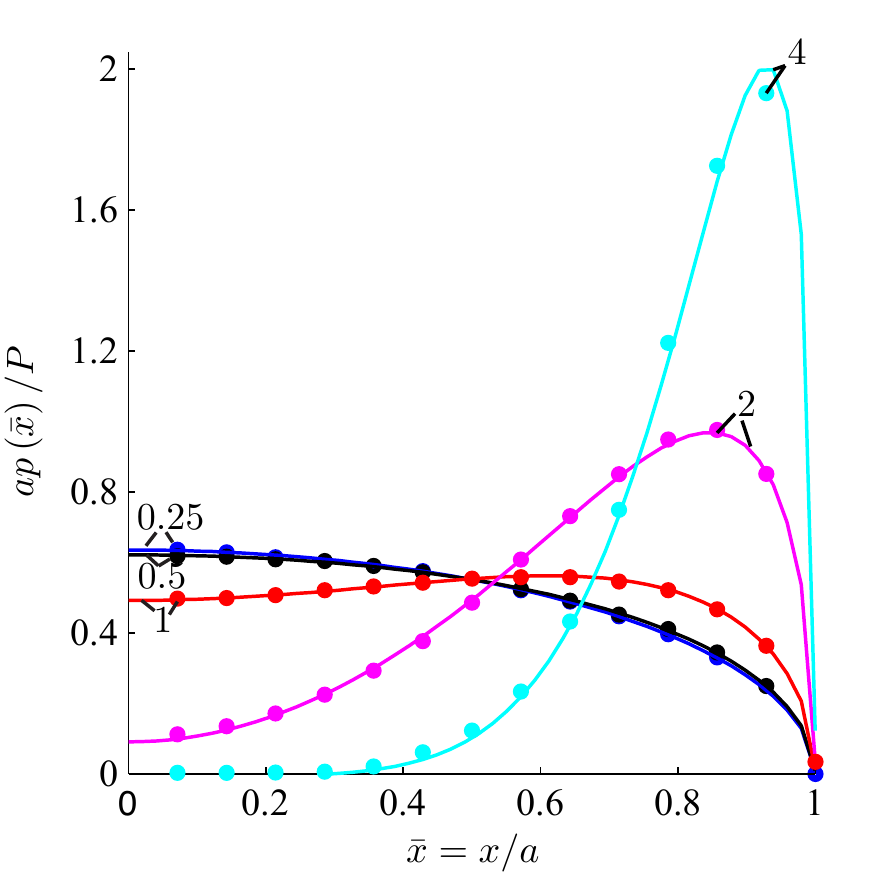}}
\subfloat[][]{\includegraphics[width=0.5\linewidth]{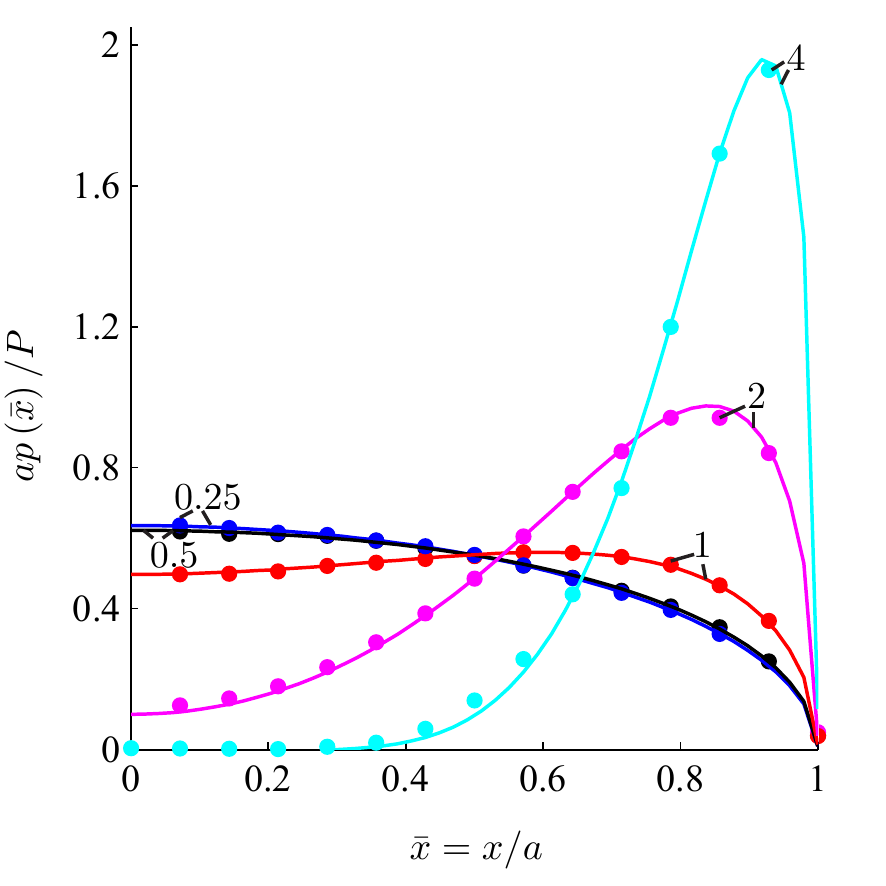} }
\caption{The non-dimensional contact pressures $a p \left( \bar{x} \right) / P$ during the non-adhesive indentation of (a) a clamped and (b) a simply beam. The slenderness ratio of the beam  $l/h=10$. Several contact areas $a$ are investigated by varying $a/h$ ratio, which are noted next to then corresponding curves. The solid lines are results obtained from the semi-analytical procedure of Sec.~\ref{sec:Thinbeam_Numerical_solution}. Dots represent the predictions of \cite{Keer1983smooth}.}
\label{Thinbeams:Hertz_pressures_SA_KM_compare}
\end{figure}

Figure~\ref{Thinbeams:Hertz_pressures_SA_abaqus_compare} plots $\varphi \left( \bar{x} \right)$ and $a p \left( \bar{x} \right) $, computed by solving  \eqref{thinbeam:toplayer_Int_eqn_Hertz} -- \eqref{thinbeam:end_pressure_Hertz} and from FE simulations. Results for both clamped and simply supported beams are shown. These pressure profiles are plotted at different  $a/h$ ratios, by varying the contact area $a$, for a beam of thickness $h$ and half-span $l$. We observe that at  low $a/h$ ratios, the contact pressure  is maximum at the center of the contact area and vanishes as we approach the contact edges. Increasing $a/h$ causes the contact pressure to decrease at the center of the contact area and increase near its ends. This behavior was also observed in Paper I.

As in Paper I, we find that, at the same $a/h$, the contact pressure $\varphi$ in a simply supported beam is smaller than that in a clamped beam, because the bending stiffness of the former is lower. However, this difference is not reflected when we plot $a p \left( \bar{x} \right) $; cf. Figs.~\ref{Thinbeams:Hertz_pressures_SA_abaqus_compare}(b) and \ref{Thinbeams:Hertz_pressures_SA_abaqus_compare}(d) . We observe a close match between our predictions and FE simulations for all $a/h$, except for a small deviation between the two at $a/h=2.5$ in the case of simply supported beams. We suspect that the latter may be due to the shear-free boundary condition at the top and bottom surface of the beam that was employed in the theoretical model but is not imposed in the FE model. We note that method of Paper I does well until $a/h \approx 1$. With greater indentation, the $a/h$ ratio increases, and the contact pressure at the center of the contact patch becomes negative. This reflects loss of contact, which is observed in both theoretical predictions and FE computations. For clamped beams, contact loss initiates at the center of the contact area when  $a/h \gtrsim 2.5$. In simply supported beams contact loss is predicted for $a/h \gtrsim 2.5$ by our analytical model but for $a/h \gtrsim 2.25$ by FE simulations.

\begin{figure}[h!]
\centering
\subfloat[][]{\includegraphics[width=0.5\linewidth]{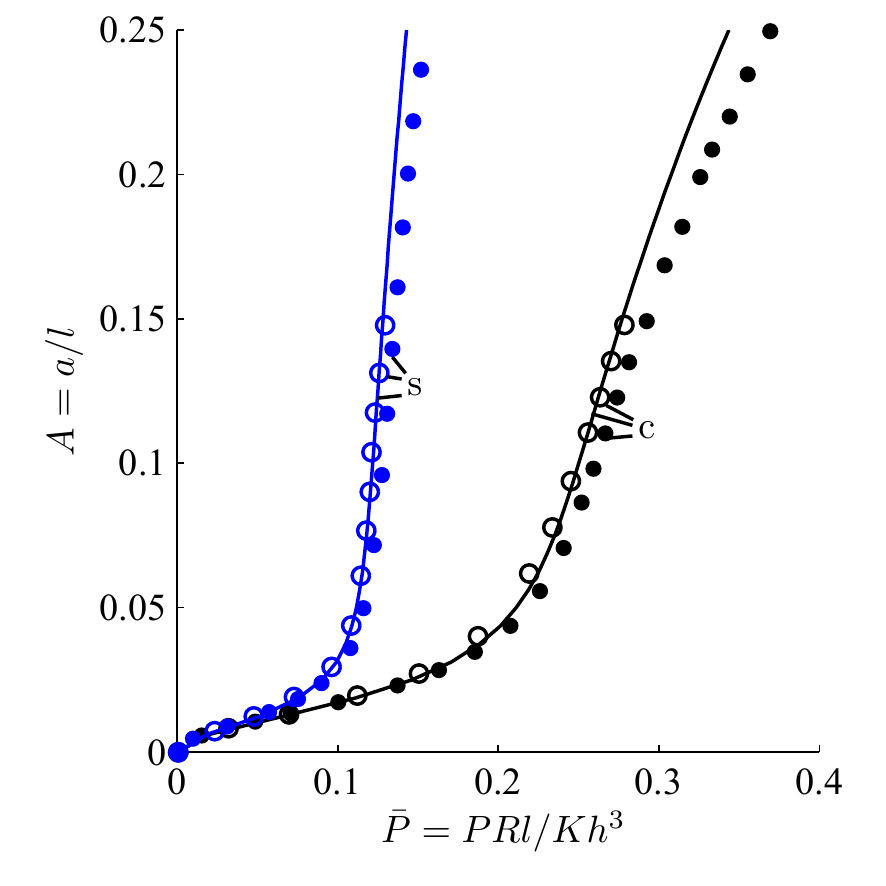}}
\subfloat[][]{\includegraphics[width=0.5\linewidth]{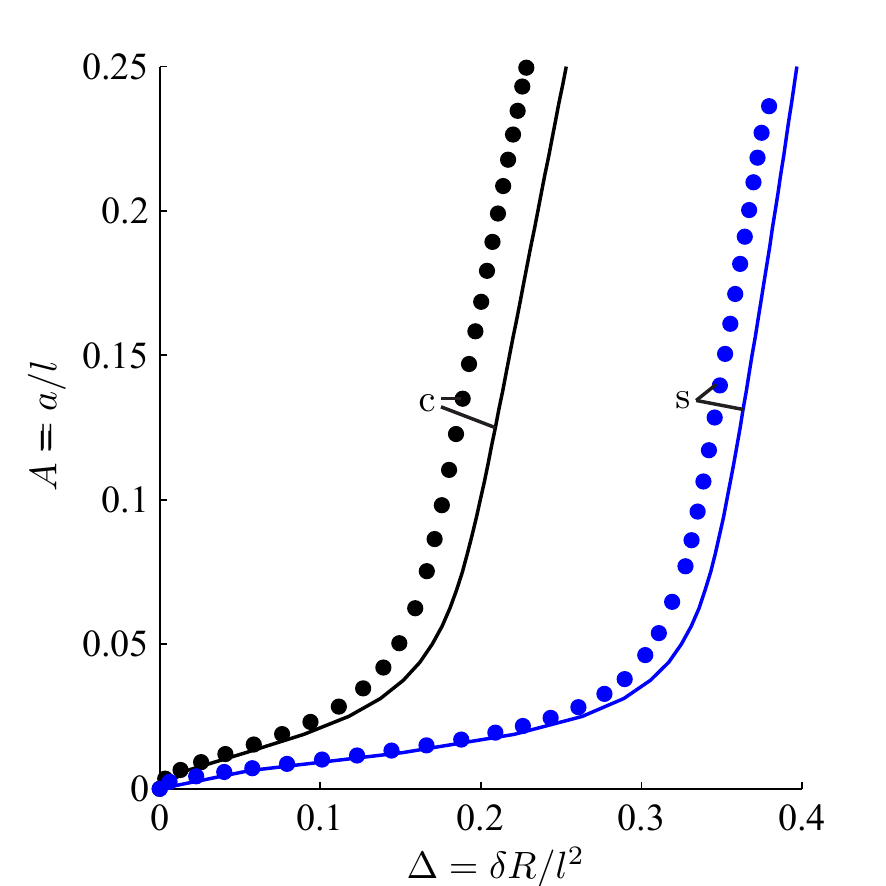} }
\caption{Non-adhesive contact of clamped (`c') and simply supported (`s') beams. The contact area $A$ is plotted as a function of (a) the total load $\bar{P}$ acting on the punch and (b) the punch's displacement $\Delta$. The beam's slenderness ratio $l/h$=10. Solid lines are results obtained from the semi-analytical procedure of Sec.~\ref{sec:Thinbeam_Numerical_solution}. Filled circles correspond to FE simulations of Sec.~\ref{sec:Thinbeam_FE_model}. Predictions of \cite{Sankar1983} are shown by open circles, when available.}
\label{Thinbeams:Hertz_SA_abaqus_BVS_compare}
\end{figure}

\begin{figure}[h!]
\centering
\includegraphics[width=0.5\linewidth]{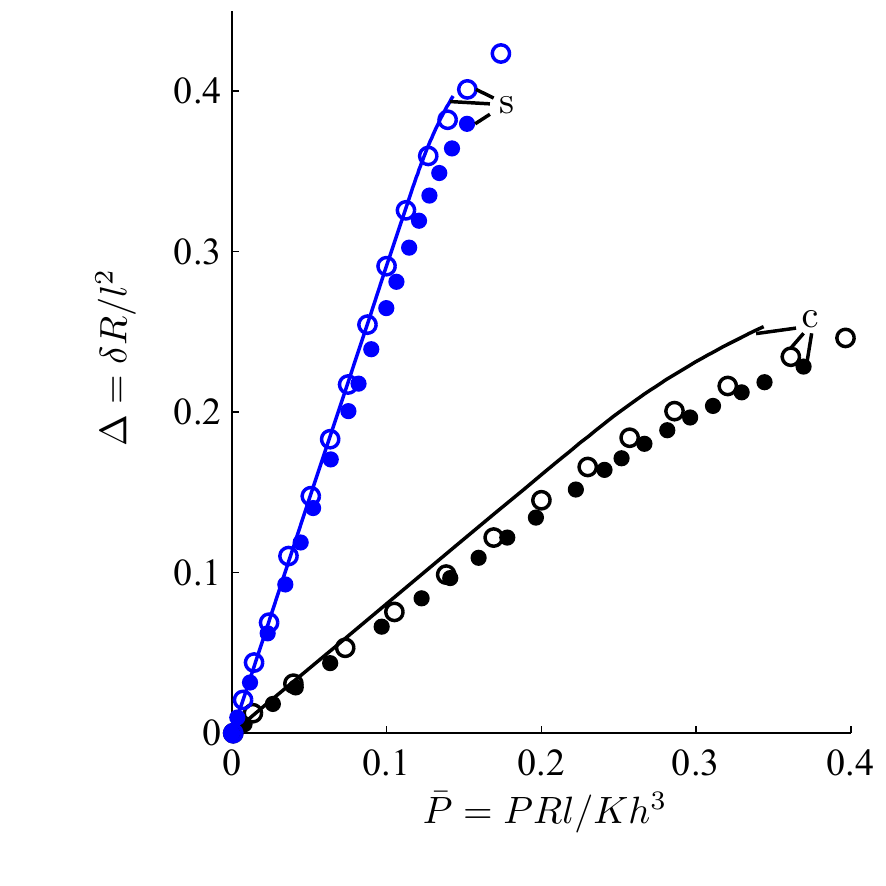}
\caption{Non-adhesive contact of clamped (`c') and simply supported (`s') beams. The displacement $\Delta$ of the punch is shown as a function of the total load $\bar{P}$. See also the caption of Fig.~\ref{Thinbeams:Hertz_SA_abaqus_BVS_compare}.}
\label{Thinbeams:Hertz_SA_abaqus_BVS_compare2}
\end{figure}

Next,  in Fig.~\ref{Thinbeams:Hertz_pressures_SA_KM_compare} we compare our results for contact pressures $a p \left( \bar{x} \right) $ with those of \cite{Keer1983smooth} for both clamped and simply supported beams. We find an extremely close match between the two untill $a/h \approx 2$. Beyond that, while the match remains close for almost the entire contact area, a deviation is observed at the center of the contact patch: we find negative pressures at the center of the contact patch, whereas \cite{Keer1983smooth} report no negative contact pressure. From this, it appears that the earlier formulations of \cite{Keer1983smooth} -- also \cite{Sankar1983}, which we discuss later -- do \textit{not} predict contact loss. Figure~\ref{Thinbeams:Hertz_pressures_SA_KM_compare} confirms our previous observation that contact pressures of clamped and simply supported beams do \textit{not} vary much, when scaled as $ap(\bar{x})/P$. 

Next, in Fig.~\ref{Thinbeams:Hertz_SA_abaqus_BVS_compare} we plot the variation of the contact area $A$ with the total load $\bar{P}$ on the punch and the punch's displacement $\Delta$ for both clamped and simply supported beams. We also compare with FE computations and results of \cite{Sankar1983}. The process is repeated for the variation of $\Delta$ with $\bar{P}$ in Fig.~\ref{Thinbeams:Hertz_SA_abaqus_BVS_compare2}. Figure~\ref{Thinbeams:Hertz_SA_abaqus_BVS_compare}(a) shows that for the same contact area $A$, the load $\bar{P}$ required for a simply supported beam is small compared to a clamped beam. This is because the simply supported beam bends more easily. This is also why we observe greater displacements $\Delta$ in these beams in Figs.~\ref{Thinbeams:Hertz_SA_abaqus_BVS_compare}(b) and \ref{Thinbeams:Hertz_SA_abaqus_BVS_compare2}. Finally, Figs.~\ref{Thinbeams:Hertz_SA_abaqus_BVS_compare} and \ref{Thinbeams:Hertz_SA_abaqus_BVS_compare2} show a close match of our theoretical predictions with FE computations and the results of \cite{Sankar1983}. 

As in Paper 1, we now set the Young's modulus and Poisson's ratio to $E=0.083$ MPa and $\nu=0.4$, respectively, as representative of typical adhesives. We also use these material parameters while studying the adhesive beams. The geometric parameters remain unchanged, i.e. $l=40$ mm, $R=225$ mm, $h=4$ mm.

\begin{figure}[h!]
\centering
\begin{multicols}{2}
\includegraphics[width=\linewidth]{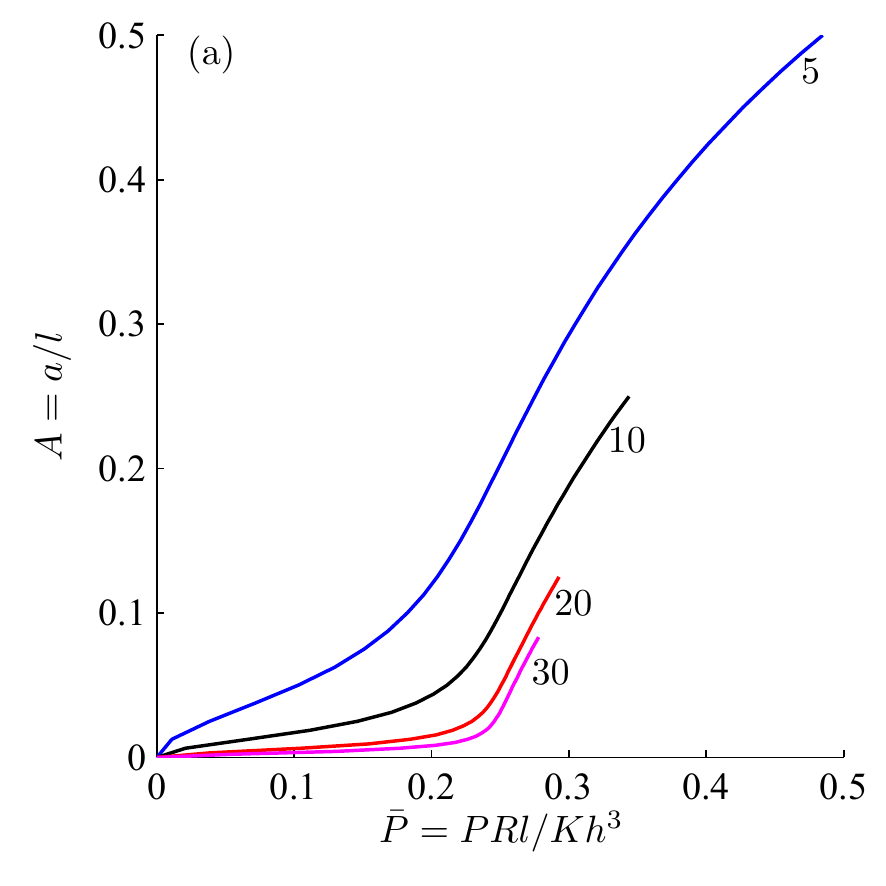} 
\includegraphics[width=\linewidth]{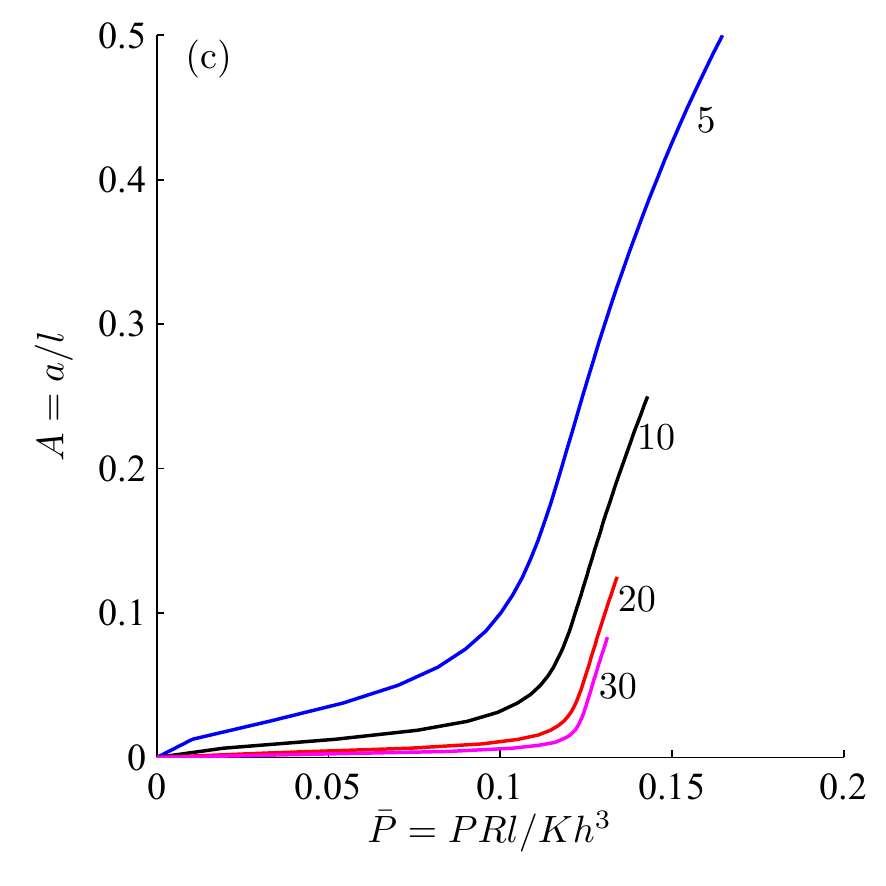} 
\includegraphics[width=\linewidth]{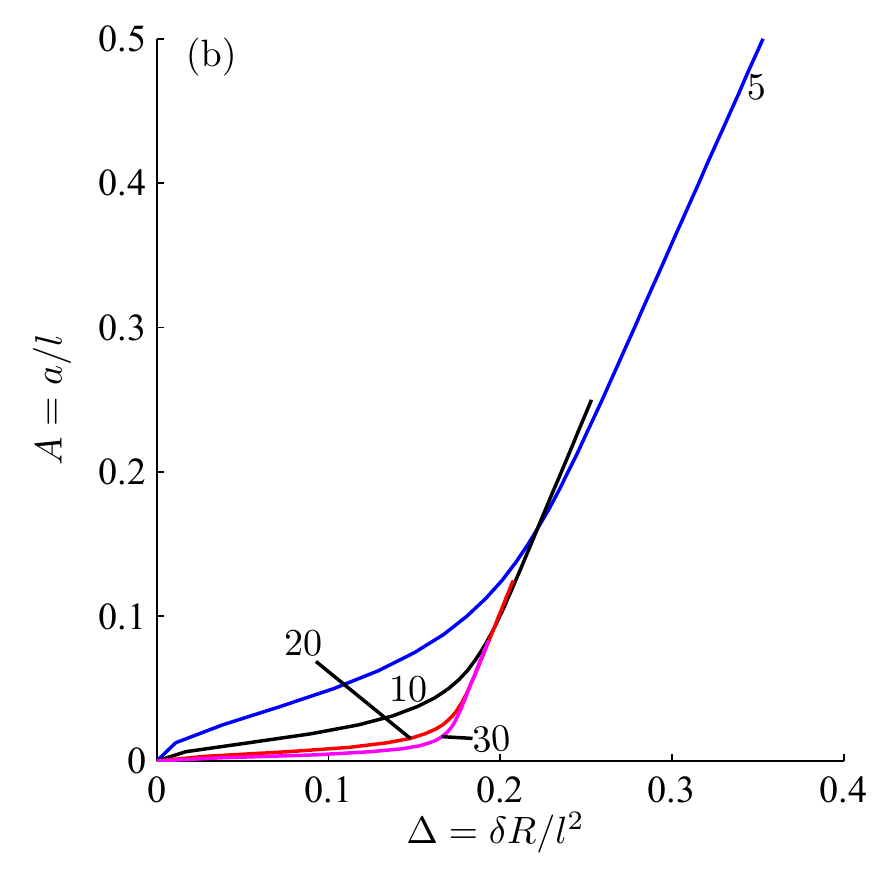} 
\includegraphics[width=\linewidth]{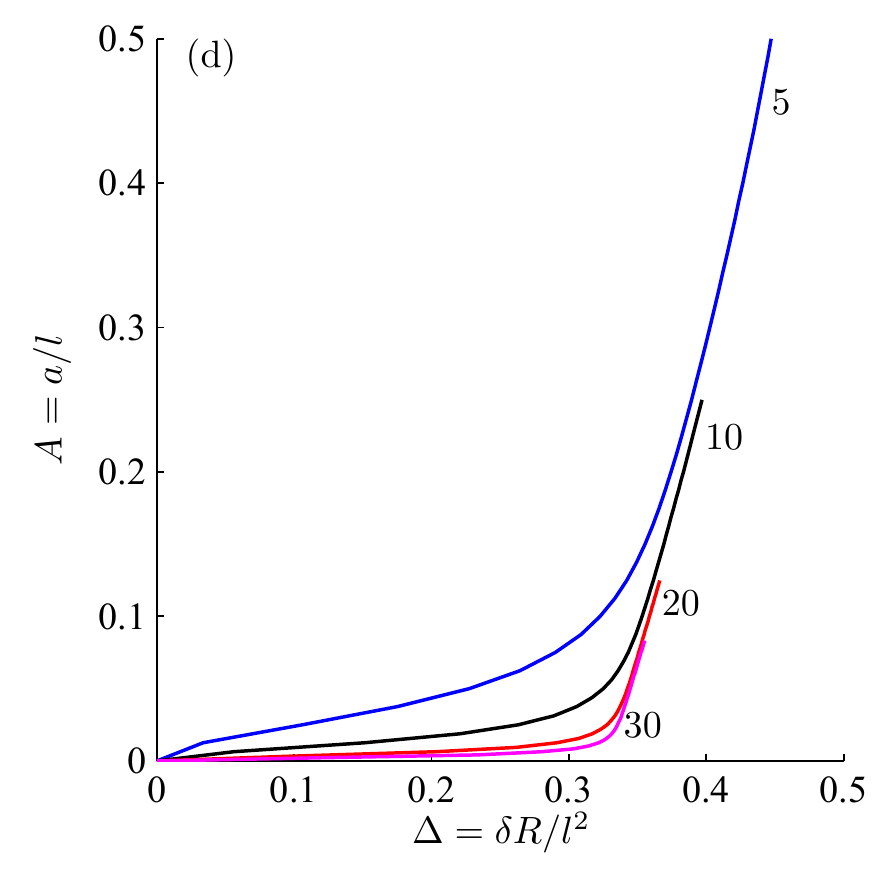} 
\end{multicols}
\caption{Non-adhesive contact of clamped (top row) and simply supported (bottom row) beams. Variation of contact area $A$ with total load $\bar{P}$ and punch's displacement $\Delta$ is shown. Different slenderness ratios $l/h$ are considered and these are noted next to their associated curves.}
\label{Thinbeams:Hertz_A_P_Delta}
\end{figure}

In Fig.~\ref{Thinbeams:Hertz_A_P_Delta}, we plot the variation of the contact area $A$ with respect to the total load $\bar{P}$ acting on the punch and the punch's displacement $\Delta$ at several slenderness ratios $l/h$ for both clamped and simply supported beams. With increasing $l/h$ the beam's resistance to bending decreases and, hence, we find smaller loads $\bar{P}$, or larger deflections $\Delta$, at the same contact area $A$. For the same reason the load $\bar{P}$  required to achieve the same $A$ in simply supported beam is smaller than the ones for clamped beams. At the same time, the displacements $\Delta$ are higher for simply supported beams.

In Fig.~\ref{Thinbeams:Hertz_A_P_Delta}, we observe that the curves change their slope abruptly. This is due to the beam wrapping around the punch rapidly with only a small increase in the load or the punch's displacement. However, the plots in Fig.~\ref{Thinbeams:Hertz_A_P_Delta} do not reflect this aspect well, as $l$ is a common parameter in $A$, $\bar{P}$, $\Delta$ and $l/h$. At the end of this section we will employ an alternative set of non-dimensional variables which will provide clearer insight.

\begin{figure}[h!]
\centering
\subfloat[][]{\includegraphics[width=0.5\linewidth]{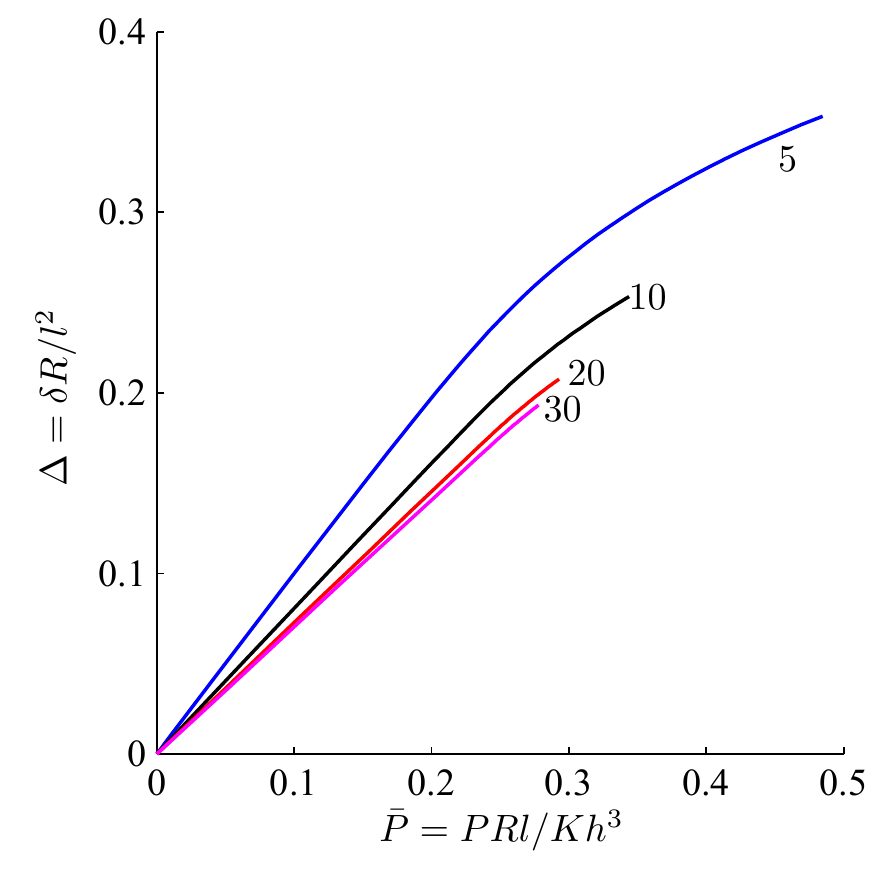}}
\subfloat[][]{\includegraphics[width=0.5\linewidth]{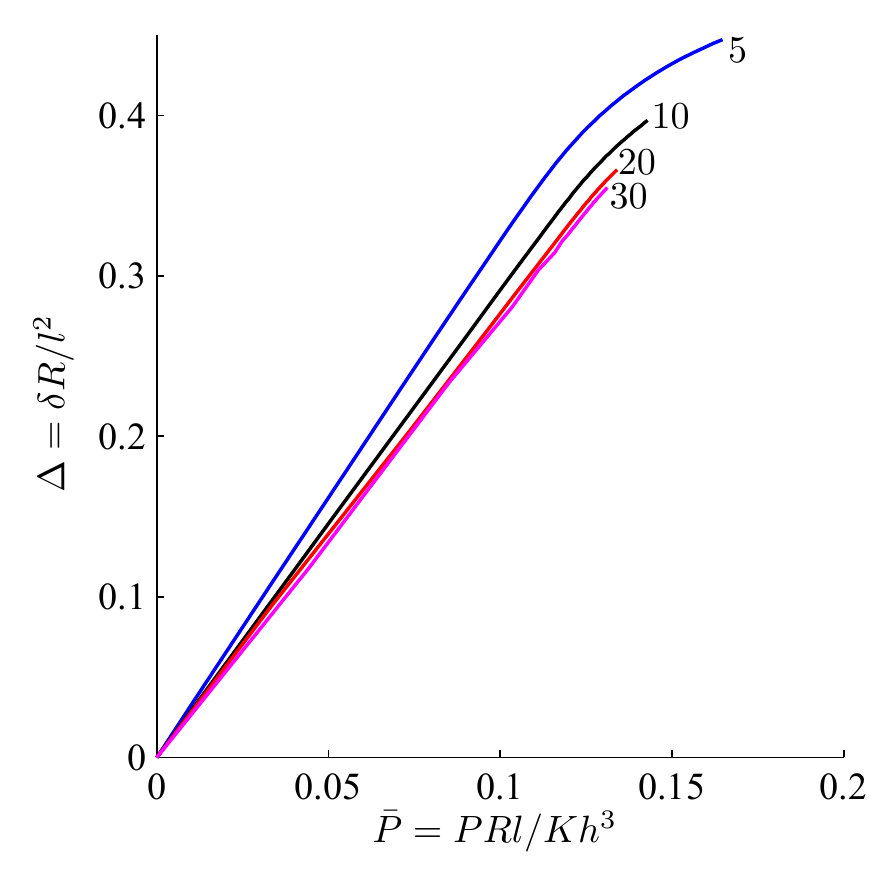} }
\caption{Non-adhesive contact of (a)  clamped and (b) simply supported beams. Variation of punch's displacement $\Delta$ with total load $\bar{P}$ is shown. Several slenderness ratios $l/h$, as noted next to their associated curves, are considered.}
\label{Thinbeams:Hertz_P_Delta}
\end{figure}

Next, in Fig.~\ref{Thinbeams:Hertz_P_Delta} we plot the variation of $\Delta$ with $\bar{P}$ for various $l/h$ for both clamped and simply supported beams. In Paper I, we found that these collapsed onto a single curve. This is not observed in Fig.~\ref{Thinbeams:Hertz_P_Delta}. This collapse observed in Paper I was driven by the assumption that displacement at the bottom surface of the beam was given by the displacement of an Euler-Bernoulli beam. The relationship between $\Delta$ and $\bar{P}$ for Euler-Bernoulli beams with different $l/h$ is self-similar. However, here, the displacement of the bottom surface is found directly as a solution to the elasticity problem and is distinct from that obtained in Paper I. 
%{\color{blue} SHOW PERHAPS?}

\begin{figure}[h!]
\centering
\begin{multicols}{2}
\includegraphics[width=\linewidth]{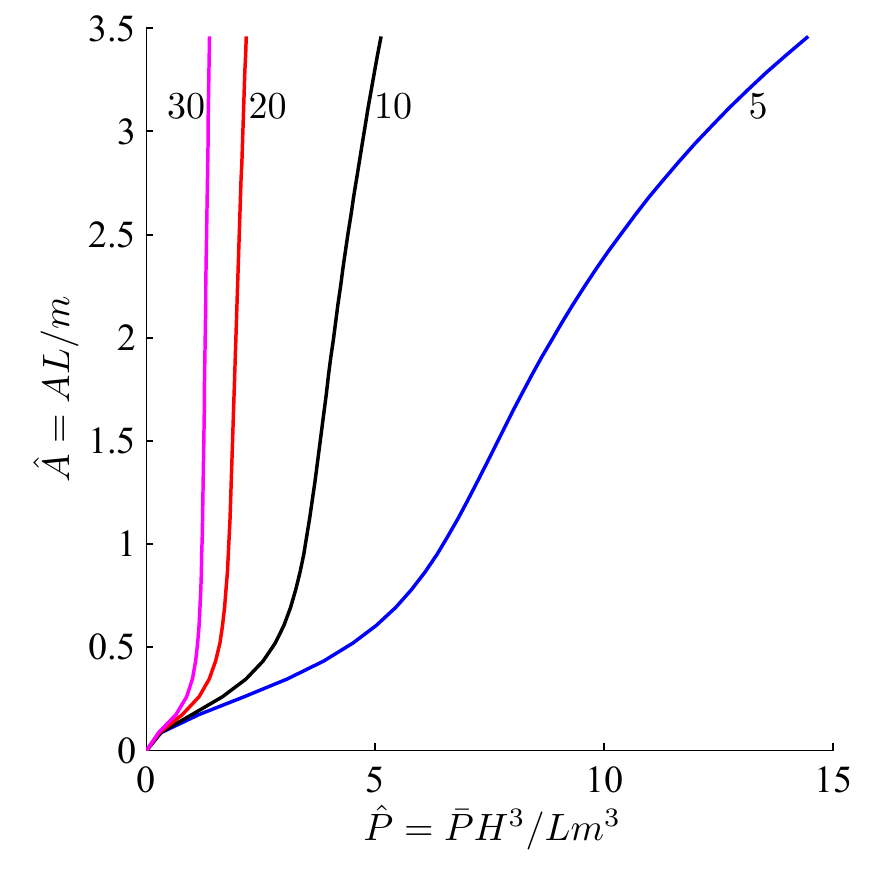} 
\includegraphics[width=\linewidth]{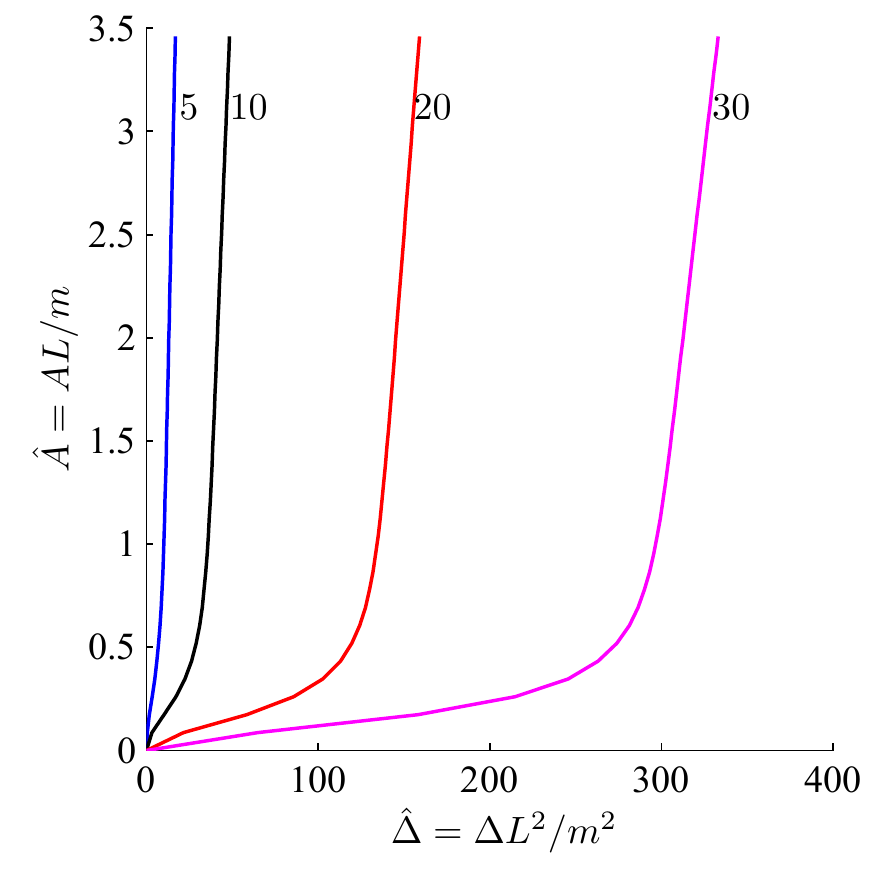} 
\end{multicols}
\caption{Non-adhesive contact of clamped  beams. Variation of contact area $\hat{A}$ with (a) total load $\hat{P}$ and (b) punch's displacement $\hat{\Delta}$ is shown. Different slenderness ratios $l/h$ are considered and these are noted next to their associated curves.}
\label{Thinbeams:Hertz_A_P_Delta2}
\end{figure}

We return to explaing the sudden change of slope observed in the curves of Fig.~\ref{Thinbeams:Hertz_A_P_Delta}. To this end, we follow \cite{Maugis1992adhesion} and employ the non-dimensionalized parameters \begin{equation}
\hat{A} = \frac{AL}{m} = a \left( \frac{K}{\pi w R^2} \right)^{1/3}, \quad \hat{P} = \frac{\bar{P} H^3}{L m^3} = \frac{P}{\pi w} \quad \text{ and} \quad \hat{\Delta} = \frac{\Delta L^2}{m^2} =  \delta \left( \frac{K^2}{\pi^2 w^2 R} \right)^{1/3}, 
\end{equation}
where $H=h/R$, instead of, respectively, $A$, $P$ and $\Delta$, to report our results. We set the adhesion energy $w=0.02 \times 10^{-3} \text{ J}/\text{mm}^2$. In the present case of non-adhesive contact $w$ serves only to facilitate non-dimensionalization. 

We plot the variation of $\hat{A}$ with $\hat{P}$ and $\hat{\Delta}$ at several $l/h$ in Fig.~\ref{Thinbeams:Hertz_A_P_Delta2}. Only clamped beams are considered. The results for simply supported beams are qualitatively similar. The rapid wrapping of the beam is reflected by the sudden increase in A in Fig.~\ref{Thinbeams:Hertz_A_P_Delta2}. As slender beams bend easily, this wrapping happens at lower  loads for such beams; see Fig.~\ref{Thinbeams:Hertz_A_P_Delta2}(a). For the same reason, we observe more displacement in these beams in Fig.~\ref{Thinbeams:Hertz_A_P_Delta2}(b). 

\section{Results: Adhesive contact - JKR approximation}
\label{sec:Results_Thinbeam_JKR}
The JKR approximation is recovered when the scaled adhesive strength $\lambda \rightarrow \infty$ and the adhesive zone vanishes, i.e. $\bar{c} \rightarrow 1$. Hence, equations \eqref{thinbeam:toplayer_Int_eqn_final} and \eqref{thinbeam:bottomlayer_Int_eqn_final} become, respectively, 
\begin{flalign}
&& \Delta  - \frac{1}{2} \, \bar{x}^2 A^2 &= \frac{8 \hat{\gamma}^3}{3 \pi} \sum_{n=0}^{N} b_{2n} \mathcal{J}^{t}_{2n} \left( \bar{x}  \right) + \frac{1}{\pi\hat{\gamma}}  \sum_{n=1}^{M} d_{n} \mathcal{Q}^{t}_{n} \left( \bar{x}  \right) \label{thinbeam:toplayer_Int_eqn_JKR} &&\\
\text{and }
&& 0 &=\frac{8 \hat{\gamma}^3}{3 \pi} \sum_{n=0}^{N} b_{2n} \mathcal{J}^{b}_{2n} \left( \hat{x}  \right) + \frac{\hat{\gamma}}{\pi} \sum_{n=1}^{M} d_{n} \mathcal{Q}^{b}_{n} \left( \hat{x}  \right).
\label{thinbeam:bottomlayer_Int_eqn_JKR} &&
\end{flalign}
The end condition on the contact pressure is determined by the Griffith criterion \eqref{fracture_Griffith_nondim}. By substituting \eqref{thinbeam:phi_cheby} in \eqref{fracture_Griffith_nondim}, we obtain
\begin{equation}
 b_{0} + b_{2} + \cdots + b_{2N}  = - \frac{m}{2 \pi L} \left( \frac{l}{h} \right)^3 \frac{\sqrt{6A m}}{L}.
\label{thinbeam:end_pressure_JKR}
\end{equation}
We now solve \eqref{thinbeam:toplayer_Int_eqn_JKR} -- \eqref{thinbeam:end_pressure_JKR} through the algorithm of Sec.~\ref{sec:Thinbeam_solution_procedure}. 

\begin{figure}[h!]
\centering
\begin{multicols}{2}
\includegraphics[width=\linewidth]{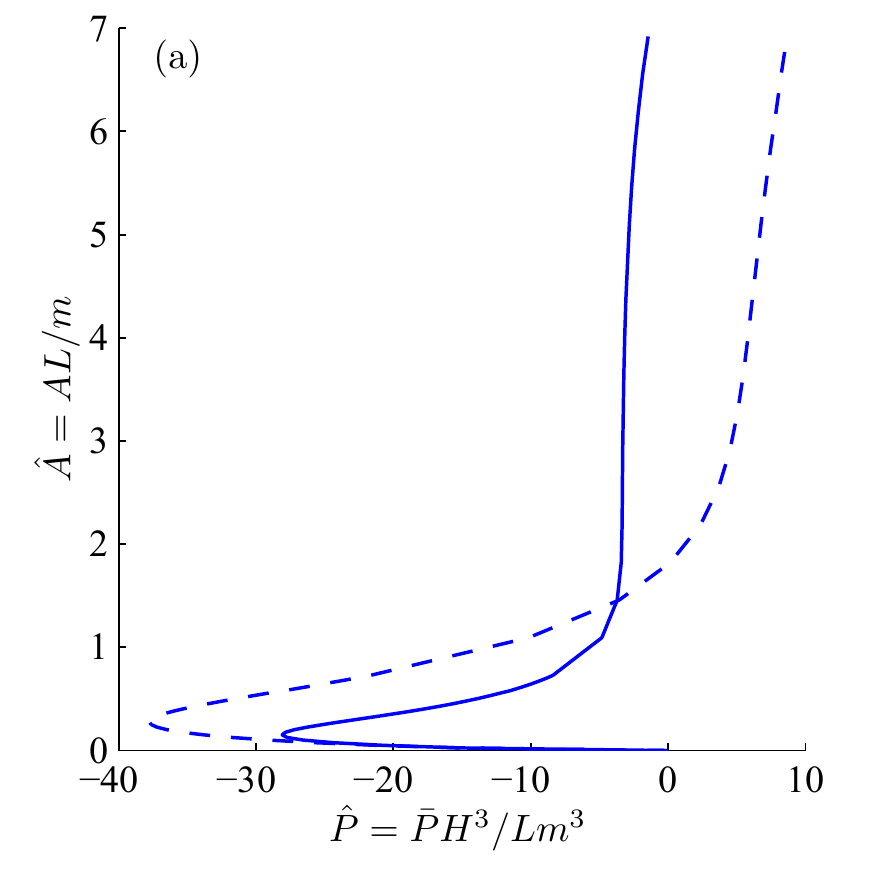}
\includegraphics[width=\linewidth]{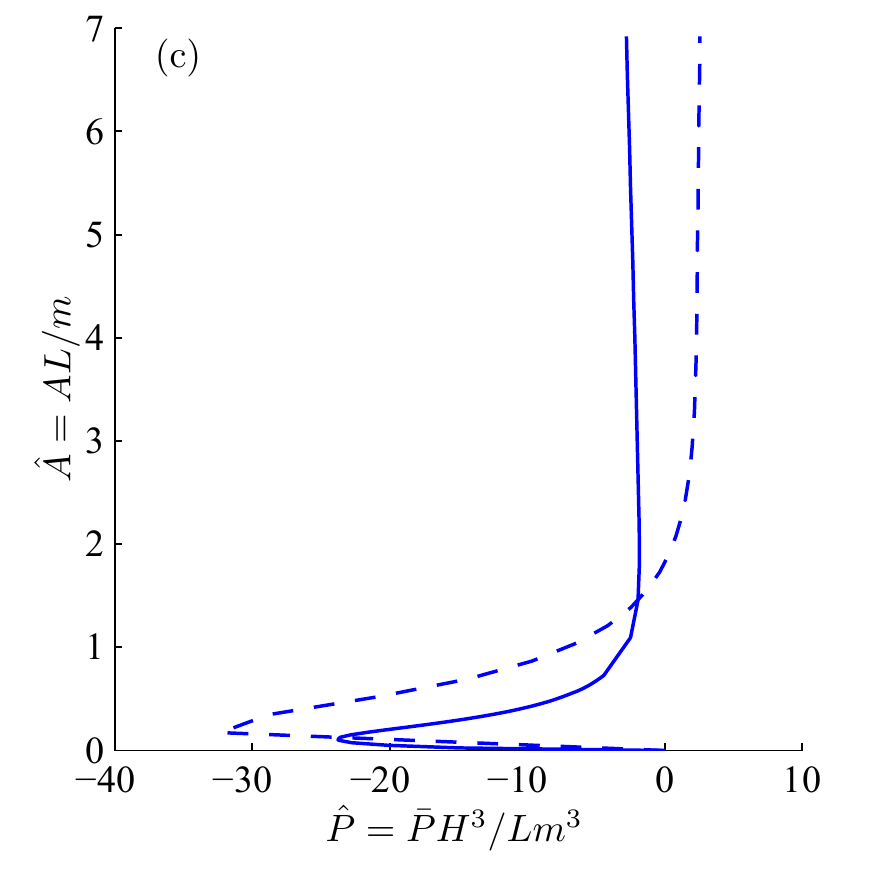}
\includegraphics[width=\linewidth]{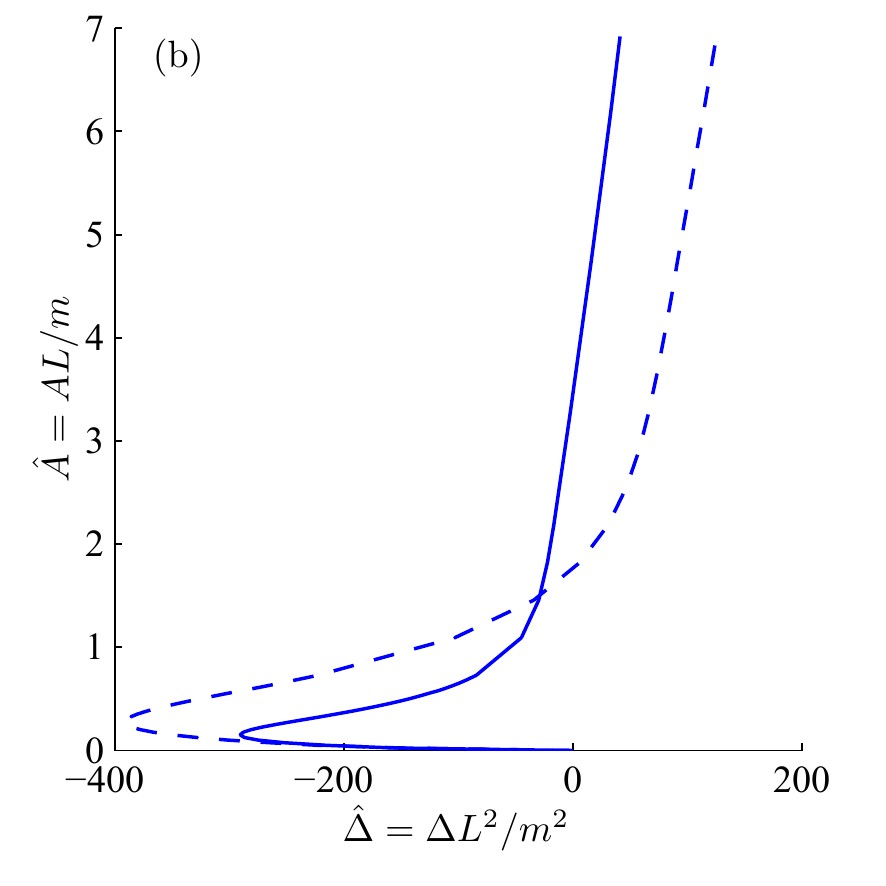}
\includegraphics[width=\linewidth]{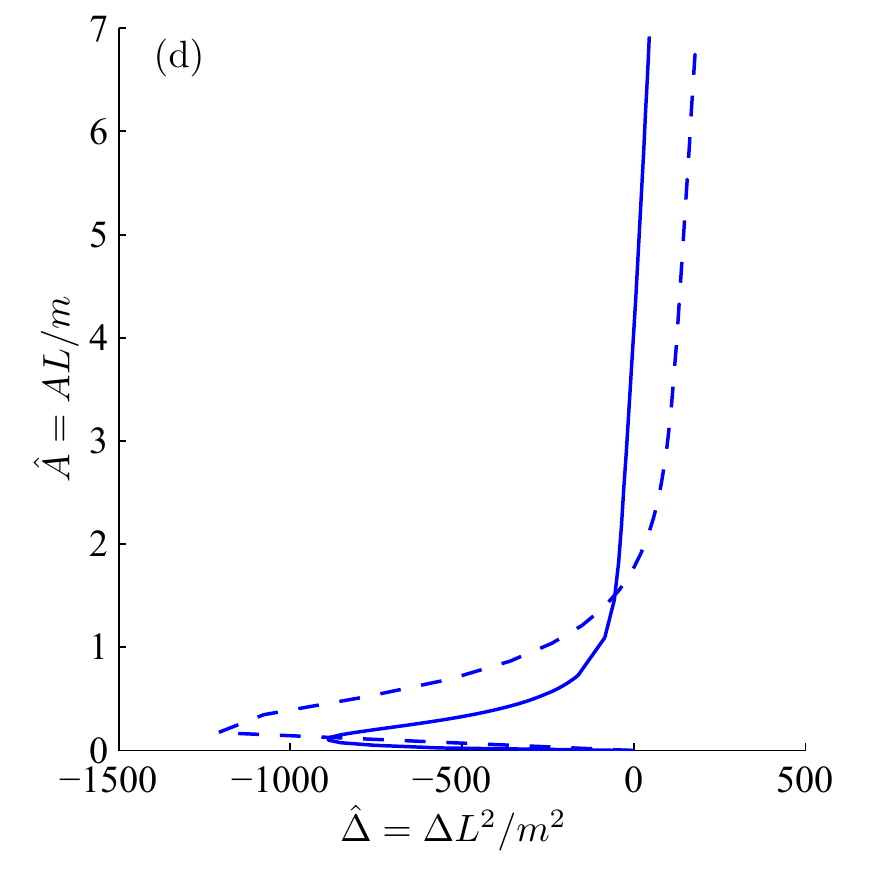}
\end{multicols}
\caption{Adhesive contact of clamped (top row) and simply supported (bottom row) beams with the JKR approximation. Variation of contact area $\hat{A}$ with the total load $\hat{P}$ are shown in (a) and (c), and  the punch's displacement $\hat{\Delta}$ are shown in (b) and (d). The beam's slenderness ratio $l/h=10$.  Solid lines correspond to $l=40$ mm and $h=4$ mm, while the dashed line is for a beam with $l=80$ mm and $h=8$ mm.}
\label{Thinbeams:JKR_A_P_Delta_samehbyl}
\end{figure}

\begin{figure}[h!]
\centering
\begin{multicols}{2}
\includegraphics[width=\linewidth]{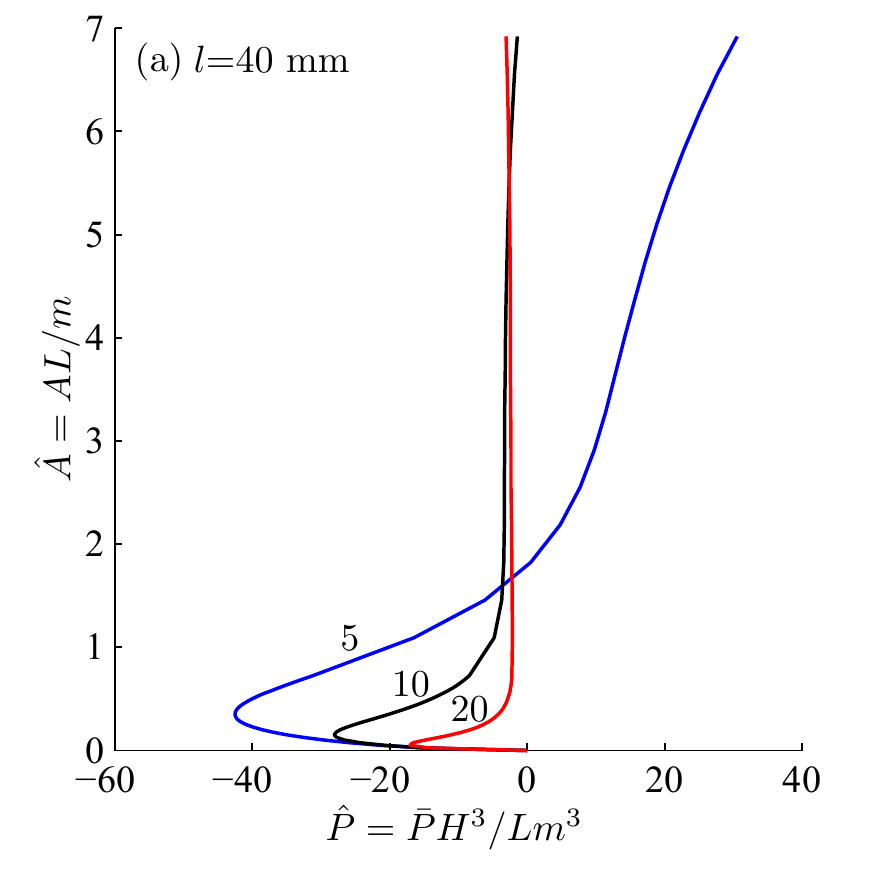}
\includegraphics[width=\linewidth]{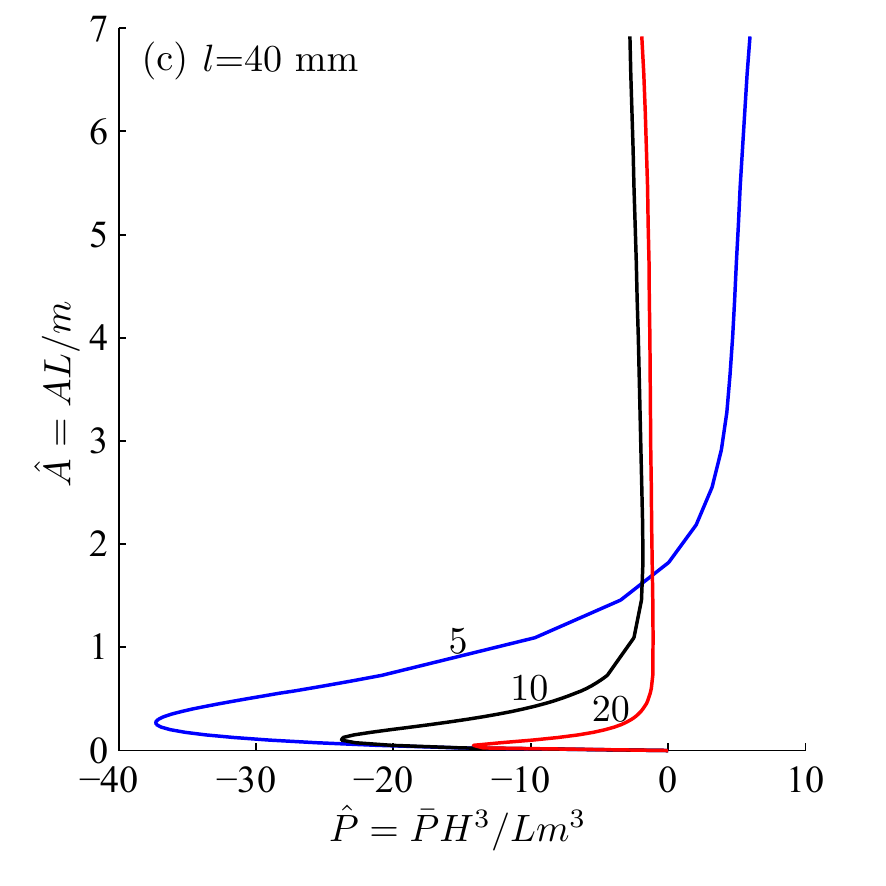}
\includegraphics[width=\linewidth]{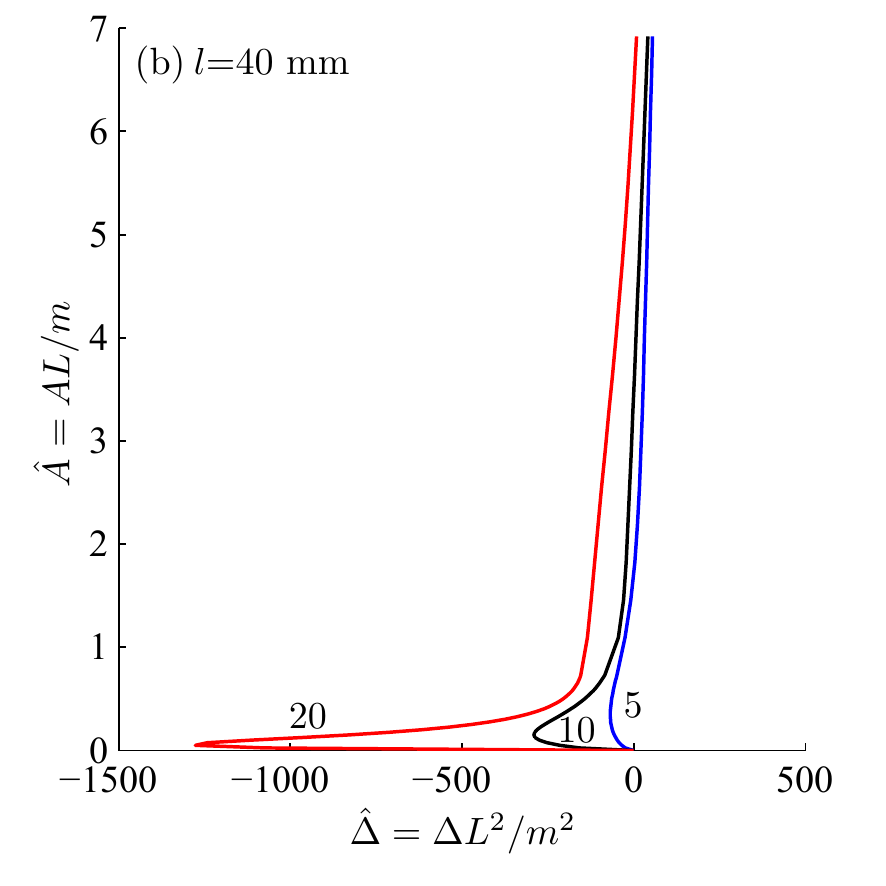} 
\includegraphics[width=\linewidth]{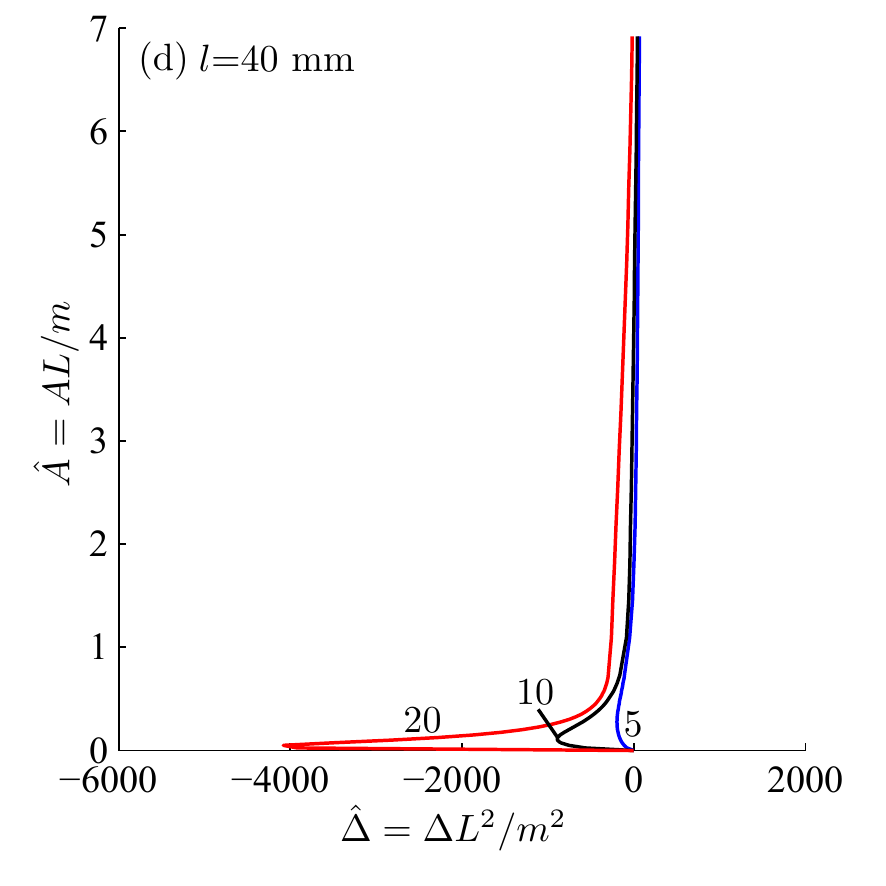} 
\end{multicols}
\caption{Adhesive contact of clamped (top row) and simply supported (bottom row) beams with the JKR approximation.  Left column, i.e. (a) and (c), reports the variation of contact area $\hat{A}$ with total load $\hat{P}$, while the right column, i.e. (b) and (d), plots the change of $\hat{A}$ with the punch's displacement $\hat{\Delta}$. Results are obtained for several slenderness ratios $l/h$ as noted next to their curves, while keeping $l=40$ mm.}
\label{Thinbeams:JKR_A_P_Delta_change_h}
\end{figure}

\begin{figure}[h!]
\centering
\begin{multicols}{2}
\includegraphics[width=\linewidth]{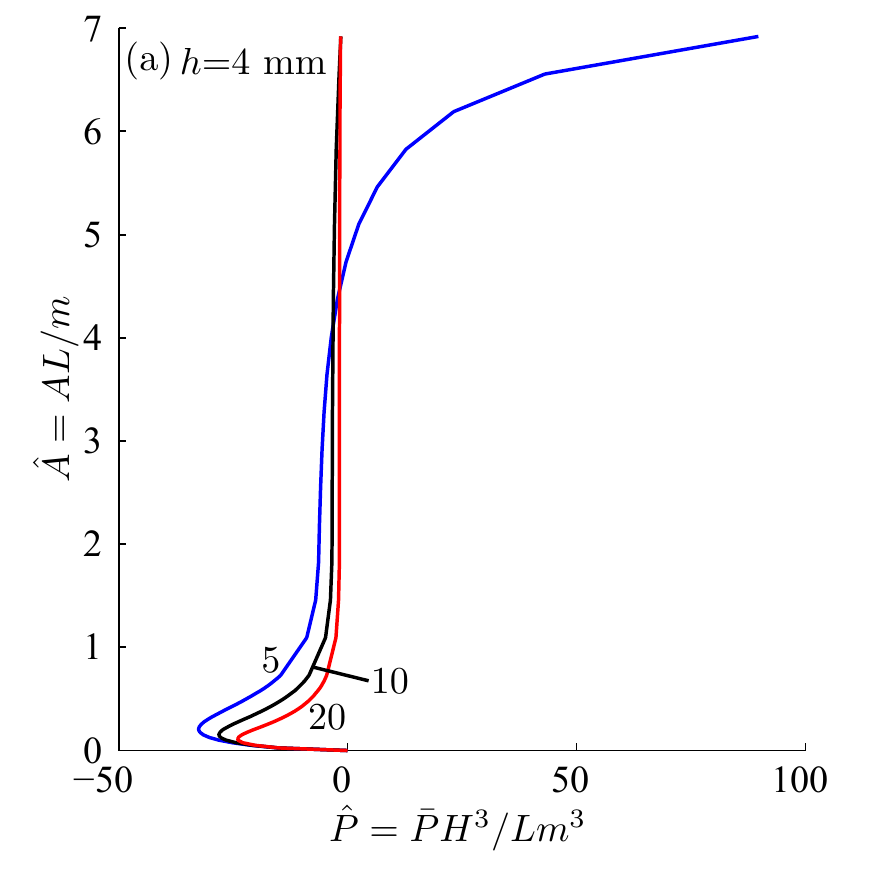}
\includegraphics[width=\linewidth]{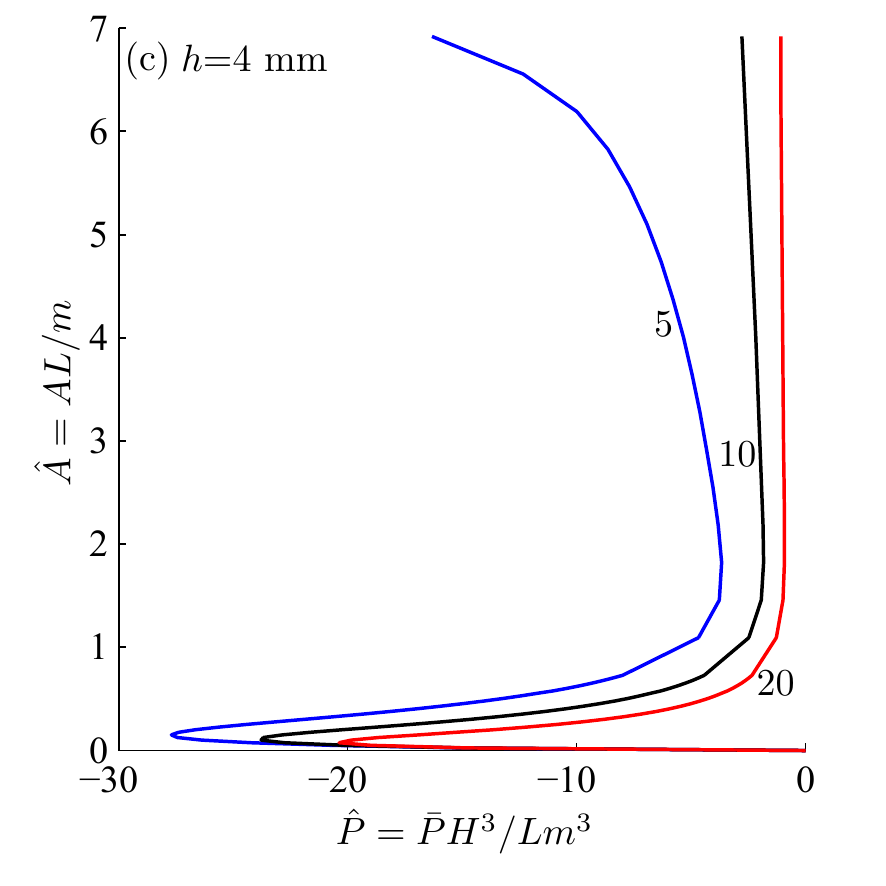}
\includegraphics[width=\linewidth]{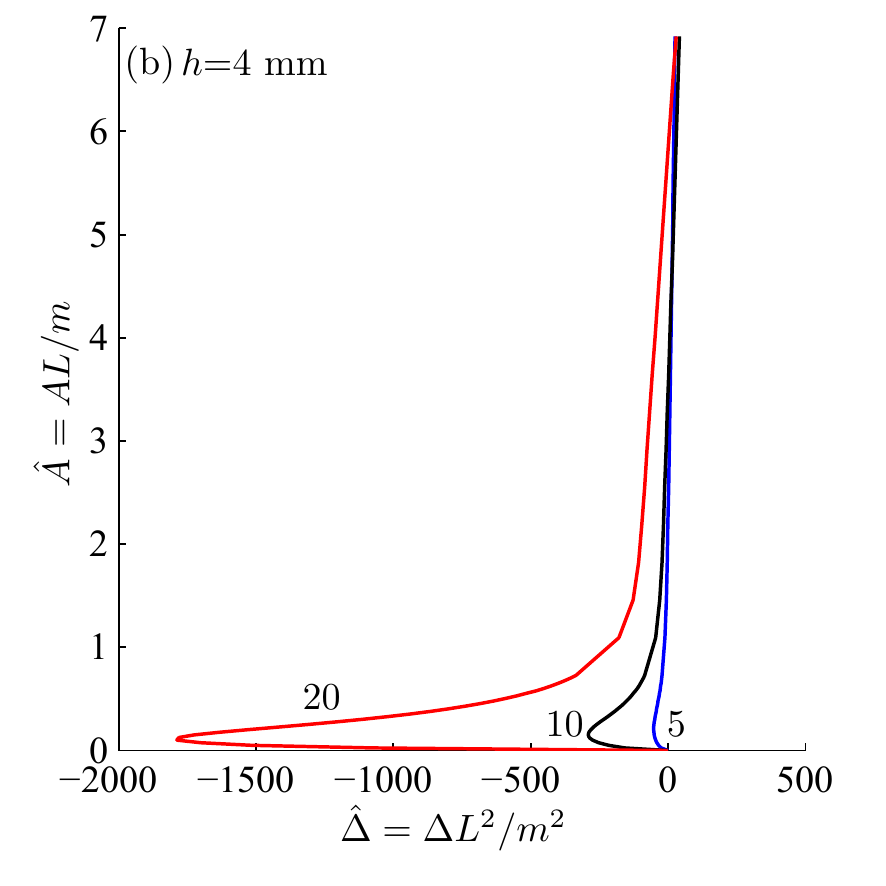} 
\includegraphics[width=\linewidth]{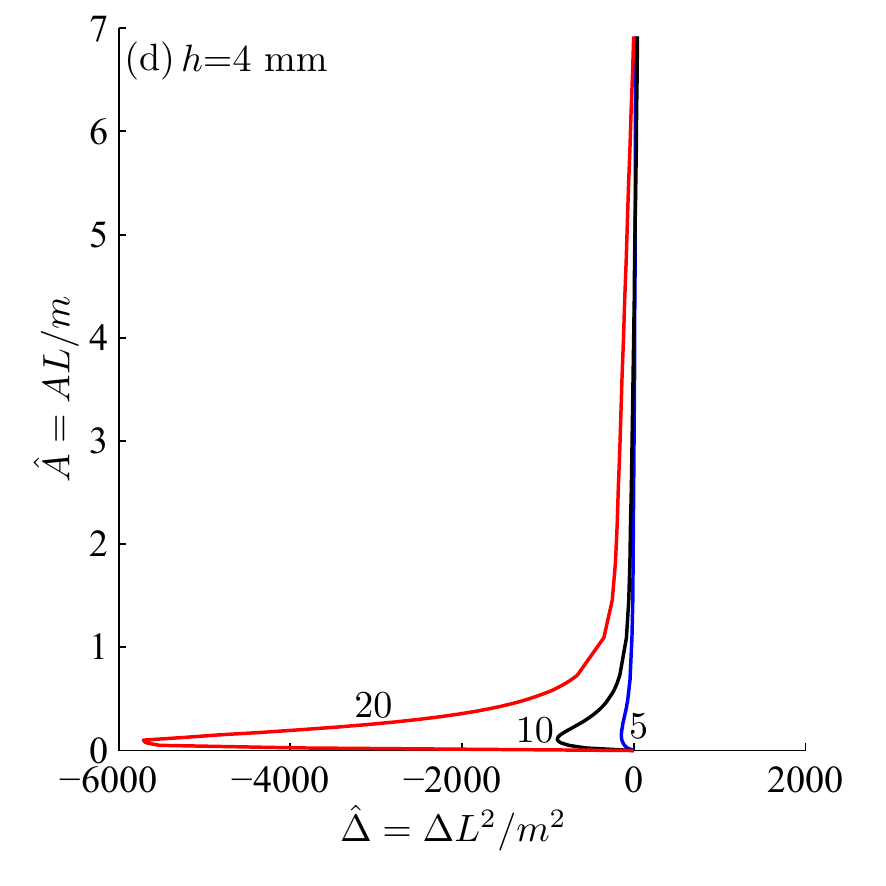} 
\end{multicols}
\caption{Adhesive contact of clamped (top row) and simply supported beams (bottom row) with the JKR approximation.  Left column, i.e. (a) and (c), reports the variation of contact area $\hat{A}$ with total load $\hat{P}$, while the right column, i.e. (a) and (c), plots the change of $\hat{A}$ with the punch's displacement $\hat{\Delta}$. Results are obtained for several slenderness ratios $l/h$ as noted next to their curves, while keeping $h=4$ mm.}
\label{Thinbeams:JKR_A_P_Delta_change_l}
\end{figure}

In Fig.~\ref{Thinbeams:JKR_A_P_Delta_samehbyl} we plot the variation of the contact area $\hat{A}$ with the load $\hat{P}$ acting on the punch and the displacement $\hat{\Delta}$ of the punch for both clamped and simply supported beams. The slenderness ratio $l/h$ is kept constant, but two different combinations of $l$ and $h$ are investigated. We observe that the curves for same $l/h$ are sensitive to $l$ and $h$ individually and depend \textit{not} only on the slenderness ratio. This was also observed in Paper I. This may be traced back to the presence of $L=l/R$ on the right hand side of \eqref{thinbeam:end_pressure_JKR}. It is easier to explore the dependence of $h$ and $l$ repeatedly by employing $\hat{A}$, $\hat{P}$ and $\hat{\Delta}$ and we do so in  Figs.~\ref{Thinbeams:JKR_A_P_Delta_change_h} and \ref{Thinbeams:JKR_A_P_Delta_change_l}. 

%Next, we plot $\hat{A}$ versus $\hat{P}$ and $\hat{\Delta}$, for both clamped and simply supported beams. 
Curves in Fig.~\ref{Thinbeams:JKR_A_P_Delta_change_h} are obtained for several $l/h$ by varying $h$ while keeping $l=40$ mm. Beams with high $l/h$ ratio bend easily due to the adhesion, and we observe smaller negative loads $\hat{P}$ and larger negative displacements $\hat{\Delta}$ at a given contact area $\hat{A}$. Note that negative loads and displacements indicate, respectively, tensile force on the punch and the upward bending of beams. From Fig.~\ref{Thinbeams:JKR_A_P_Delta_change_h} we observe that these slender beams wrap around -- as indicated by sudden slope changes in $\hat{A}$ versus $\hat{P}$ and $\hat{\Delta}$ curves -- the punch earlier, i.e. at smaller $\hat{P}$. For sufficiently slender beams, the wrapping occurs even when there is no compressive (positive) load on the punch.  In these beams the bending resistance is unable to counterbalance adhesive forces.  The above features, viz. extent of wrapping and the response to adhesive forces, are, expectedly, heightened in the case of simply supported beams, whose bending resistance is lower. 

Finally, we plot the variation of the contact area $\hat{A}$ with the load $\hat{P}$ acting on the punch and the displacement $\hat{\Delta}$ for several slenderness ratios $l/h$. We change $l$ and set $h=4$ mm. The results for both clamped and simply supported beams are shown in Fig.~\ref{Thinbeams:JKR_A_P_Delta_change_l}. Qualitatively Fig.~\ref{Thinbeams:JKR_A_P_Delta_change_l} is similar in many respects to Fig.~\ref{Thinbeams:JKR_A_P_Delta_change_h}.  From Fig.~\ref{Thinbeams:JKR_A_P_Delta_change_l}(c), we observe that, for $\hat{A} \gtrsim 1.5$  in a simply supported beam of $l/h=5$, the load $\hat{P}$  decreases with the increase in contact area $\hat{A}$. At the same time, the punch's displacement $\hat{\Delta}$ increases; see Fig.~\ref{Thinbeams:JKR_A_P_Delta_change_l}(d). This is explained by the presence of negative (tensile) stresses at the center of the contact area in addition to the very large negative stresses allowed in the contact pressure distribution at the contact edges. The contact area over which these tensile stresses act also increases with the increase in contact area. Hence, the load $\hat{P}$ decreases with the increase in contact area $\hat{A}$.

\section{Results: Adhesive contact with an adhesive zone model}
\label{sec:Results_Thinbeam_Maugis}
Finally, we study the behaviour of adhesive beams indented by a rigid cylindrical punch with the help of adhesive-zone models. In these models an adhesive force acts over an adhesive zone of length $d=c-a$ outside the contact area. Here, we model the distribution of the adhesive forces through the Dugdale-Barenblatt model, so that the normal traction on the (extended) beam's top surface is given by \eqref{thinbeam:Pfn}. The contact problem is resolved by solving \eqref{thinbeam:toplayer_Int_eqn_numerical} -- \eqref{thinbeam:Griffith_Eqn_numerical} following the solution procedure of Sec.~\ref{sec:Thinbeam_solution_procedure}. 
\begin{figure}[h!]
\centering
\begin{multicols}{2}
\includegraphics[width=\linewidth]{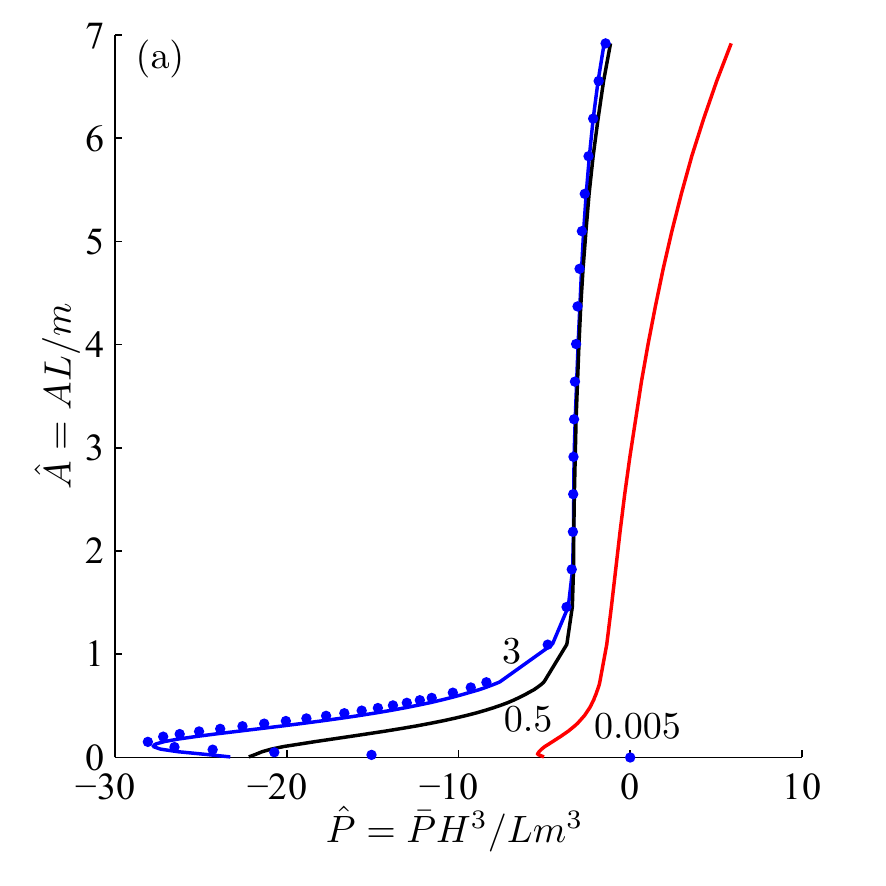}
\includegraphics[width=\linewidth]{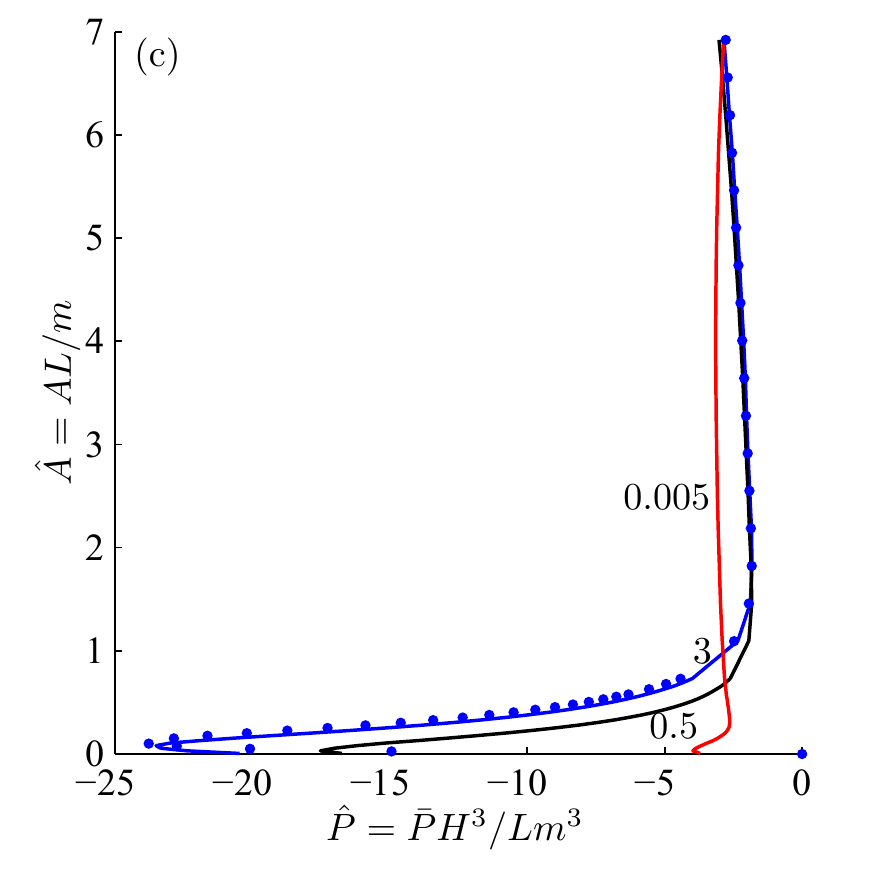}
\includegraphics[width=\linewidth]{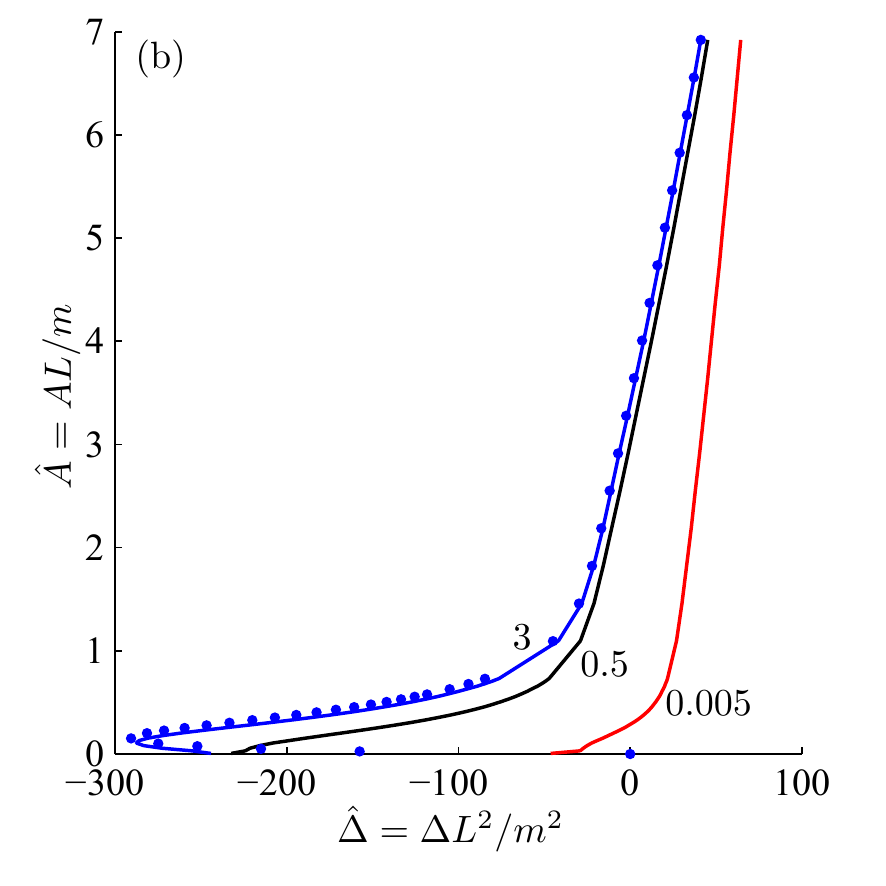} 
\includegraphics[width=\linewidth]{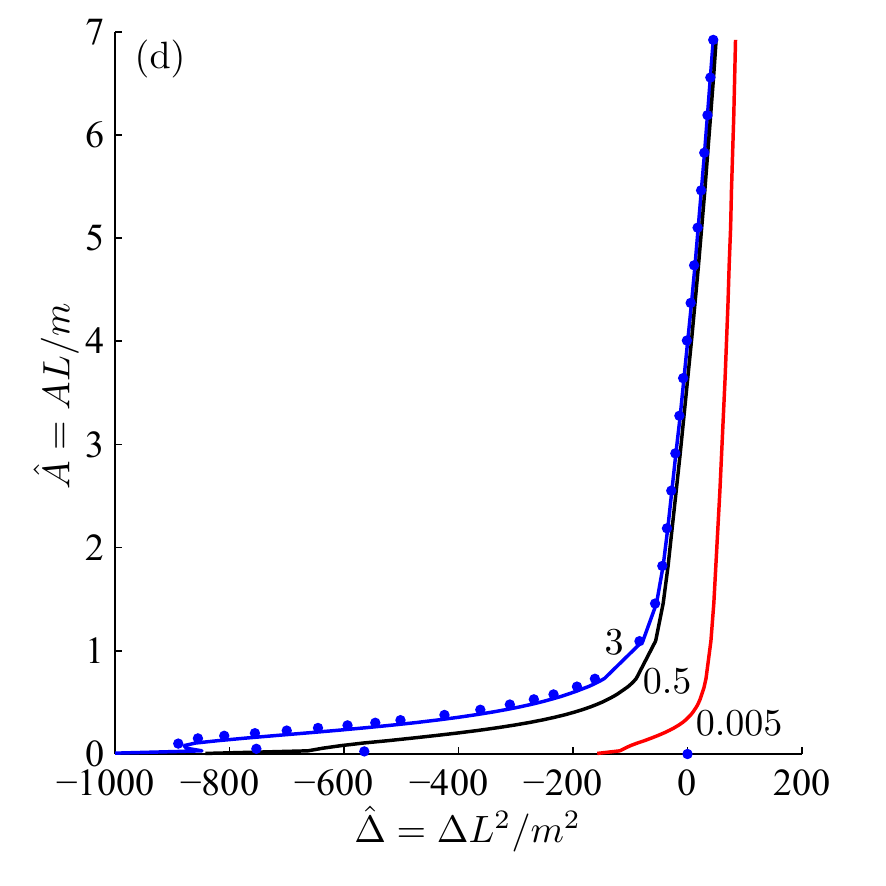} 
\end{multicols}
\caption{Adhesive contact of clamped (top row) and simply supported (bottom row) beams with an adhesive zone model. Variation of contact area $\hat{A}$ with the total load $\hat{P}$ are shown in (a) and (c), and the punch's displacement $\hat{\Delta}$ are shown in (b) and (d). Different adhesive strengths $\lambda$ are considered and these are indicated next to their associated curves. The beam's thickness $h=4$ mm and $l=40$ mm. Filled circles represent the JKR solution for the corresponding beam; cf. Sec.~\ref{sec:Results_Thinbeam_JKR}}
\label{Thinbeams:Maugis_A_P_Delta_diffF}
\end{figure}

For brevity, we report here the effect of only the adhesive strength $\lambda$ at a given $l$ and $h$, as the response to varying $l/h$ is found to be the same as in Sec.~\ref{sec:Results_Thinbeam_JKR}. 
% for clamped and simply supported beams.

In Fig.~\ref{Thinbeams:Maugis_A_P_Delta_diffF} we plot the variation of the contact area $\hat{A}$ with the total load $\hat{P}$ and the displacement $\hat{\Delta}$ for several adhesive strengths $\lambda$. We observe that as $\lambda \rightarrow 0$ the results approach those obtained for non-adhesive interaction as in Sec.~\ref{sec:Results_thinbeam_Hertz}. At the same time, increasing adhesive strength pushes our results towards those obtained for the JKR approximation in Sec.~\ref{sec:Results_Thinbeam_JKR}.

\begin{figure}[h!]
\centering
\includegraphics[width=0.5\linewidth]{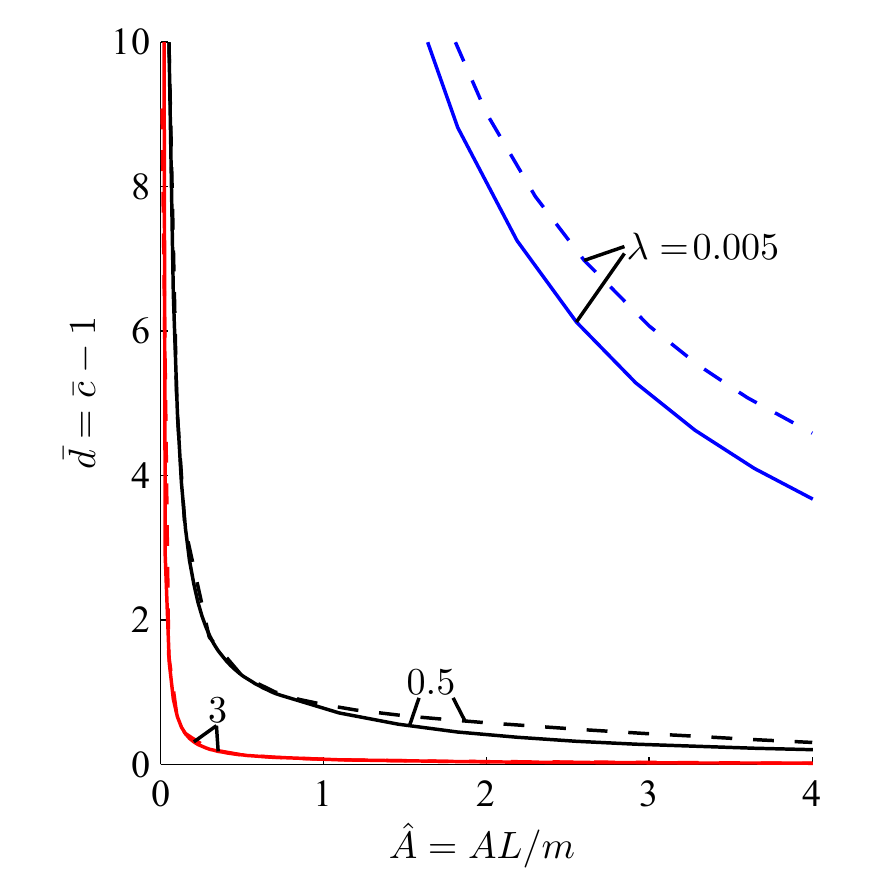}
\caption{Variation of the adhesive zone size $\bar{d}$ with the the contact area $\hat{A}$ at different adhesive strengths $\lambda$ for a clamped beam. We set l=$40$ mm and $h=4$ mm. Solid lines indicate the predictions of the formulation mentioned in this article, while the dotted lines are those obtained from the formulation of the Paper I.}
\label{Thinbeam_lambda_dbar}
\end{figure}

Next, we study the effect of adhesive strength $\lambda$ on the adhesive zone size $\bar{d} = \bar{c}-1$. For this, we plot $\bar{d}$ by varying $\hat{A}$ for several adhesive strengths in Fig.~\ref{Thinbeam_lambda_dbar}. From Fig.~\ref{Thinbeam_lambda_dbar}, we observe that with increasing $\lambda$, $\bar{d}$ goes to zero. We also find that, due to the difference in formulations, the results obtained in this article are quite different from those of Paper 1 at low to moderate adhesive strengths. At high $\lambda$ there is not much difference between the two. Finally, as seen in Paper I, varying the slenderness ratio $l/h$ and constraints imposed by end supports do not effect the adhesive zone size much. %, also seen in Paper 1.

%{\color{red} Finally, studying the effect of adhesive strength $\lambda$, slenderness ratio $l/h$ and constraints imposed by end supports on the adhesive zone size $\bar{d}=\bar{c}-1$  is similar to what we observed in part 1. Hence we do not describe this in this article.} {\color{blue} WOULD BE NICE TO REPORT IMPROVEMENTS COMPARED TO PAPER I}

\section{Comparisons with Paper~I}
\label{comparisons_with_paper1}
\begin{figure}[h!]
\centering
\subfloat[][]{\includegraphics[width=0.5\linewidth]{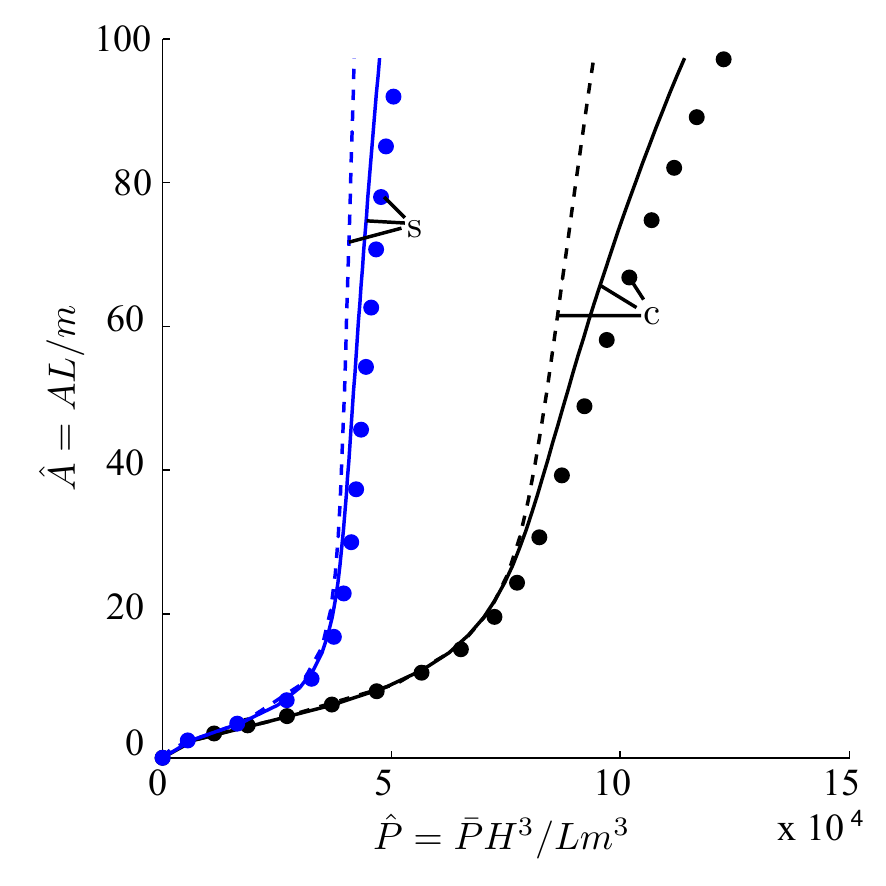}}
\subfloat[][]{\includegraphics[width=0.5\linewidth]{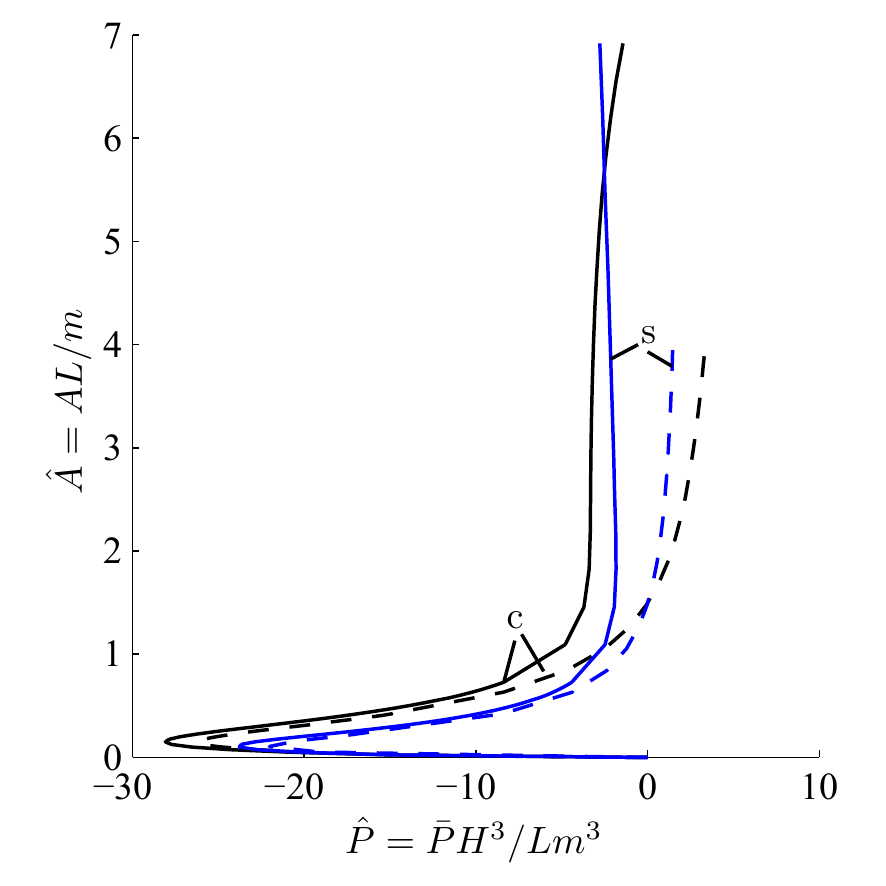} }
\caption{ The contact area $A$ is plotted as a function of the total load $\bar{P}$ acting on the punch. (a) Non-adhesive contact of clamped (`c') and simply supported (`s') beams. The beam's slenderness ratio $l/h$=10. We set $E=2000$ MPa  and $\nu=0.3$. (b) Adhesive contact of clamped (`c') and simply supported (`s') beams with JKR approximation. We set $h=4$ mm, $l=40$ mm, $E=0.083$ MPa and $\nu=0.4$.  Solid lines are results obtained from the semi-analytical procedure of Sec.~\ref{sec:Thinbeam_Numerical_solution}, while the dashed line are results from Paper~I. Filled circles correspond to FE simulations of Sec.~\ref{sec:Thinbeam_FE_model}.}
\label{comparison_with_Paper1}
\end{figure}

We now compare predictions of the formulation of this paper with Paper~I for both non-adhesive and adhesive contacts. For this, we plot the variation of the contact area $\hat{A}$ with variation in the total load $\hat{P}$ acting on the punch for clamped and simply supported beams as shown in Fig.~\ref{comparison_with_Paper1}. For the non-adhesive beams, we also plot FE results to show the comparison better; see Fig.~\ref{comparison_with_Paper1}(a). From Fig.~\ref{comparison_with_Paper1}(a), we observe that both the semi-analytical formulations predicts the behavior of the beams correctly for small $\hat{A}$.  However, with increasing $\hat{A}$, predictions of the current formulation are closer to the FE simulations than those of Paper~I. Finally, from Fig.~\ref{comparison_with_Paper1}(b), we observe that both the formulations predicts the same behavior in the beams before wrapping in the JKR approximation of the contact. However, for a large portion of the contact area the results of both the formulations predicts the behaviour of the beams differently. Thus, to establish the correctness of these formulations, we need to study the behaviour of the beams experimentally. So, in the next section, we check the experimental feasibility for the indentation in adhesive beams. 

\section{Experimental validation feasibility for the adhesive contact}
\begin{figure}[h!]
\centering
\includegraphics[width=0.5\linewidth]{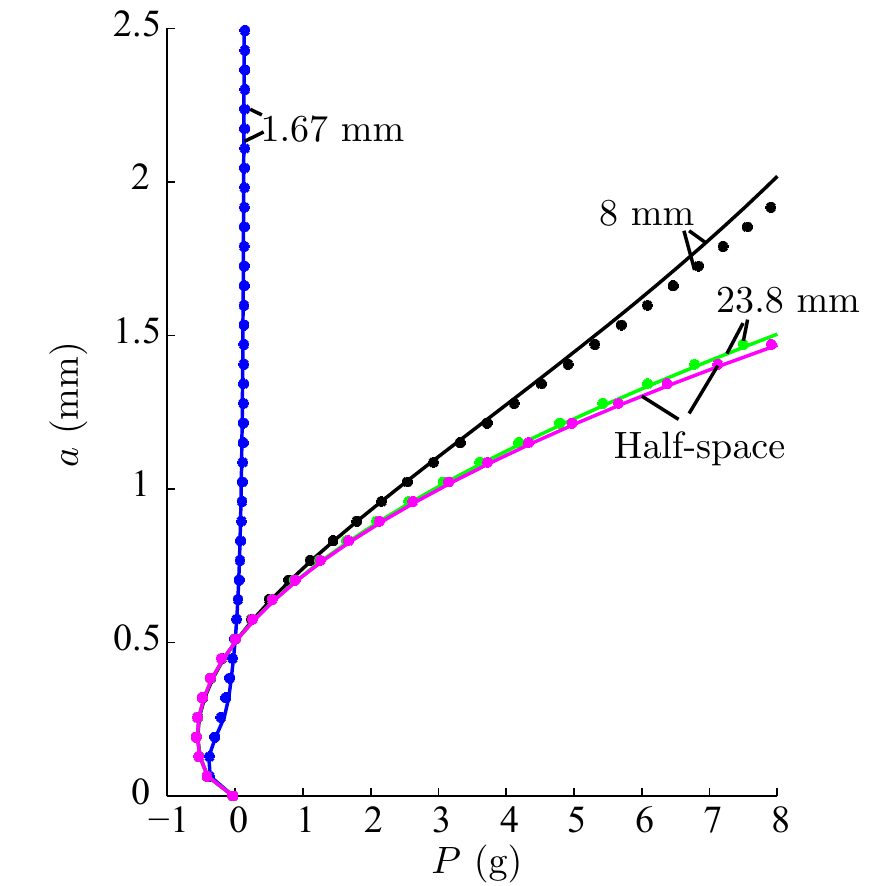}
\caption{ Variation of the contact area a (in mm) with the total load P (in g) for adhesive clamped beams with half-span l = 50 mm at several thicknesses $h$, as noted next to their curves. We set $E =1.237$ MPa, $W =27.941$ N/mm. The contact is modeled by the JKR approximation of the contact. Solid lines correspond to theoretical predictions of Paper I, while the asterisk represent those of current approach.}
\label{thinbeam_thick beam_experimental_comparison}
\end{figure}

In this section, we see the feasibility of conducting the JKR experiments for the beams with our experimental set-up mentioned in Paper I.  For this purpose, we first obtain predictions for PDMS (poly-dimethyl- siloxane) samples employed in Paper I utilizing the formulations of this paper and Paper I. The PDMS samples were prepared using a 10:1 weight ratio mixture of Sylgard 184 silicone elastomer base and curing agent. We consider clamped beams with fixed half-span $l=50$ mm and vary the thickness $h$ between $1$ mm and $25$ mm. The results are shown in Fig.~\ref{thinbeam_thick beam_experimental_comparison}. 

In our experiments, the maximum value of the load $P$ is limited by our micro-weighing balance. From Fig.~\ref{thinbeam_thick beam_experimental_comparison}, we observe that with a 10:1 PDMS material, we can \textit{not} clearly distinguish the results obtained by the formulation of this paper and that of Paper I. The limitations of the load measuring capability of our micro-weighing balance used in the experimental set-up makes differentiating between the predictions of our current and previous (Paper I) formulations hard. To achieve this with our current experimental set-up  requires a softer material with strong adhesive characteristics than PDMS. 
%Currently, we are in search of a suitable material with which we may identify the extent of applicability of our two formulations.

\section{Conclusions} 
\label{sec:Thinbeam_conclusions}
The assumption made in the Paper I about the vertical displacement of the beam's bottom surface limits the range of the corresponding semi-analytical analysis. In this paper we removed this constraint by solving for the vertical displacement directly, thereby extended the applicability of the formulation. The current formulation modeled the indention of adhesive beam in terms of two coupled Fredholm integral equations of first kind. These dual integral equations were solved through a collocation technique employing series approximations for the unknown contact pressure and bottom surface's vertical displacement. Care should be exercised when selecting the approximating series for the displacement. We then compared our predictions with FE simulations and previously published results for non-adhesive indentation, and found a satisfactory match for a wide range of indentation. Next, we presented the results for JKR approximation and adhesive zone model approximation of the contacts. Finally, we studied the experimental feasibility to establish the varacity of the current approach. However, for this we need a softer adhesive material than the PDMS -- which we are in search of. 

In the future we aim to extend the present approach to three-dimensional axi-symmetric indentation. More immediately though, in part III of this work, we will present experimental data for the adhesive indentation of beams.

%\pagebreak
\appendix
\section{Vertical displacement of the beam's top surface ($V \left( \xi, 0 \right)$) and normal traction at its bottom surface ($S_{yy} \left( \xi, h \right)$) in Fourier space}
\label{sec:Fourier_tranforms}
In Fourier space the transformed horizontal and vertical displacements may be solved as, respectively, 
\begin{align}
\label{Gen_disp_1}
U \left( \xi, y \right) &= \left(\kappa a_1 + \xi y \left( a_1+ i a_3 \right) \right) e^{\xi y}+\left( \kappa b_1 - \xi y \left( b_1 - i b_3 \right) \right) e^{-\xi y}, \\  
%\intertext{and}
\text{and } \quad
\label{Gen_disp_2}
V \left( \xi, y \right) &= \left(\kappa a_3 + i \xi y \left( a_1+ i a_3 \right) \right) e^{\xi y}+\left( \kappa b_3 + i \xi y \left( b_1 - i b_3 \right) \right) e^{-\xi y},
\end{align}
with
\begin{equation}
U \left( \xi, y \right) = \int\limits_{-\infty}^{\infty} u \left( x, y\right) e^{i \xi x} dx, \quad 
V \left( \xi, y \right) = \int\limits_{-\infty}^{\infty} v \left( x, y\right) e^{i \xi x} dx, \nonumber
\end{equation}
and unknown constants $a_1$, $a_3$, $b_1$, and $b_3$. These constants are obtained by applying the boundary conditions,  which in Fourier space are 
\begin{subequations}
\label{BCS_for_contact_xi}
\begin{alignat}{3}
\text{at} \; \;  y=0: &\quad S_{\xi y} &= 0,  &\quad S_{yy} &= \bar{P} (\xi), \label{BC_y0_xi}\\
\text{at} \; \;  y=h: &\quad S_{\xi y} &= 0,  &\quad V &= \bar{v} (\xi), \label{BC_yh_xi}
\end{alignat}
\end{subequations}
with
\begin{eqnarray}
S_{\xi y} &=& \frac{E}{2 \left( 1+ \nu \right)} \left( \frac{d}{dy} U - i \xi V\right),  \nonumber\\
S_{yy} &=& \frac{E}{ \left( 1+ \nu \right)} \left( \frac{d}{dy} V + \frac{\nu}{1-2 \nu} \left(-i \xi U + \frac{d}{dy} V \right) \right),  \nonumber \\
\bar{P} (\xi) &=& \int\limits_{-\infty}^{\infty} -P\left( x \right) e^{i \xi x} dx  \quad \text{and } \quad
\bar{v} (\xi) = \int\limits_{-\infty}^{\infty} v \left( x \right) e^{i \xi x} dx. \nonumber 
\end{eqnarray}
With this, we find the vertical displacement of the top surface in Fourier space to be
\begin{align}
%\label{eqn:V_xi}
V \left( \xi, 0 \right)=&  - \frac{2 \, \bar{P} \left( \xi \right)}{E^*} \, \frac{ \sinh^2{\xi \, h}}{\xi \left( \xi \, h + \sinh{\xi \, h } \cosh{\xi \, h} \right)} + \bar{v} \left( \xi \right) \, \frac {\sinh{ \xi \, h} + \xi\,h \cosh{ \xi\,h } }{ \xi \, h + \sinh{ \xi \, h} \cosh{ \xi \, h }}, \nonumber
\end{align}
and the normal traction at the bottom layer in Fourier space to be
\begin{align}
S_{yy} \left( \xi, h \right) &= \bar{P} \left( \xi \right) \, \frac {\sinh{ \xi \, h} + \xi\,h \cosh{ \xi\,h } }{ \xi \, h + \sinh{ \xi \, h} \cosh{ \xi \, h }}  + E^* \bar{v} \left( \xi \right) \, \frac{\xi}{2} \,  \frac {\sinh^2{ \xi \, h} - \xi^2 \,h^2 }{ \xi \, h + \sinh{ \xi \, h} \cosh{ \xi \, h }}, \nonumber
\end{align}
where $E^* = {E}/{\left(1-\nu^2 \right)}$.

\section{Displacement calculations}
\label{sec:Appendix_thinbeam_displacements}
\subsection{Clamped beam}
We approximate the vertical displacement of the bottom surface in a beam that is clamped at the ends as
\begin{equation}
\vartheta^{c}_b \left( \hat{\tau} \right) = d^{c} _0 + \sum_{n=1}^{M} d^{c}_n \cos \left( n \pi \hat{\tau} \right).
\label{clampedbeam_disp_guess}
\end{equation}
The above displacement filed is symmetric in $\hat{\tau}$, i.e., $\vartheta^{c}_b  \left( \hat{\tau} \right) = \vartheta^{c}_b  \left( -\hat{\tau}\right)$. The slope conditions at the ends, i.e., $d \vartheta^{c}_b / d \hat{\tau} =0$ at $\hat{\tau} = \pm 1 $, are automatically satisfied. We find the unknown coefficient $d^c_0$ in the displacement employing $\vartheta^{c}_b  \left( \pm 1 \right) =0$, to obtain
\begin{equation}
d^{c} _0 = \sum_{n=1}^{M}  d^{c} _n \left( -1 \right)^{n+1}.
\label{clampedbeam_disp_const}
\end{equation}
Combining \eqref{clampedbeam_disp_guess} and \eqref{clampedbeam_disp_const} yields
\begin{equation}
\vartheta^{c}_b  \left( \hat{\tau} \right) = \sum_{n=1}^{M}  d^{c} _n \left[ \left ( -1 \right) ^{n+1} + \cos \left( n \pi \hat{\tau} \right)  \right].
\end{equation}
Differentiating the above twice with respect to $\hat{\tau}$ provides 
%$\kappa^{c}  \left( \hat{\tau} \right)$ as
\begin{equation}
\kappa^{c}_b  \left( \hat{\tau} \right) = - \pi^2 \sum_{n=1}^{M} d^{c} _n n^2 \cos \left( n \pi \hat{\tau} \right).
\end{equation}

\subsection{Simply supported beam}
For a simply supported beam, we approximate the vertical displacement of the elastic layer's bottom surface in $-1 \le \hat{\tau} \le 1$, which should be symmetric in $ \hat{\tau} $, as
\begin{equation}
\vartheta^{s}_b  \left( \hat{\tau} \right) =  d^{s} _0 + \sum_{n=1}^{M} d^{s} _{2n-1} \sin \left\{ \frac{\left( 2n-1 \right) \pi \left( \hat{\tau}+1 \right)}{2} \right\} .
\end{equation}
The above displacement field also satisfies $d^2 \vartheta^{s} /d \hat{\tau}^2 =0$ at $\hat{\tau} = \pm 1 $, as appropriate for a simply supported beam. Satisfying $\vartheta^{s}  \left( \pm 1 \right) =0$, makes the unknown constant $d^s_0 = 0$. Hence, $\vartheta_b^{s}  \left( \hat{\tau} \right)$ becomes
\begin{equation}
\vartheta^{s}_b  \left( \hat{\tau} \right) = \sum_{n=1}^{M} d^{s} _{2n-1} \left( -1  \right)^{n-1} \cos \left\{ \frac{\left( 2n-1 \right) \pi \hat{\tau}}{2} \right\}.
\end{equation}
Differentiating the above twice we obtain the curvature $\kappa^{s}_b \left( \hat{\tau} \right) $ in $-1 \le \hat{\tau} \le 1$ as
\begin{eqnarray}
\kappa^{s}_b \left( \hat{\tau} \right) &=& - \sum_{n=1}^{M} d^{s}_{2n-1} \left( -1  \right)^{n-1} \left( \frac{ \left( 2n-1 \right) \pi }{2} \right)^2 \cos \left( \frac{\left( 2n-1 \right) \pi \hat{\tau}}{2} \right).
\end{eqnarray}
%and the slope is given by
%\begin{equation}
%\frac{d \vartheta^{s}_b  \left( \hat{\tau} \right)}{d \hat{\tau}} = \sum_{n=1}^{M} d^{s} _{2n-1} \frac{\left( 2n-1 \right) \pi}{2} \left( -1  \right)^{n} \sin \left( \frac{\left( 2n-1 \right) \pi \hat{\tau}}{2} \right)
%\end{equation}
Calculating the slope at the ends of the beam yields
\begin{equation}
\left. \frac{d \vartheta^{s}_b  \left( \hat{\tau} \right)}{d \hat{\tau}} \right|_{\hat{\tau}=1} = - \sum_{n=1}^{M} d^{s} _{2n-1} \frac{\left( 2n-1 \right) \pi}{2}.  
\end{equation}
Finally, finding the displacement in $1< \hat{\tau} < \infty $, by extending the beam along its slope at the end supports, we obtain
\begin{equation}
\vartheta^{s}_b  \left( \hat{\tau} \right) = \left. \frac{d \vartheta^{s}  \left( \hat{\tau} \right)}{d \hat{\tau}} \right|_{\hat{\tau}=1} \left( \hat{\tau} - 1 \right).
\end{equation}

%The curvature in $ \left| \hat{\tau} \right| \in \left[ 1, \hat{l}_1 \right] $ is zero. The curvature in $ \left| \hat{\tau} \right| \in \left[ \hat{l}_1, \hat{l}_2 \right]$ is given by
%\begin{eqnarray}
%\kappa^{s}_b \left( \hat{\tau} \right) &=&  \left. \frac{ d \vartheta^{s}_b \left( \hat{\tau} \right)}{d \hat{\tau}} \right|_{\hat{\tau}=1} \left[ 2  \frac{ d w_2 \left( \hat{\tau} \right)}{d \hat{\tau}} +  \left( \hat{\tau} - 1 \right) \frac{ d^2 w_2 \left( \hat{\tau} \right)}{d \hat{\tau}^2} \right]
%\end{eqnarray}

\section{Calculating $\bar{\vartheta}_b \left( \hat{\omega} \right)$ and $\bar{\kappa}_b \left( \hat{\omega} \right)$}
\label{sec:Appendix_thinbeam_vomega_calc}
As in Paper I, here too we consider the beams as linear elastic layers of infinite length. Thus, we employ mollifiers to find  $\hat{\vartheta}_b \left( \hat{\omega} \right)$ and $\hat{\kappa}_b \left( \hat{\omega} \right) $.  Therefore,  we write
\begin{flalign}
&& \hat{\vartheta} _b\left( \hat{\omega} \right)  &=  \int\limits_{-\infty}^{\infty} \vartheta_b \left( \hat{\tau} \right) W \left( \hat{\tau} \right) \cos{ \hat{\omega} \hat{\tau} } \, d \hat{\tau} \label{four_defn_v} && \\
\text{and} &&
\hat{\kappa}_b \left( \hat{\omega} \right)  &=  \int\limits_{-\infty}^{\infty} \frac{d^2}{d \hat{\tau}^2} \left[ \vartheta_b \left( \hat{\tau} \right) W \left( \hat{\tau} \right) \right] \cos{ \hat{\omega} \hat{\tau} } \, d \hat{\tau}, \label{four_defn_kappa} &&
\end{flalign}
where the mollifier \citep{Muthukumar}
\begin{subequations}
\label{defn:mollifiers}
\begin{flalign}
&& W \left( \hat{\tau} \right) &= \left\{ 
\begin{array}{ll}
1 &  \text{for } \left| \hat{\tau} \right| \leq  \left| \hat{l}_1 \right|, \\
 w_2 \left( \hat{\tau} \right)  & \text{for } \left|  \hat{l}_1 \right|< \left| \hat{\tau} \right| <  \left| \hat{l}_2 \right| \\
0 & \text{for } \left|  \hat{\tau} \right| \geq \left|  \hat{l}_2 \right|,
\end{array}
,\right. && \\
\text{and}
&& w_2 \left( \hat{\tau} \right) &= 
\frac{ \exp \left\{ {-1/\left( \hat{l}_2 - \hat{\tau} \right)^2} \right\} }{\exp \left\{ {-1/\left( \hat{l}_2 - \hat{\tau} \right)^2} \right\} + \exp \left\{ {-1/\left( \hat{\tau} - \hat{l}_1 \right)^2} \right\} }, &&
\end{flalign}
\end{subequations}
with $\hat{l}_1$ and $\hat{l}_2$ locating portions of the beam that are far away from the ends with $\hat{l}_2>\hat{l}_1>>1$. The mollifier $W \left( \hat{\tau} \right)$ is infinitely differentiable, and alters the displacement far from the beam's ends and makes it integrable. 
%where $\hat{l}_1$ and $\hat{l}_2$ are the portions of the beam that are far away from the ends with $\hat{l}_2>\hat{l}_1>>1$. These values are chosen in such a way that the displacements  of the overhangs will not effect the displacements over $\hat{\tau} \in \left[ 0,1 \right]$. Here $W \left( \hat{\tau} \right)$ is a mollifier \citep{Muthukumar} that alters the displacement, and is infinitely differentiable. 

\subsection{Clamped beam}
\label{sec:Appendix_thinfixedbeam_vomega_calc}
For a clamped beam, finding $\bar{\vartheta}^{c}_b  \left( \hat{\omega} \right)$ and $\bar{\kappa}^{c}_b  \left( \hat{\omega}\right) $, we obtain
\begin{equation}
\hat{\vartheta}^{c}_b  \left( \hat{\omega} \right) = \sum_{n=1}^{M}  d^{c} _n \hat{\beta}^{c} _{n} \left( \hat{\omega} \right)  \quad
\text{and} \quad
\hat{\kappa}^{c}_b  \left( \hat{\omega} \right) = \sum_{n=1}^{M} d^{c} _n \hat{\kappa}^{c} _{n} \left( \hat{\omega} \right), \label{thinbeam:V_omega_clampedbeam} 
\end{equation}
where
\begin{equation}
\hat{\beta}^{c} _{n} \left( \hat{\omega} \right) =  \frac{2 \left( -1 \right)^{n+1} \sin{ \hat{\omega} }} {\hat{\omega}} +  \frac{2 \hat{\omega}  \left( -1 \right)^{n+1} \sin{ \hat{\omega} }}{n^2 \pi^2 - \hat{\omega}^2}  \quad
\text{and} \quad
\hat{\kappa}^{c}_{n} \left(  \hat{\omega}  \right) =  - n^2 \pi^2 \, \frac{2 \hat{\omega}  \left( -1 \right)^{n+1} \sin{ \hat{\omega} }}{n^2 \pi^2 - \hat{\omega}^2}. \nonumber
\end{equation}
%\begin{flalign}
%&& \hat{\vartheta}^{c}_b  \left( \hat{\omega} \right) &= \int\limits_{-1}^{1} \vartheta^{c}_b  \left( \hat{\tau} \right) \cos { \hat{\omega} \hat{\tau} } \, d\hat{\tau} = \sum_{n=1}^{M}  d^{c} _n \hat{\beta}^{c} _{n} \left( \hat{\omega} \right)  \label{thinbeam:V_cheby_omega} && \\
%\text{and} &&
%\hat{\kappa}^{c}_b  \left( \hat{\omega} \right) &= \int\limits_{-1}^{1} \kappa^{c}_b  \left( \hat{\tau} \right) \cos{\hat{\omega} \hat{\tau}} \, d\hat{\tau} = \sum_{n=1}^{M} d^{c} _n \hat{\kappa}^{c} _{n} \left( \hat{\omega} \right), \label{thinbeam:ddV_cheby_omega}  &&
%\end{flalign}
%respectively, where
%\begin{flalign}
%&& \hat{\beta}^{c} _{n} \left( \hat{\omega} \right) &=  \frac{2 \left( -1 \right)^{n+1} \sin{ \hat{\omega} }} {\hat{\omega}} +  \frac{2 \hat{\omega}  \left( -1 \right)^{n+1} \sin{ \hat{\omega} }}{n^2 \pi^2 - \hat{\omega}^2}  &&
%\\
%\text{and} 
%&&  \hat{\kappa}^{c}_{n} \left(  \hat{\omega}  \right) &=  - n^2 \pi^2 \, \frac{2 \hat{\omega}  \left( -1 \right)^{n+1} \sin{ \hat{\omega} }}{n^2 \pi^2 - \hat{\omega}^2}. &&
%\end{flalign}

\subsection{Simply supported beam}
\label{sec:Appendix_thinsimplebeam_vomega_calc}
Evaluating $\bar{\vartheta}^{s}_b \left( \hat{\omega} \right)$ and $\hat{\kappa}{s}_b \left( \hat{\omega}\right)$  for a simply supported beam yields
\begin{equation}
\hat{\vartheta}^{s}_b \left( \hat{\omega} \right) = \sum_{n=1}^{M}  \tilde{d}^{s}_n \, \hat{\beta}^{s}_{n} \left( \hat{\omega} \right)  \quad
\text{and} \quad
\hat{\kappa}^{s}_b \left( \hat{\omega} \right) = \sum_{n=1}^{M} \tilde{d}^{s}_{n} \hat{\kappa}^{s}_{n} \left( \hat{\omega} \right),
\label{thinbeam:V_omega_simplebeam} 
\end{equation}
%\begin{eqnarray}
%\hat{\vartheta}^{s}_b \left( \hat{\omega} \right) &=& \sum_{n=1}^{M}  \tilde{d}^{s}_n \, \hat{\beta}^{s}_{n} \left( \hat{\omega} \right),  \\
%\text{and} \quad
%\hat{\kappa}^{s}_b \left( \hat{\omega} \right) &=& \sum_{n=1}^{M} \tilde{d}^{s}_{n} \hat{\kappa}^{s}_{n} \left( \hat{\omega} \right),
%\end{eqnarray}
where $\tilde{d}^{s}_n = d^{s}_{2n-1}$,
\begin{flalign}
&& \hat{\beta}^{s}_{n} \left( \hat{\omega} \right) = & \int\limits_{-1}^{1} \left( -1  \right)^{n-1} \cos \left( \frac{\left( 2n-1 \right) \pi \hat{\tau}}{2} \right) \, \cos{ \hat{\omega} \hat{\tau}  } \, d \hat{\tau} - \left( 2n-1 \right) \pi \int\limits_{1}^{\hat{l}_2} \left( \hat{\tau} - 1 \right) \, W \left( \hat{\tau} \right) \, \cos{ \hat{\omega}  \hat{\tau} } \, d \hat{\tau} \nonumber &&\\
\text{and}  && 
\hat{\kappa}^{s}_{n} \left( \hat{\omega}  \right) = &  -  \left( \frac{ \left( 2n-1 \right) \pi }{2} \right)^2 \, \int\limits_{-1}^{1} \left( -1  \right)^{n-1} \cos \left( \frac{\left( 2n-1 \right) \pi \hat{\tau}}{2} \right) \, \cos{ \hat{\omega}  \hat{\tau} } \, d\hat{\tau}  \nonumber && \\ && & - 2 \left( 2n-1 \right) \pi  \int\limits_{\hat{l}_1}^{\hat{l}_2} \frac{ d w_2 \left( \hat{\tau} \right)}{d \hat{\tau}} \cos{ \hat{\omega}  \hat{\tau} } \, d\hat{\tau} 
 - \left( 2n-1 \right) \pi  \int\limits_{\hat{l}_1}^{\hat{l}_2} \left( \hat{\tau} - 1 \right) \frac{ d^2 w_2 \left( \hat{\tau} \right)}{d \hat{\tau}^2} \, \cos{ \hat{\omega}  \hat{\tau} } \, d\hat{\tau}. \nonumber &&
\end{flalign}

%Finally, evaluating $\hat{\kappa} \left( \hat{\omega}\right)$ gives
%where
%\begin{eqnarray}
% \hat{\kappa}^{s}_{n} \left( \hat{\omega}  \right) &=&  -  \left( \frac{ \left( 2n-1 \right) \pi }{2} \right)^2 \, \int\limits_{-1}^{1} \left( -1  \right)^{n-1} \cos \left( \frac{\left( 2n-1 \right) \pi \hat{\tau}}{2} \right) \, \cos{ \hat{\omega}  \hat{\tau} } \, d\hat{\tau}  - 2 \left( 2n-1 \right) \pi  \int\limits_{\hat{l}_1}^{\hat{l}_2} \frac{ d w_2 \left( \hat{\tau} \right)}{d \hat{\tau}} \cos{ \hat{\omega}  \hat{\tau} } \, d\hat{\tau}, \nonumber \\
% && - \left( 2n-1 \right) \pi  \int\limits_{\hat{l}_1}^{\hat{l}_2} \left( \hat{\tau} - 1 \right) \frac{ d^2 w_2 \left( \hat{\tau} \right)}{d \hat{\tau}^2} \, \cos{ \hat{\omega}  \hat{\tau} } \, d\hat{\tau}, \nonumber
%\end{eqnarray}

Finally, employing \eqref{thinbeam:V_omega_clampedbeam}  and \eqref{thinbeam:V_omega_simplebeam}, we may write $\bar{\vartheta}_b \left( \hat{\omega} \right)$ and $\hat{\kappa}_b \left( \hat{\omega} \right) $, respectively,  as
\begin{gather}
\hat{\vartheta}_b  \left( \hat{\omega} \right) = \sum_{n=1}^{M}  d_n \hat{\beta}_{n} \left( \hat{\omega} \right)  \quad \text{and} \quad
\bar{\kappa}_b  \left( \hat{\omega} \right) =  \sum_{n=1}^{M} d_n \hat{\kappa}_{n} \left( \hat{\omega} \right). \label{thinbeams:v_omega_derivation} 
\end{gather}

%\newpage
% Bibliography
\singlespacing
\providecommand{\noopsort}[1]{}\providecommand{\singleletter}[1]{#1}%

%%\bibliographystyle{apa}
%\bibliographystyle{plainnat}
%\bibliography{refs_short,refs}

%\bibliography{mybib}% Produces the bibliography via BibTeX.

\begin{thebibliography}{}

%Type = Book
\bibitem[\protect\astroncite{Alexandrov and Pozharskii}{2001}]{alexandrov2001three}
Alexandrov, A., Pozharskii, D. (2001).
\newblock \textit{Three-dimensional contact problems}.
\newblock {Kluwer Academic Publishers, Dordrecht, The
  Netherlands}.
  
%Type = Article
\bibitem[\protect\astroncite{Arul and Ghatak}{2008}]{Arul2008bioinspired}
Arul, E.~P., Ghatak, A. (2008).
\newblock {Bioinspired design of a hierarchically structured
  adhesive}.
\newblock {\em Langmuir}, \textbf{25}: 611--617.
  
%Type = Article
\bibitem[\protect\astroncite{Barthel and Perriot} {2007}]{barthel2007adhesive}
Barthel, E., Perriot, A. (2007).
\newblock {Adhesive contact to a coated elastic substrate}.
\newblock {\em J. Phys. D: Appl. Phys.}, \textbf{40}: 1059--1067.
  
%Type = Misc
\bibitem[\protect\astroncite{Chatterjee}{2002}]{Anindya}
Chatterjee, A. (2002).
\newblock {\em Lecture notes: An elementary continuation technique}.
{\url{http://home.iitk.ac.in/~anindya/continuation.pdf}}.
  
%Type = Article
\bibitem[\protect\astroncite{Chaudhury et~al.}{1996}]{chaudhury1996adhesive}
Chaudhury, M.~K., Weaver, T., Hui, C., Kramer, E. (1996).
\newblock {Adhesive contact of cylindrical lens and a flat
  sheet}.
\newblock {\em J. Appl. Phys.}, \textbf{80}: 30--37.
  
%Type = Book
\bibitem[\protect\astroncite{Crandall et~al.}{2008}]{Crandall2008mech}
Crandall, S.~H., Dahl, N.~C., Lardne, T.~J. (2008).
\newblock {\em Mechanics of Solids}.
\newblock {Tata McGraw-Hill Education Pvt. Ltd., New Delhi,
  India}.
  
%Type = Article
\bibitem[\protect\astroncite{Dalmeya et~al.}{2012}]{Dalmeya2012contact}
Dalmeya, R., Sharma, I., Upadhyay, C., Anand, A. (2012).
\newblock {Contact of a rigid cylindrical punch with an adhesive
  elastic layer}.
\newblock {\em J. Adhes.}, \textbf{88}: 1--31.

%Type = Article
\bibitem[\protect\astroncite{Derjaguin}{1934}]{derjaguin1934untersuchungen}
Derjaguin, B.~V. (1934).
\newblock {Untersuchungen {\"u}ber die reibung und adh{\"a}sion, iv}.
\newblock {\em Colloid. Polym. Sci.}, 
\textbf{69}(2): 155--164.

%Type = Article
\bibitem[\protect\astroncite{Derjaguin et~al.}{1975}]{derjaguin1975effect}
Derjaguin, B.~V., Muller, V.~M., Toporov, Y.~P. (1975).
\newblock {Effect of contact deformations on the adhesion of particles}.
\newblock {\em J. Colloid Interface Sci.}, \textbf{53}: 314--326.
  
%Type = Book
\bibitem[\protect\astroncite{Galin and Gladwell}{2008}]{galin2008contact}
Galin, L.~A., Gladwell, G.~M.~L. (2008).
\newblock {\em Contact Problems: The legacy of L.A. Galin}.
\newblock {Solid Mechanics and Its Applications, Springer
  Netherlands}.
  
%Type = Book
\bibitem[\protect\astroncite{Gladwell}{1980}]{Gladwell1980contact}
Gladwell, G.~M.~L.(1980).
\newblock {\em Contact Problems in the Classical Theory of
  Elasticity}.
\newblock {Sijthoff \& Noordhoff Publishers, Alphen aan den
  Rijn, The Netherlands}.
  
%Type = Book
\bibitem[\protect\astroncite{Goryacheva}{1998}]{goryacheva1998contact}
Goryacheva, I.~G. (1998).
\newblock {\em Contact Mechanics in Tribology}.
\newblock {Springer, Netherlands}.
  
%Type = Article
\bibitem[\protect\astroncite{Hiller}{1976}]{Hiller1976}
Hiller, U.~J. (1976).
\newblock {Comparative studies on the functional morphology of
  two gekkonid lizards}.
\newblock {\textit Bombay Nat. Hist. Soc.}, \textbf{73}: 278 -- 282.
  
%Type = Book
\bibitem[\protect\astroncite{Hills et~al.}{1993}]{Hills1993mechanics}
Hills, D.~A., Nowell, D., Sackfield, A. (1993).
\newblock {\em Mechanics of Elastic Contacts}.
\newblock {Butterworth-Heinemann, Oxford, UK}.

%Type = Book
\bibitem[\protect\astroncite{Johnson}{1985}]{Johnson1985contact}
Johnson, K.~L. (1985).
\newblock {\em Contact Mechanics}.
\newblock {Cambridge U. Press, Cambridge, UK}.

%Type = Article
\bibitem[\protect\astroncite{Johnson et~al.}{1971}]{johnson1971surface}
Johnson, K.~L., Kendall, K., Roberts, A.~D. (1971).
\newblock {Surface energy and the contact of elastic solids}.
\newblock {\em Proc. R. Soc. Lond., A}, \textbf{324}: 301--313.
  
%Type = Article
\bibitem[\protect\astroncite{Johnston et~al.}{2014}]{johnston2014mechanical}
Johnston, I.~D., McCluskey, D.~K., Tan, C.~K.~L., Tracey, M.~C. (2014).
\newblock {Mechanical characterization of bulk {S}ylgard 184 for
  microfluidics and microengineering}.
\newblock {\em J. Micromech. Microeng.}, \textbf{24}: 035017.
  
%Type = Book
\bibitem[\protect\astroncite{Kanninen and Popelar}{1985}]{kanninen1985advanced}
Kanninen, M.~F., Popelar, C.~L. (1985).
\newblock {\em Advanced Fracture Mechanics}.
\newblock {Oxford U. Press, New York, USA}.

%Type = Article
\bibitem[\protect\astroncite{Keer and Miller}{1983}]{Keer1983smooth}
Keer, L.~M., Miller, G.~R. (1983).
\newblock {Smooth indentation of finite layer}.
\newblock {\em J. Eng. Mech.}, \textbf{109}: 706--717.
  
%Type = Article
\bibitem[\protect\astroncite{Keer and Schonberg}{1986}]{keer1986smooth}
Keer, L.~M., Schonberg, W.~P. (1986).
\newblock {Smooth indentation of an isotropic cantilever beam}.
\newblock {\em Int. J. Solids Struct.}, \textbf{22}: 87--106.
  
%Type = Article
\bibitem[\protect\astroncite{Keer and Silva}{1970}]{keer1970bending}
Keer, L.~M., Silva, M.~A.~G. (1970).
\newblock {Bending of a cantilever brought gradually into
  contact with a cylindrical supporting surface}.
\newblock {\em Int. J. Mech. Sci.}, \textbf{12}: 751--760.
  
%Type = Article
\bibitem[\protect\astroncite{Kim et~al.}{2014}]{Kim2014}
Kim, J.~H., Ahn, Y.~J., Jang, Y.~H., Barber, J.~R. (2014).
\newblock {Contact problems involving beams}.
\newblock {\em Int. J. Solids Struct.}, \textbf{51}: 4435--4439.
  
%Type = Book
\bibitem[\protect\astroncite{Mason and Handscomb}{2003}]{mason2003book}
Mason, J.C., Handscomb, D.C. (2003).
\newblock {\em Chebyshev Polynomials}.
\newblock {Chapman \& Hall/CRC, Boca Raton, Florida, USA}.

%Type = Article
\bibitem[\protect\astroncite{Maugis}{1992}]{Maugis1992adhesion}
Maugis, D. (1992). 
\newblock {Adhesion of spheres: the {JKR-DMT} transition using a
  {D}ugdale model}.
\newblock {\em J. Colloid Interface Sci.}, \textbf{150}: 243--269.
  
%Type = Misc
\bibitem[\protect\astroncite{Muthukumar}{2016}]{Muthukumar}
Muthukumar, T. (2016). 
\newblock {\em Lecture notes: Sobolev spaces and applications}.
{\url{http://home.iitk.ac.in/~tmk/courses/mth656/main.pdf}}.
  
%Type = Book
\bibitem[\protect\astroncite{Polyanin and Manzhirov}{2008}]{polyanin2008handbook}
Polyanin, A., Manzhirov, A. (2008).
\newblock {\em Handbook of Integral Equations: Second Edition}.
\newblock {Chapman \& Hall/CRC, Boca Raton, Florida, USA}.

%Type = Book
\bibitem[\protect\astroncite{Press et~al.}{1992}]{press1992numerical}
Press, W.~H., Teukolsky, S.~A., Vetterling, W.~T., Flannery, B.~P. (1992).
\newblock {\em Numerical Recipes in C: The Art of Scientific
  Computing}.
\newblock {Cambridge U. Press India Pvt. Ltd., New Delhi,
  India}.
 
%Type = Article
\bibitem[\protect\astroncite{Punati et~al.}{2017}]{Punati2017}
Punati, V.~S., Sharma, I., Wahi, P. (2017).
\newblock {Contact mechanics of adhesive beams. Part I: Moderate indentation}.
\newblock {{\url{arXiv:1707.07543} [cond-mat.soft]} }%, \textbf{35}: 379--386.
%\newblock {\em J. Appl. Mech.}, \textbf{35}: 379--386.
   
%Type = Article
\bibitem[\protect\astroncite{Rice}{1968}]{rice1968path}
Rice, J.~R. (1968).
\newblock {A path independent integral and the approximate
  analysis of strain concentration by notches and cracks}.
\newblock {\em J. Appl. Mech.}, \textbf{35}: 379--386.
  
%Type = Book
\bibitem[\protect\astroncite{Sadd}{2005}]{Sadd2010elasticity}
Sadd, M.~H. (2005).
\newblock {\em Elasticity: Theory, Applications, and Numerics}.
\newblock {Elsevier India, New Delhi, India}.

%Type = Article
\bibitem[\protect\astroncite{Sankar and Sun}{1983}]{Sankar1983}
Sankar, B.~V., Sun, C.~T. (1983).
\newblock {Indentation of a beam by a rigid cylinder}.
\newblock {\em Int. J. Solids Struct.}, \textbf{19}: 293--303.
  
%Type = Book
\bibitem[\protect\astroncite{Sneddon}{1995}]{Sneddon1995fourier}
Sneddon, I.~N., (1995).
\newblock {\em Fourier Transforms}.
\newblock {Dover Publications, New York, USA}.

%Type = Article
\bibitem[\protect\astroncite{Sun and Sankar}{1985}]{sun1985smooth}
Sun, C.~T., Sankar, B.~V. ( 1985).
\newblock {Smooth indentation of an initially stressed
  orthotropic beam}.
\newblock {\em Int. J. Solids Struct.}, \textbf{21}: 161--176.
  
%Type = Book
\bibitem[\protect\astroncite{Timoshenko and Goodier}{1970}]{Timoshenko1970}
Timoshenko, S.~P., Goodier, J.~N. (1970).
\newblock {\em Theory of Elasticity}.
\newblock {McGraw-Hill, Singapore}.

\end{thebibliography}
% End document
\end{document}